\documentclass[aps,amsmath,twocolumn,amssymb,floatfix,showpacs,superscriptaddress,longbibliography]{revtex4-1}
\pdfoutput=1
\usepackage[dvips]{graphics}
\usepackage{bm}
\usepackage{float}
\usepackage{epsfig}
\usepackage{enumerate}
\usepackage{subfigure}
\usepackage{amsmath}
\usepackage{color}
\usepackage{braket}
\usepackage{graphicx}
\usepackage{float}
\usepackage[colorlinks=true,linktoc=page,linkcolor=blue,citecolor=magenta,urlcolor=red]{hyperref}

\newcommand{\vect}[1]{\bm{\mathrm{#1}}}

\newcommand{\ie}{{\it i.e.,\,\,}}
\newcommand{\eg}{{\it e.g.,~}}

\newcommand\bea{\begin{eqnarray}}
\newcommand\eea{\end{eqnarray}}
\newcommand\beq{\begin{equation}}  
\newcommand\eeq{\end{equation}}

\newcommand{\non}{\nonumber}

\begin{document} 

\title{Floquet Second Order Topological Superconductor based on Unconventional Pairing}  
 \author{Arnob Kumar Ghosh}
\email{arnob@iopb.res.in}
\affiliation{Institute of Physics, Sachivalaya Marg, Bhubaneswar-751005, India}
\affiliation{Homi Bhabha National Institute, Training School Complex, Anushakti Nagar, Mumbai 400094, India}
\author{Tanay Nag}
\email{tnag@physik.rwth-aachen.de}
\affiliation{SISSA, via Bonomea 265, 34136 Trieste, Italy}
\affiliation{Institut f\"ur Theorie der Statistischen Physik, RWTH Aachen University, 52056 Aachen, Germany}
\author{Arijit Saha}
\email{arijit@iopb.res.in}
\affiliation{Institute of Physics, Sachivalaya Marg, Bhubaneswar-751005, India}
\affiliation{Homi Bhabha National Institute, Training School Complex, Anushakti Nagar, Mumbai 400094, India}

\begin{abstract}
We theoretically investigate the Floquet generation of second-order topological superconducting (SOTSC) phase in the high-temperature platform both in two-dimension (2D) and three-dimension (3D). Starting from a $d$-wave superconducting pairing gap, we periodically kick the mass term to engineer the dynamical SOTSC phase within a specific range of the strength of the drive. Under such dynamical breaking of time-reversal symmetry (TRS), we show the emergence of the \textit{weak} SOTSC phase, harboring eight corner modes \ie two zero-energy Majorana per corner, with vanishing Floquet quadrupole moment. On the other hand, our study interestingly indicates that upon the introduction of an explicit TRS breaking Zeeman field, the \textit{weak} SOTSC phase can be transformed into \textit{strong} SOTSC phase, hosting one zero-energy Majorana mode per corner, with quantized quadrupole moment. We also compute the Floquet Wannier spectra that further establishes the \textit{weak} and \textit{strong} nature of these phases. We numerically verify our protocol computing the exact Floquet operator in open boundary condition and then analytically validate our findings with the low energy effective theory (in the high-frequency limit). The above protocol is applicable for 3D as well where we find one dimensional (1D) hinge mode in the SOTSC phase. We then show that these corner modes are robust against moderate disorder and the topological invariants continue to exhibit quantized nature until disorder becomes substantially strong. The existence of zero-energy Majorana modes in these higher-order phases is guaranteed by the anti-unitary spectral symmetry. 
\end{abstract}

\maketitle

\section{Introduction}{\label{sec:I}}

The advent of topological superconductors (TSCs), harboring Majorana zero modes (MZMs) at their boundary, have generated immense interest in the quantum condensed matter community from both theoretical and experimental perspectives in the last few years~\cite{Kitaev_2001,qi2011topological,hasan2010colloquium,das2012zero,Deng1557}. Due to the non-Abelian statistics of the MZMs, they are proposed to be suitable candidates for the topological quantum computation~\cite{Ivanov2001,nayak08}. There have been multitudinous proposals based on heterostructure setup of materials with strong spin-orbit coupling such as topological insulator, semiconductor thin films, and nanowires with the proximity induced superconductivity and Zeeman field, that provide an efficient platform to realize the MZMs~\cite{Fu2008,Sau2010,Lutchyn10,Hughes2010,Oreg2010}.

In recent times, the concept of conventional bulk-boundary correspondence of topological phases in various topological systems such as topological insulators (TIs), Dirac semimetals (DSMs), Weyl semimetals (WSMs), TSCs, etc. have been generalized to higher-order topological (HOT) phases~\cite{benalcazar2017,benalcazarprb2017,Song2017,Langbehn2017,schindler2018,Khalaf2018,Geier2018,Franca2018,Zhu2018,Liu2018,Yan2018,wang2018higher,WangWeak2018,Ezawakagome,Roy2019,Trifunovic2019,ZengPRL2019,Zhang2019,Volpez2019,YanPRB2019,Ghorashi2019,GhorashiPRL2020,Sumathi2020,Wu2020,jelena2020HOTSC,BitanTSC2020,SongboPRR12020,SongboPRR22020,SongboPRB2020,kheirkhah2020vortex,PlekhanovArxiv2020}. To be precise, an $n^{\rm th}$ order $d$-dimensional TI/ TSC exhibit gapless modes on their $(d-n)$-dimensional boundary ($n \le d$). In particular, a 3D second (third) order TIs/TSCs  
are characterized by the presence of one (zero)-dimensional hinge (corner) modes, whereas, 2D second order topological insulators (SOTIs)/SOTSCs exhibit zero-dimensional (0D) corner modes only. 
There have been a few experimental proposals to realize 2D SOTIs hosting corner modes, based on acoustic materials~\cite{XueAcousticKagome}, photonic crystals~\cite{PhotonicChen,PhotonicXie}, 
and electrical circuit~\cite{Imhof2018} setups.

Given the growing interest of the research community in this field, the non-equilibrium Floquet engineering emerges as a fertile hunch to generate the dynamical analog of the HOT phases. It has been shown that a trivial phase can be made topologically non-trivial with suitable periodic driving~\cite{lindner2011floquet,Dora12,Rudner2013,Thakurathi2013,rechtsman2013photonic,maczewsky2017observation,Eckardt2017}. The time translational symmetry of the problem causes the Floquet topological phase to host dissipationless dynamical topological boundary modes. The resulting bulk-boundary correspondence here becomes intriguing in the presence of the extra-temporal dimension. 
This non-equilibrium version of generating topological phase have been applied to HOTIs/HOTSCs 
resulting in Floquet HOTIs (FHOTIs)~\cite{Bomantara2019,Nag19,YangPRL2019,Seshadri2019,chaudhary2019phononinduced,Martin2019,Jelena2019,Ghosh2020,Huang2020,HuPRL2020,BomantaraPRB2020,YangPRR2020,Nag2020,
ApoorvTiwari2020,ZhangYang2020,ghosh2020floquet,bhat2020equilibrium,GongArxiv2020}. However, the search for Floquet HOTSCs (FHOTSCs) is still at its initial stage \cite{chaudhary2019phononinduced,Jelena2019,PhysRevLett.123.167001,ghosh2020floquet,RWBomantaraPRR2020} even from theoretical point of view.

In a very recent work, we show that the Floquet SOTSC phase can be engineered by kicking the time-reversal symmetry (TRS) breaking perturbation while the underlying static, $s$-wave proximitized parent system is in a trivial phase~\cite{ghosh2020floquet}. In this article, we aim to generate Floquet SOTSC phase, hosting MZMs at the corner (hinges) for a 2D (3D) system, 
by suitably tuning some other parameter of the underlying first order TI such as the on-site mass term. Given the recent experimental advancements in Floquet systems based on 
solid state systems~\cite{Wang453}, meta-materials \cite{rechtsman2013photonic,peng2016experimental,Experiment2014,fleury2016floquet,maczewsky2017observation}, we believe that our question regarding the Floquet generation of MZMs in HOTSC phase is timely and pertinent. More importantly, the kicking of on-site term has been able to demonstrate a variety of interesting theoretical observation such as, Floquet topological insulator and superconductor~\cite{Thakurathi2013,Seshadri2019,mass_drive_FTS,nag20} dynamical localization~\cite{mass_kick_DL1,mass_kick_DL2}, survival probability of initial state~\cite{refId0, ps2} and thus motivates us to consider the on-site mass kicking dynamical protocol in order to obtain the desired outcome. However, the dynamical manipulation of mass term is 
yet to be experimentally realized. Moreover, the 2D TIs have been experimentally realized at high-temperature $100$ K~\cite{Wu76, Qian1344}, paving the way to explore the high-temperature platform 
of MZMs.

To investigate the above mentioned possibility, we begin with a 2D TI in close proximity to a $d$-wave superconductor with unconventional pairing.  Here, we consider the external drive to be periodically kicking the onsite mass term of the first order TI. This results in a \textit{weak} Floquet SOTSC (FSOTSC) phase where two MZMs are localized at each corner of the 2D sample. Here the TRS is dynamically broken in the effective Floquet Hamiltonian. Remarkably, this degeneracy of two Majorana corner modes (MCMs) per corner can be lifted by incorporating an explicit TRS broken Zeeman field leading to a~\textit{strong} FSOTSC phase with one MZM per corner. We numerically study the exact Floquet operator to obtain the above mentioned results that are further verified by the low energy effective theory
in the high-frequency approximation. We characterize the topological nature of these phases by the appropriate topological invariants as Floquet quadrupole moment (FQM) and Floquet Wannier spectra (FWS). We then extend our proposal to 3D where also we find the \textit{weak} (\textit{strong}) FSOTSC phase in absence (presence) of a homogeneous Zeeman field. Here, the corrresponding HOT
phase hosts 1D Majorana hinge modes (MHMs). Furthermore, we investigate the effect of disorder on these Majorana corner modes (MCMs) and we find that these modes are robust against moderate
disorder strength. The existence of the MZMs are protected by anti-unitary spectral symmetry in all of the above cases. 

\begin{figure}
	\centering
	\subfigure{\includegraphics[width=0.48\textwidth]{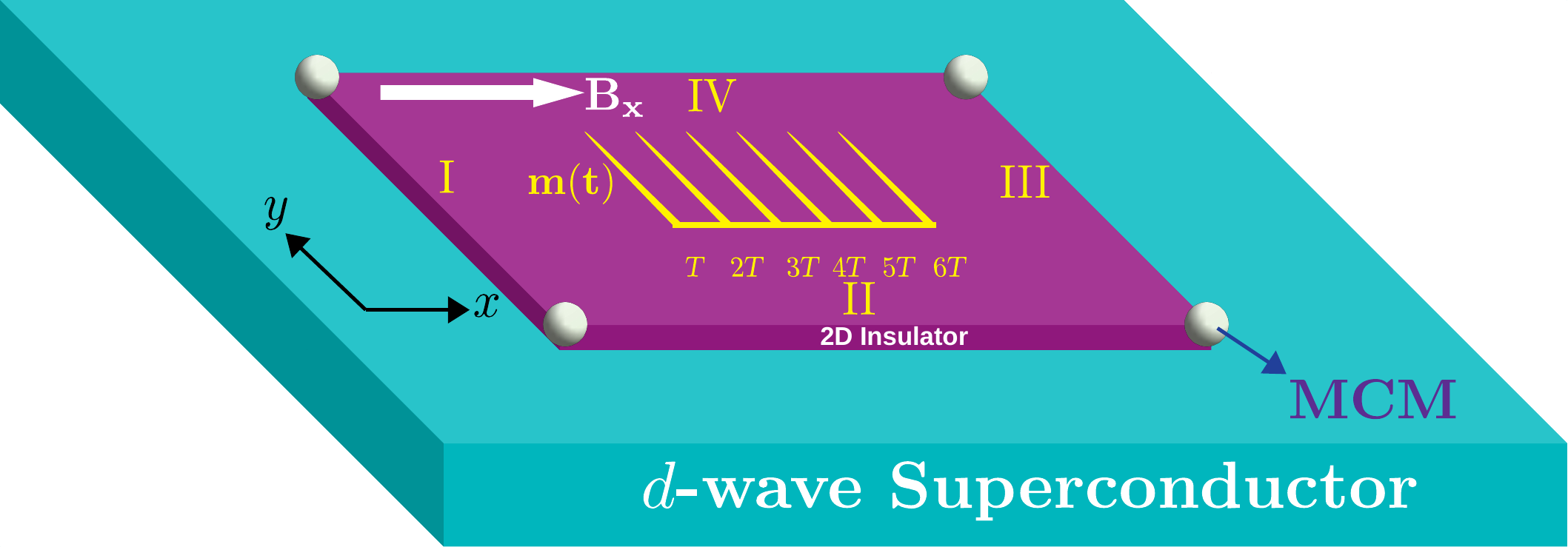}}
	\caption{(Color online) Schematic of our setup is demonstrated in presence of a periodic kick (yellow, light grey) in the mass term as an external drive. Here, a 2D TI (violet, dark grey) is placed 
	in close proximity to a bulk high-temperature $d$-wave superconductor (cyan, light grey). MCMs are shown by circular dots (grey) at the four corners of the 2D sample and the four edges of the    
	TI are denoted by I, II, III, IV. 
	}
	\label{Schematics}
\end{figure}
The remainder of the paper is organized as follows. In Sec. \ref{sec:II}, we introduce our model and the driving protocol. We illustrate the emergence of Floquet MCMs in the local density of states (LDOS) 
behavior. We resort to low energy edge theory to understand the emergence of the corner mode solutions. First, we solve these edge Hamiltonian to derive the corner mode solution. We characterize the MCMs using FQM and FWS. In Sec. \ref{sec:III}, we provide a protocol to generate 3D FSOTSC. We show the appearance of MHMs in rod geometry. We use low energy surface theory to confirm the 
existence of these hinge modes. Then we provide the analytical solutions for the hinge modes by solving the surface Hamiltonian. In Sec. \ref{sec:IV}, we study the effect of disorder on MZMs and 
show that they are robust against a finite amount of disorder.  Finally in Sec. \ref{sec:V}, we summarize and conclude our results.

\section{Majorana Corner modes in 2D}\label{sec:II}
 In this section, we discuss in detail the model Hamiltonian of our set-up along with the driving protocol, the emergence of Floquet MCMs by numerically computing the exact Floquet evolution operator,
 low energy effective edge theory in the high-frequency regime and analytical understanding of the MCMs solutions therein. 

\subsection{Model Hamiltonian and Driving Protocol}

\subsubsection{Model}

We consider the model of a 2D TI proximitized with $d$-wave superconductivity on a square lattice~\cite{Yan2018}. The experimental realization of high-temperature 2D TI allows us to consider proximity induced high-temperature superconductor~\cite{Wu76, Qian1344}. It acquires the following form while written in Bogoliubov-de Gennes (BdG) basis $H_{\rm BdG}=\frac{1}{2}\sum_{{\vect{k}}}\Psi^\dagger_{\vect{k}} \mathcal{H}_0({\vect{k}})\Psi_{\vect{k}}$, with $\Psi_{\vect{k}}=\left(c_{{\vect{k}},a\uparrow}, c_{-{\vect{k}},a\uparrow}^\dagger, c_{{\vect{k}},a\downarrow}, c_{-{\vect{k}},a\downarrow}^\dagger, c_{{\vect{k}},b\uparrow}, c_{-{\vect{k}},b\uparrow}^\dagger, c_{{\vect{k}},b\downarrow}, c_{-{\vect{k}},b\downarrow}^\dagger\right)^{\tilde{T}}$ and $\mathcal{H}_0({\vect{k}})$, is given by
\begin{eqnarray}
\mathcal{H}_0({\vect{k}})&=&\epsilon(\vect{k}) \Gamma_1+ \lambda_x \sin k_x \Gamma_2+ \lambda_y \sin k_y \Gamma_3+\Delta(\vect{k}) \Gamma_4 \non \\
&&+B_x\Gamma_5 \equiv {\vect{N}}(\vect{k})\cdot {\bm \Gamma}\ ,
\label{ham1}
\end{eqnarray}
Here, $t_{x,y}$ and $\lambda_{x,y}$ represent the nearest-neighbour hopping and spin-orbit coupling respectively, $\epsilon(\vect{k})=\left(m_0-t_x \cos k_x -t_y \cos k_y \right)$, $\Delta(\vect{k})=\Delta_0 \left(\cos k_x-\cos k_y \right)$ is the $d$-wave superconducting paring term, $m_0$ is the crystal-field splitting energy. In Eq.(\ref{ham1}), $\Gamma_1=\sigma_z \tau_z$, $\Gamma_2=\sigma_x s_z$, 
$\Gamma_3=\sigma_y \tau_z$, $\Gamma_4= s_y \tau_y$ and $\Gamma_5= s_x \tau_z$. The three Pauli matrices ${\bm \sigma}$, ${\bm s}$ and ${\bm \tau}$ act on orbital $(a,b)$, spin $(\uparrow, \downarrow)$ and particle-hole degrees of freedom respectively. When $B_x=0$, the system respects TRS \ie $\mathcal{T}^{-1} \mathcal{H}_0(\vect{k}) \mathcal{T}=\mathcal{H}_0(-\vect{k})$ with $\mathcal{T}=is_y \mathcal{K}$, $\mathcal{K}$ being the complex-conjugation operator. The  Hamiltonian continues to preserve the  particle-hole symmetry (PHS) $\mathcal{C}^{-1} \mathcal{H}_0 (\vect{k}) \mathcal{C}=-\mathcal{H}_0(-\vect{k})$ with $\mathcal{C}= \tau_x \mathcal{K}$ for $B_x \ne 0$. Below we discuss the properties of our model at length. 

First, we discuss the topological nature associated with the Hamiltonian (Eq.(\ref{ham1})) for $\Delta(\vect{k})=0$ and $B_x=0$. This model supports topological phase, hosting gapless helical edge modes,  when $|m_0|<t_x +t_y$. For $|m_0|>t_x +t_y$, the model becomes trivially gapped. Upon the introduction of $d$-wave pairing only \ie $\Delta(\vect{k})\ne 0$ and $B_x=0$, the 1D massless Dirac fermions representing the edge modes of TI phase, become gapped by the induced unconventional superconducting pairing. However, the specific nature of pairing symmetry causes the Dirac mass to change signs at the corners.  This in turn generates MCMs, a signature of SOTSC phase for TRS invariant TSC system, as domain-wall excitations when $|m_0|<t_x +t_y$. Here, one can observe Majorana Kramers pair (MKPs) (\ie two Majoranas per corner are time-reversal partners of each other) pinned at zero-energy which are protected by the TRS. When $|m_0|>t_x +t_y$, the underlying TI becomes non-topological and $d$-wave pairing cannot lead to MKPs anymore as the system becomes gapped. Now, in presence of TRS breaking Zeeman field $B_x\ne 0$, the degeneracy of MKPs can be lifted by destroying one mode in the pair at the corner. Therefore, the Hamiltonian supports MZMs with one Majorana per corner for $\Delta(\vect{k})\ne 0$ and $B_x\ne 0$ while $|m_0|<t_x +t_y$. The magnetic field allows us to tune the bulk gap which is introduced by the superconducting order parameter $\Delta(\vect{k})$. Given this background, it would be interesting to study the emergence of MCMs starting from the underlying non-topological phase $|m_0|>t_x +t_y$ by periodically kicking the mass with a finite amplitude and frequency.

\subsubsection{Driving Protocol and Floquet Operator}
Here we introduce our driving protocol in the form of periodic kick in the mass term as discussed before
\begin{eqnarray}
m(t)&=& m_1 \sum_{r=1}^{\infty} \delta(t-rT) \ , 
\label{kick1}
\end{eqnarray}
with $r$ being an integer. This driving protocol allows one to write the exact Floquet operator in the following way using the time ordered (${\rm \bf TO}$) notation as
\begin{eqnarray}
U(T)&=&{\rm \bf TO} \exp \left[-i\int_{0}^{T}dt\left(\mathcal{H}_0({\vect{k}})+m(t)\Gamma_1\right)\right] \nonumber \\
&=& \exp(-i \mathcal{H}_0({\vect{k}}) T)~\exp(-i m_1 \Gamma_1)\ .
\label{fo}
\end{eqnarray}
Here, $T$, $m_{1}$ are the period and amplitude of the drive respectively. The above decomposition essentially means that the system is freely evolved between two subsequent kicks. 
The Floquet operator $U(T)$ can be written in a more compact form as follows
\begin{eqnarray}
U(T)&=&C_T\left(p-i q \Gamma_1\right)-i S_T\sum_{j=1}^{5}\left(n_j\Gamma_j+m_j\Gamma_{j1}\right)\ ,
\end{eqnarray}
where, $C_T=\cos(\left|{\vect{N}}(\vect{k})\right| T)$, $S_T=\sin(\left|{\vect{N}}(\vect{k})\right| T)$, $p=\cos m_1$, $q=\sin m_1$, $n_j=\frac{N_j(\vect{k})\cos m_1}{\left|{\vect{N}}(\vect{k})\right|}$, $m_j=\frac{N_j(\vect{k})\sin m_1}{\left|{\vect{N}}(\vect{k})\right|}$ and $\Gamma_{j1}=\frac{1}{2i}\left[\Gamma_j,\Gamma_1\right]$ with $j=2,3,4,5$. One can thus obtain the general form of the effective Hamiltonian as
\begin{widetext}
\begin{eqnarray}
	H_{\rm Flq} =\frac{\xi_{\vect{k}}}{\sin\xi_{\vect{k}}T}\Bigg[\sin(\left|{\vect{N}}(\vect{k})\right| T) \cos m_1 \sum_{j=1}^{5}r_j \Gamma_j +\cos(\left|{\vect{N}}(\vect{k})\right| T) \sin m_1 \Gamma_1 +\sin(\left|{\vect{N}}(\vect{k})\right| T) \sin m_1 \sum_{j=2}^{5} r_j \Gamma_{j1} \Bigg]\ ,
\end{eqnarray}
\end{widetext}
with $\xi_{\vect{k}}=\frac{1}{T}\cos^{-1} \left[\cos(\left|{\vect{N}}(\vect{k})\right| T) \cos m_1\right]$, $r_j=\frac{N_j(\vect{k})}{\left|{\vect{N}}(\vect{k})\right|}$. In the high-frequency limit \ie~$T \rightarrow 0$ and small amplitude of drive \ie~$m_1 \rightarrow 0$, one can neglect the higher order terms in $T$ and $m_1$. Thus, the effective Hamiltonian in that limit reads as
\begin{equation}
H_{\rm Flq}\approx \mathcal{H}_0(\vect{k})+\frac{m_1}{T} \Gamma_1+ m_1 \sum_{j=2}^{5} r_j \Gamma_{j1} \ .
\label{Highfreq2d}
\end{equation}

In Eq.(\ref{Highfreq2d}), terms associated with $\Gamma_{j1}$ are originated due to the driving and not present in the static Hamiltonian (Eq.(\ref{ham1})). It is noteworthy that these extra $r_j\Gamma_{j1}$ terms break TRS $\mathcal{T}$ in the system. Consequently, the above Hamiltonian breaks TRS $\mathcal{T}$ even when $B_x=0$. This reflects the fact that the TRS can be broken dynamically by mass kicking while the static perturbation respects the TRS. Therefore, it would be important to study the effect of these terms in the dynamics with and without the magnetic field. The former situation can be referred to as the explicit breaking of TRS while the later is related to the dynamical breaking of TRS.  At the outset, we would like to comment from  Eq.(\ref{Highfreq2d}) that the mass term gets renormalized by the driving $m_0 \to m_0 + m_1/T$. Therefore, the topological phase boundary is thus got renormalized. Hence, MCMs can be found when $|m_0 + m_1/T|< t_x +t_y$ with $\Delta(\vect{k})\ne 0$ and $B_x\ne 0$. Similar to the static case, here also for the Floquet case the magnetic field allows us to tune the bulk gap which is introduced by the superconducting order parameter 
$\Delta(\vect{k})$. This would lead to analytical handling which we describe below in terms of the low energy effective model. In the presence of $\Delta$, interestingly, we find $H_{\rm Flq}$ continues 
to preserve the anti-unitary PHS generated by $\mathcal{C}$. This anti-unitary symmetry is essential to localize the MCMs at zero-energy associated with the effective Hamiltonian (Eq.(\ref{Highfreq2d})). Here, we would like to emphasize that we make use of the High-frequency Hamiltonian (Eq.(\ref{Highfreq2d})) only to corroborate the existence of the \textit{dressed} corner modes which we obtain from the exact diagonalization of the Floquet Operator $U(T)$ (Eq.(\ref{fo})). 

\subsection{Floquet Corner mode}
\begin{figure}
	\centering
	\subfigure{\includegraphics[width=0.48\textwidth]{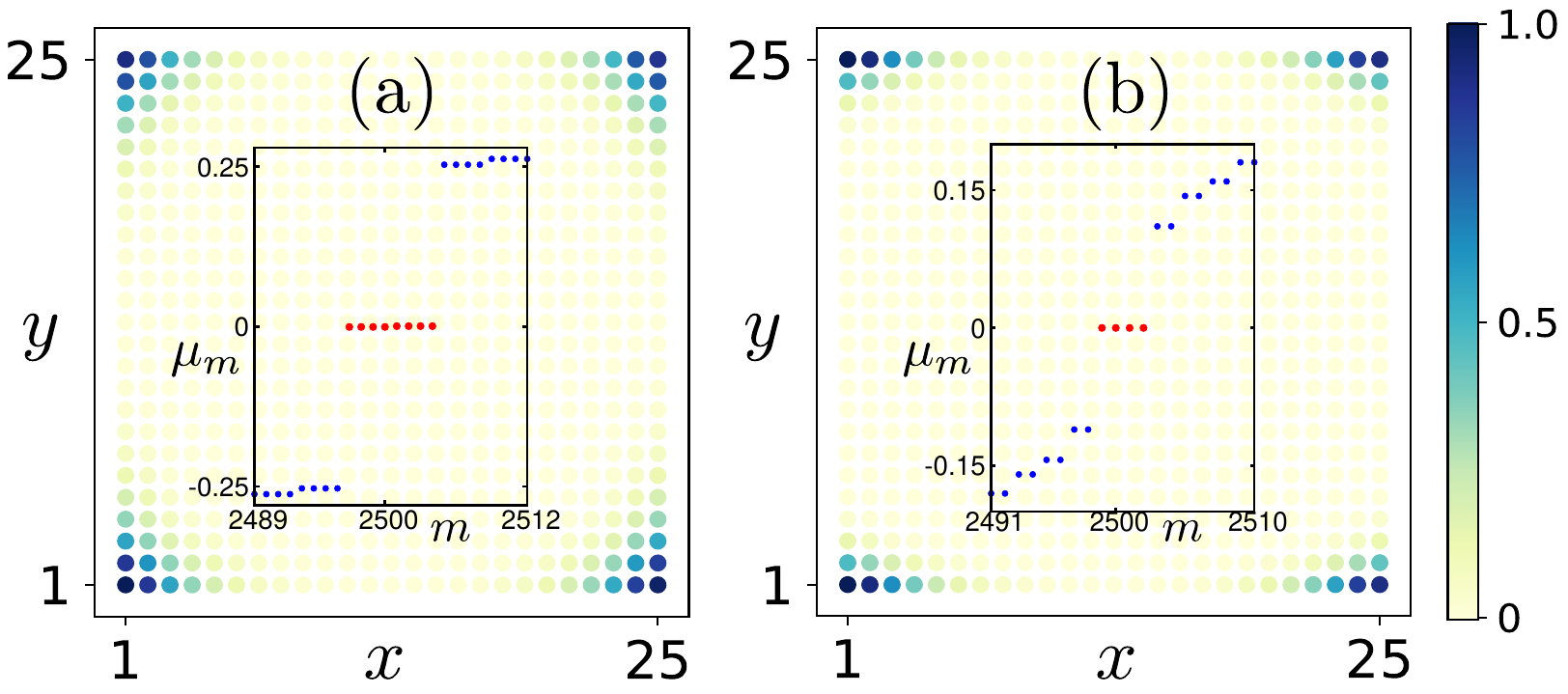}}
	\caption{(Color online) (a) The behavior of LDOS, associated with the MCMs at $\mu_m = 0$, is demonstrated for $B_x=0$ considering finite geometry. The inset exhibits the Floquet quasi-energy spectrum $\mu_m$ as a function of $m$ where eight MZMs are observed.  Here, $L_x=L_y=25, t_x=t_y=\lambda_x=\lambda_y=1.0, m_0=2.5$, $\Delta=0.6, m_1=-0.4, T=0.419$. (b) LDOS in finite geometry and eigenvalue spectrum (inset) are depicted for $B_x =0.3$. Here four MZMs appear at quasi-energy $\mu_m = 0$. The number of MCMs at $\mu_m =0$ can be reduced to half upon the introduction of the Zeeman field. 
	}
	\label{LDOS}
\end{figure}

Having established the possible route to an analytical understanding of the emergence of MZMs, we here numerically show their existence in the 2D FSOTSC phase that is characterized by the presence of Floquet MCMs. To be precise, we find the signature of dynamical MCMs that appear in the Floquet LDOS as shown in Fig.\ \ref{LDOS}. We know $U(T)|\phi_m \rangle=\exp(-i \mu_m T)|\phi_m\rangle$, where $|\phi_m \rangle$ is the Floquet quasi-energy states corresponding to the Floquet quasi-energy $\mu_m$. To calculate the LDOS, we numerically diagonalize the Floquet operator, $U(T)$ (in Eq.(\ref{fo})) 
in the open boundary condition (OBC) . Here, we consider the Floquet eigenmodes $|\phi_m \rangle$'s associated with  $\mu_m \approx 0$ (within numerical accuracy) to compute the LDOS of these 
zero-energy states. These MCMs can be easily distinguished from the gapped bulk modes. Note that, the 2D SOTSC phase hosting MCMs has been very recently studied in static~ \cite{Wu2020,Volpez2019,Yan2018,YanPRB2019} and Floquet driven~\cite{Jelena2019,ghosh2020floquet} cases with different driving protocols.
 
 We consider two cases, in the first case, we set the amplitude of the in-plane magnetic field, $B_x=0$, and the corresponding LDOS is shown in Fig.~\ref{LDOS}(a). One can identify from the inset of 
 Fig. \ref{LDOS}(a) that there exist eight MCMs \ie two MZMs per corner. While for $B_x \neq 0$, we obtain four MCMs \ie one MZM per corner as depicted in the inset of Fig. \ref{LDOS}(b). One can thus infer from Fig. \ref{LDOS}(a) that two MZMs sharing the same corner annihilate each other to give rise to an electronic corner mode which may lead to a \textit{weak} FSOTSC phase. Interestingly, in Fig. \ref{LDOS}(b) the applied in-plane magnetic field yields only one MZM per corner carrying a signature of \textit{strong} FSOTSC phase~\cite{ghosh2020floquet}. Moreover, the localization properties of these MCMs are significantly modified with the introduction of the Zeeman field. From the weight structure of MCMs as observed in LDOS, it is evident that the localization length remains the same in 
 $x$ and $y$ direction for $B_x=0$. Finite $B_x$ breaks this symmetry of localization length and the weight becomes less along the $y$ direction as compared to the $x$ direction. The introduction of 
 $B_y$ instead of $B_x$ does not change these properties. We discuss more regarding this in the later part of the paper.


\subsection{Low energy edge theory}

Here, we turn ourself to the low energy edge theory to corroborate the findings of the MCMs. We begin by expanding the effective Hamiltonian in the high-frequency limit (Eq.(\ref{Highfreq2d})) around 
$\Gamma=(0,0)$ point as
\begin{eqnarray}\label{model2lowenergy}
H_{\rm Flq,\Gamma}&\approx&\left(m'-\frac{t_x}{2}k_x^2-\frac{t_y}{2}k_y^2 \right) \Gamma_1+\lambda_xk_x\Gamma_2+\lambda_yk_y\Gamma_3 \non \\ &&-\frac{\Delta}{2} \left(k_x^2-k_y^2\right) \Gamma_4+B_x \Gamma_5 + \frac{m_1}{T} \Gamma_1+m_1 \lambda_x k_x \Gamma_{21}\non \\
&&+m_1 \lambda_y k_y \Gamma_{31}-\frac{m_1\Delta}{2} \left(k_x^2-k_y^2\right) \Gamma_{41} \ ,
\end{eqnarray}
where, $m'=\left(m_0-t_x-t_y\right)$, $\Gamma_{21}=-\sigma_y s_z \tau_z$, $\Gamma_{31}=\sigma_x$ and $\Gamma_{41}=\sigma_z s_y \tau_x$. For a demonstrative example, we present the analytical solution for edge-I. We consider here open (periodic) boundary condition along $x$ ($y$) direction. We rewrite $H_{\rm Flq, \Gamma}=H_0 (-i \partial_x) + H_p(-i \partial_x,k_y)$ by replacing $k_x\rightarrow -i\partial_x$ and neglecting $k_y^2$ term. We thus obtain 
\begin{eqnarray}
H_0&=&\left(m-\frac{t_x}{2} \partial_x^2 \right)\Gamma_1-i\lambda_x\partial_x \  \Gamma_2\ , \non \\
H_p&=&\lambda_yk_y\Gamma_3+\frac{\Delta}{2} \partial_x^2 \ \Gamma_4+B_x \Gamma_5-im_1\lambda_x \partial_x \ \Gamma_{21} \non \\
&&+ m_1\lambda_y k_y \Gamma_{31}+\frac{m_1 \Delta}{2} \partial_x^2 \ \Gamma_{41} \ .
\end{eqnarray}
Here, we consider the pairing amplitude $\Delta$ and the amplitude of the in-plane magnetic field $B_x$ to be small and treat them as small perturbation~\cite{Yan2018,YanPRB2019,ghosh2020floquet}. The mass term $m=\left(m'+\frac{m_1}{T}\right)$ is considered to be less than zero. Assuming $\Psi$ to be the zero-energy eigenstate of $H_0$ and following the boundary condition $\Psi(x=0)=\Psi(x=\infty)=0$, we obtain
\begin{equation}
\Psi_{\alpha}=\left|\mathcal{N}_x\right|^2e^{-\mathcal{K}_1x}\sin \mathcal{K}_2x \ e^{ik_yy} \Phi_{\alpha}\ ,
\end{equation}
where, $\mathcal{K}_1=\frac{\lambda_x}{t_x}$, $\mathcal{K}_2=\sqrt{\left|\frac{m}{t_x}\right|-\mathcal{K}_1^2}$, $\left|\mathcal{N}_x\right|^2=\frac{4\mathcal{K}_1\left(\mathcal{K}_1^2+\mathcal{K}_2^2\right)}{\mathcal{K}_2^2}$ and $\Phi_{\alpha}$ is a 8-component spinor satisfying $\sigma_ys_z\tau_z\Phi_{\alpha}=-\Phi_{\alpha}$. Our chosen basis reads

\begin{eqnarray}
\Phi_1&=&\ket{\sigma_y=+1}\otimes\ket{s_z=+1}\otimes\ket{\tau_z=-1},\non\\
\Phi_2&=&\ket{\sigma_y=-1}\otimes\ket{s_z=+1}\otimes\ket{\tau_z=+1},\non \\
\Phi_3&=&\ket{\sigma_y=-1}\otimes\ket{s_z=-1}\otimes\ket{\tau_z=-1},\non \\
\Phi_4&=&\ket{\sigma_y=+1}\otimes\ket{s_z=-1}\otimes\ket{\tau_z=+1}\ .
\end{eqnarray}

The matrix element of $H_p$ in this basis can be written as
\begin{equation}
H_{\rm I,\alpha\beta}^{\rm Edge}=\int_{0}^{\infty} dx \ \Psi_{\alpha}^\dagger(x) H_p(-i \partial_x,k_y)\Psi_{\beta}(x)\ ,
\end{equation} 
Thus, we obtain the effective Hamiltonian for the edge-I as
\begin{equation}
H_{\rm I}^{\rm Edge}=- \lambda_y k_y s_z + M_{\rm I} s_y \tau_y \ ,
\end{equation}
where, $M_{\rm I}=\lvert \frac{m \Delta}{t_x}\rvert$. Similarly, for edge-II, III and IV, one can obtain the effective Hamiltonian as
\begin{eqnarray}
H_{\rm II}^{\rm Edge}&=& \lambda_x k_x s_z - M_{\rm II} s_y \tau_y -B_x s_x \tau_z \ , \non \\
H_{\rm III}^{\rm Edge}&=&- \lambda_y k_y s_z + M_{\rm III} s_y \tau_y \ , \non \\
H_{\rm IV}^{\rm Edge}&=& \lambda_x k_x s_z - M_{\rm IV} s_y \tau_y -B_x s_x \tau_z \ .
\end{eqnarray}
where, $M_{\rm II}=\lvert \frac{m \Delta}{t_y}\rvert$, $M_{\rm III}=M_{\rm I}$ and $M_{\rm IV}=M_{\rm II}$. Therefore, the low energy effective Hamiltonian written in the edge co-ordinate $j$ is given 
by the compact form as
\begin{equation}
\label{Model1edge}
H_{j}^{\rm Edge}=-i\lambda(j)s_z\partial_{j}+M(j) s_y\tau_y-B(j)s_x\tau_z\ ,
\end{equation} 
with $\lambda(j)=\left\{-\lambda_y,\lambda_x,-\lambda_y,\lambda_x\right\}$, $M(j)=\left\{M_{\rm I},-M_{\rm II},M_{\rm III},-M_{\rm IV}\right\}$ and $B(j)=\left\{0,B_x,0,B_x\right\}$.
Note that, one can consider $B_y$ instead of $B_x$; however, the above results do not change qualitatively. Now to proceed further we study two cases. 
\subsubsection{Case I: $B_x=0$}
First, we turn-off the in-plane magnetic field $B_{x}$. The edge Hamiltonian $H_j^{\rm Edge}$ can be decomposed into two independent blocks as
\begin{eqnarray}
H^{\rm Edge}_j=H_{\tau_y=+1}\oplus H_{\tau_y=-1}\ ,
\end{eqnarray}
where,
\begin{eqnarray}
H_{\tau_y=+1}&=& -i\lambda(j)s_z\partial_{j}+M(j) s_y\ , \non \\H_{\tau_y=-1}&=& -i\lambda(j)s_z\partial_{j}-M(j) s_y \ .
\label{ham1_surface}
\end{eqnarray}
We obtain domain walls for both these blocks $\tau_y=\pm 1$ as the mass term $M(j)$ changes its sign between two adjacent edges; from edge-I, III to edge-II, IV.  In terms of the system parameters 
the Dirac mass changes from $\lvert \frac{m \Delta}{t_x}\rvert$ to $-\lvert \frac{m \Delta}{t_y}\rvert$. Consequently, one finds two MZMs per corner (see Fig.\ \ref{LDOS}(a)); each block is giving rise to
one MZM per corner. Therefore, one finds the origin of MCMs at each corner as obtained from the above low-energy edge theory. The dynamical breaking of TRS is only thus able to generate \textit{weak} FSOTSC phase (see text for discussion).

\subsubsection{Case II: $B_x\neq0$}
Incorporating the Zeeman field $B_x\neq0$, in $\tau_x= s_z$ subspace, the last term in Eq.(\ref{Model1edge}) can be written as $B(j)s_x\tau_z\rightarrow \mp B(j)s_y$ for $\tau_y =\pm 1$ block.  
The edge Hamiltonian $H_j^{\rm Edge} =H_{\tau_y=+1}\oplus H_{\tau_y=-1}$ thus takes the following form 
upon decomposing into two independent blocks as
\begin{eqnarray}
H_{\tau_y=+1}&=& -i\lambda(j)s_z\partial_{j}+ D(j) s_y \ , \non \\H_{\tau_y=-1}&=& -i\lambda(j)s_z\partial_{j}- D'(j) s_y \ ,
\label{ham2_surface}
\end{eqnarray}
with $D(j)=\{D_{\rm I},D_{\rm II},D_{\rm III},D_{\rm IV} \}=M(j) + B(j)$ and $D'(j)= \{D'_{\rm I},D'_{\rm II},D'_{\rm III},D'_{\rm IV} \}=M(j) - B(j)$. Hence, one can eventually obtain two decoupled diagonal blocks with Dirac masses in different edges as $D_{\rm I}=\lvert \frac{m \Delta}{t_x}\rvert$, $D_{\rm II}=-\lvert \frac{m \Delta}{t_y}\rvert + B_x$, $D_{\rm III}=\lvert \frac{m \Delta}{t_x}\rvert $ and  $D_{\rm IV}=-\lvert \frac{m \Delta}{t_y}\rvert + B_x$;  $D'_{\rm I}= D_I$, $D'_{\rm II}=-\lvert \frac{m \Delta}{t_y}\rvert - B_x$, $D'_{\rm III}=D_{III} $ and  $D'_{\rm IV}=-\lvert \frac{m \Delta}{t_y}\rvert - B_x$. Therefore, we  observe that $B_x$ can change the gap along edge-II and IV leaving edge-I and III unaltered for both of these blocks. It is now quite evident from Eq.(\ref{ham2_surface}) that $B_x=\lvert \frac{m \Delta}{t_y}\rvert$ ($B_x=-\lvert \frac{m \Delta}{t_y}\rvert$) refers to a special situation where edge-II and IV become gapless for $\tau_y=+1$ ($\tau_y=-1$) block. This is in sharp contrast to the edge Hamiltonian without magnetic field as described in Eq.(\ref{ham1_surface}) where all the four edges are massive. In the present case with magnetic field $B_x=\lvert \frac{m \Delta}{t_y}\rvert$ ($B_x=-\lvert \frac{m \Delta}{t_y}\rvert$), the remaining two edges become massive for $\tau_y=+1$ ($\tau_y=-1$) block. We consider only positive values of $B_x$ and due to that reason we investigate two instances 
where $B_x >\lvert \frac{m \Delta}{t_y}\rvert$ and $B_x<\lvert \frac{m \Delta}{t_y}\rvert$. 

When we consider $B_x > \lvert \frac{m \Delta}{t_y}\rvert $, $D'(j)$ changes its sign from edge-I (III) to edge-II (IV) while $D(j)$ remains positive in all the four edges. Therefore, $\tau_y =+1$ block becomes inactive and remains always massive while $\tau_y =-1$, being the only active block, can lead to Jackie-Rebbi localized MCMs at zero quasi-energy. One can thus observe one MZM per corner as depicted 
in Fig.\ \ref{LDOS}(b). On the other hand, for $B_x < \lvert \frac{m \Delta}{t_y}\rvert $, both the blocks turn out to be active i.e., mass changes its sign between two adjacent edges. Hence the MZMs are supported by both of these blocks. This would result in two MZMs per corner similar to the LDOS as shown in Fig.~\ref{LDOS}(a). Therefore, the explicit breaking of TRS by applying the magnetic field appears to be more efficient to obtain the FSOTSC phase as both \textit{weak} (two MZM per corner) and \textit{strong} (one MZM per corner) phases can be explored simultaneously. By contrast, the dynamical breaking of TRS without applying the magnetic field only allows us to explore the \textit{weak} FSOTSC phase. We explain these phases more elaborately while discussing the topological invariants.

\subsection{Corner Mode solution}
\subsubsection{Case I: $B_x=0$}
To obtain the analytical solution of the MCMs, residing at the intersection between edge-I and II, we solve the corresponding edge Hamiltonians for the zero-energy solution. At edge-I, we assume a solution of the form  
\begin{equation}
\Psi_C \sim e^{-\xi y} \left(a_1, a_2, a_3, a_4\right)^{\tilde{T}} \ , 
\end{equation}
Here, $\tilde{T}$ denotes the transpose.
The eigenvalue equation for $\Psi_C$ acquires the following form
\begin{equation}
\begin{pmatrix}
-i\lambda_y \xi & 0 & 0 & -M_{\rm I} \\
0&-i\lambda_y \xi&M_{\rm I} &0\\
0 & M_{\rm I} & i\lambda_y \xi & 0 \\
-M_{\rm I} & 0 & 0 & i\lambda_y \xi
\end{pmatrix}
\begin{pmatrix}
a_1 \\
a_2 \\
a_3 \\
a_4
\end{pmatrix}
=0.
\end{equation}
The secular equation for $\Psi_C$ then reads 
\begin{equation}
\det \begin{pmatrix}
-i\lambda_y \xi & 0 & 0 & -M_{\rm I} \\
0&-i\lambda_y \xi&M_{\rm I} &0\\
0 & M_{\rm I} & i\lambda_y \xi & 0 \\
-M_{\rm I} & 0 & 0 & i\lambda_y \xi
\end{pmatrix}=0\ .
\label{secular}
\end{equation}
Solving Eq.(\ref{secular}), we find four solutions for $\xi$ as
\begin{equation}
\xi=\left\{-\frac{M_{\rm I}}{\lambda_y},-\frac{M_{\rm I}}{\lambda_y},\frac{M_{\rm I}}{\lambda_y},\frac{M_{\rm I}}{\lambda_y}\right\}\ .
\end{equation}

Given the fact that $\Psi_C$ must vanish at $y\rightarrow\infty$, therefore, we obtain two linearly independent solutions for edge-I, $\Phi_C^{\rm I,1}=(1,1,i,-i)^{\tilde{T}}$ and 
$\Phi_C^{\rm I,2}=(1,-1,-i,-i)^{\tilde{T}}$. Thus, $\Psi_C$ can be expanded as
\begin{equation}
\Psi_C \sim \alpha_{\rm I} \  e^{-\frac{M_{\rm I}}{ \lambda_y} y}\Phi^{\rm I,1}_C \ + \ \beta_{\rm I} \  e^{-\frac{M_{\rm I}}{ \lambda_y} y}\Phi^{\rm I,2}_C\ ,
\end{equation}
Here, $\alpha_{\rm I}$ and $\beta_{\rm I}$ are the normalization factors for edge-I. 
Similarly, for edge-II with $\alpha_{\rm II}$ and $\beta_{\rm II}$ being the normalization factors, we obtain 
\begin{equation}
\Psi_C \sim \alpha_{\rm II} \  e^{-\frac{M_{\rm II}}{ \lambda_x} x}\Phi^{\rm II,1}_C \ + \ \beta_{\rm II} \  e^{-\frac{M_{\rm II}}{ \lambda_x} x}\Phi^{\rm II,2}_C\ ,
\end{equation}
where, $\Phi_C^{\rm II,1}=(1,1,i,-i)^{\tilde{T}}$ and $\Phi_C^{\rm II,2}=(1,-1,-i,-i)^{\tilde{T}}$. Considering the wavefuction, $\Psi_C$ to be continuous at the interface \ie at $x=y=0$, we obtain $\alpha_{\rm I}=\alpha_{\rm II}=\alpha$ and $\beta_{\rm I}=\beta_{\rm II}=\beta$. 
Hence, the wavefunction for the MCMs becomes
\begin{eqnarray}
\Psi_C &\sim&  \alpha \  e^{-\frac{M_{\rm I}}{ \lambda_y} y}\Phi^{\rm I,1}_C \ + \ \beta \  e^{-\frac{M_{\rm I}}{ \lambda_y} y}\Phi^{\rm I,2}_C\  \quad \ {\rm :edge-I}\ , \non\\
\Psi_C &\sim&  \alpha \  e^{-\frac{M_{\rm II}}{ \lambda_x} x}\Phi^{\rm I,1}_C \ + \ \beta \  e^{-\frac{M_{\rm II}}{ \lambda_x} x}\Phi^{\rm I,2}_C\  \quad  {\rm :edge-II}\ . \non \\ 
\end{eqnarray}

The localization length in the $x$ ($y$)-direction becomes $\lambda_x/M_{\rm II}$ ($\lambda_y/M_{\rm I}$). This clearly suggests that localization lengths of MCMs are dependent on the strength of hopping, spin-orbit coupling, mass, proximity induced superconducting gap function. In the present case, choice of $t_x=t_y$ and $\lambda_x=\lambda_y$ leads to the fact the localization length becomes uniform in $x$ and $y$-direction as observed in the Floquet LDOS (see Fig.~\ref{LDOS}(a)). More importantly, one can observe that there exist two MZMs at each corner corroborating our numerical findings as shown in the inset of Fig.\ \ref{LDOS}(a). The presence of two MCMs suggests that they would annihilate each other leaving an electronic state at the corner. The dynamical breaking of TRS 
thus leads to a \textit{weak} FSOTSC phase as each corner is occupied by two MCMs.

\subsubsection{Case II: $B_x\neq0$}
Here we investigate the solutions for the MCMs in presence of $B_{x}$. To begin with, we assume $B_x > M_{\rm II}$. We proceed as before and obtain the following solutions for edge-I and II as
\small
\begin{eqnarray}
\Psi_C &\sim& \alpha_{\rm I} \  e^{-\frac{M_{\rm I}}{ \lambda_y} y}\Phi^{\rm I,1}_C \ + \ \beta_{\rm I} \  e^{-\frac{M_{\rm I}}{ \lambda_y} y}\Phi^{\rm I,2}_C\  \qquad \qquad \quad \ {\rm :edge-I}, \non \\
\Psi_C &\sim& \alpha_{\rm II} \  e^{-\frac{B_x-M_{\rm II}}{ \lambda_x} x}\Phi^{\rm II,1}_C \ + \ \beta_{\rm II} \  e^{-\frac{B_x+M_{\rm II}}{ \lambda_x} x}\Phi^{\rm II,2}_C\  \ {\rm :edge-II}, \non \\
\end{eqnarray}
\normalsize
where, $\Phi_C^{\rm I,1}=(1,1,i,-i)^{\tilde{T}}$, $\Phi_C^{\rm I,2}=(1,-1,-i,-i)^{\tilde{T}}$, $\Phi_C^{\rm II,1}=(1,-1,i,i)^{\tilde{T}}$ and $\Phi_C^{\rm II,2}=(1,1,i,-i)^{\tilde{T}}$. Upon matching $\Psi_C$ at the boundary, we obtain $\alpha_{\rm I}=\beta_{\rm II}=\alpha$ and $\beta_{\rm I}=\alpha_{\rm II}=0$. The final solution becomes
\begin{eqnarray}
\Psi_C &\sim&  \alpha~  e^{-\frac{M_{\rm I}}{ \lambda_y} y}\Phi^{\rm I,1}_C  \qquad \qquad \ {\rm :edge-I}, \non \\
\Psi_C &\sim&  \alpha~  e^{-\frac{B_x+M_{\rm II}}{ \lambda_x} x}\Phi^{\rm I,1}_C\  \qquad \ {\rm :edge-II}. \non \\
\end{eqnarray}
In this case, the localization lengths are given by $\lambda_x/(B_x+M_{\rm II})$ ($\lambda_y/M_{\rm I}$) along $x$ ($y$)-direction. Thus the MCMs decay differently along the two directions into the bulk. This is also evident from the Floquet LDOS where localization of MZMs at the corners are stronger in $y$-direction as compared to $x$-direction (see Fig.\ \ref{LDOS}(b)). Moreover, there exists one MCM per corner, as shown in the inset of Fig.\ \ref{LDOS}(b).  The presence of one MCMs suggests that it corresponds to a \textit{strong} FSOTSC phase when $B_x\neq0$. This phase cannot be realized in
presence of only periodic kick drive \ie when TRS is broken dynamically. 

Furthermore, when $B_x < M_{\rm II}$, we continue to obtain two MZMs per corner like the previous case with $B_x=0$, except for the modification in localization length that is modulated by $B_x$. 
The final solution for this case reads
\small
\begin{eqnarray}
\Psi_C &\sim& \alpha \  e^{-\frac{M_{\rm I}}{ \lambda_y} y}\Phi^{\rm I,1}_C \ + \ \beta \  e^{-\frac{M_{\rm I}}{ \lambda_y} y}\Phi^{\rm I,2}_C\  \qquad \qquad \quad  {\rm :edge-I}, \non \\
\Psi_C &\sim& \alpha \  e^{-\frac{B_x+M_{\rm II}}{ \lambda_x} x}\Phi^{\rm I,1}_C \ + \ \beta \  e^{-\frac{M_{\rm II}-B_x}{ \lambda_x} x}\Phi^{\rm I,2}_C\  \quad \ {\rm :edge-II}, \non \\
\end{eqnarray}
\normalsize
Therefore, the 
incorporation of Zeeman field allows one to explore both the \textit{weak} and \textit{strong} phases depending on the values of $B_x$. 

\subsection{Topological characterization of MCMs}
Having understood the wave-functions associated with the MCMs, we would now like to characterize these FSOTSC phases with appropriate topological invariants. We compute two invariants, 
namely FWS and FQM to identify the underlying topological nature of these MCMs. For the calculation of FWS, we construct the Wilson loop operator~\cite{benalcazarprb2017} as
\begin{equation}
{\mathcal W}_x=F_{x,k_x + (N_x -1) \Delta k_x } \cdots F_{x,k_x + \Delta k_x } F_{x,k_x} \ ,
\label{WS}
\end{equation} 
with $\left[F_{x,k_x}\right]_{mn}=\langle \phi_{n, k_x + \Delta k_x} | \phi_{m,k_x} \rangle$, where $\Delta k_x= 2\pi /N_x$ ($N_x$ being the number of discrete points considered inside the Brillouin zone (BZ) along $k_x$) and $|\phi_{m,k_x} \rangle $ is the $m^{\rm th}$ occupied Floquet quasi-state in the semi-infinite geometry (considering periodic boundary condition (PBC) and open boundary condition 
(OBC) along $x$ and $y$ direction respectively). One can obtain $|\phi_{m,k_x} \rangle $ by diagonalizing the effective Floquet Hamiltonian in the high-frequency limit (Eq.(\ref{Highfreq2d})). 
Thus we obtain the Wannier Hamiltonian as
\begin{equation}
{\mathcal H}^{\rm Flq}_{{\mathcal W}_x}= -i \ln {\mathcal W}_x\ ,
\end{equation}
 whose eigenvalues $2\pi \nu^{\rm Flq}_x$ correspond to the FWS. Similarly, one can find $\nu^{\rm Flq}_y$ by considering PBC (OBC) along $y$ ($x$) direction.
 
\begin{figure}
	\centering
	\subfigure{\includegraphics[width=0.48\textwidth]{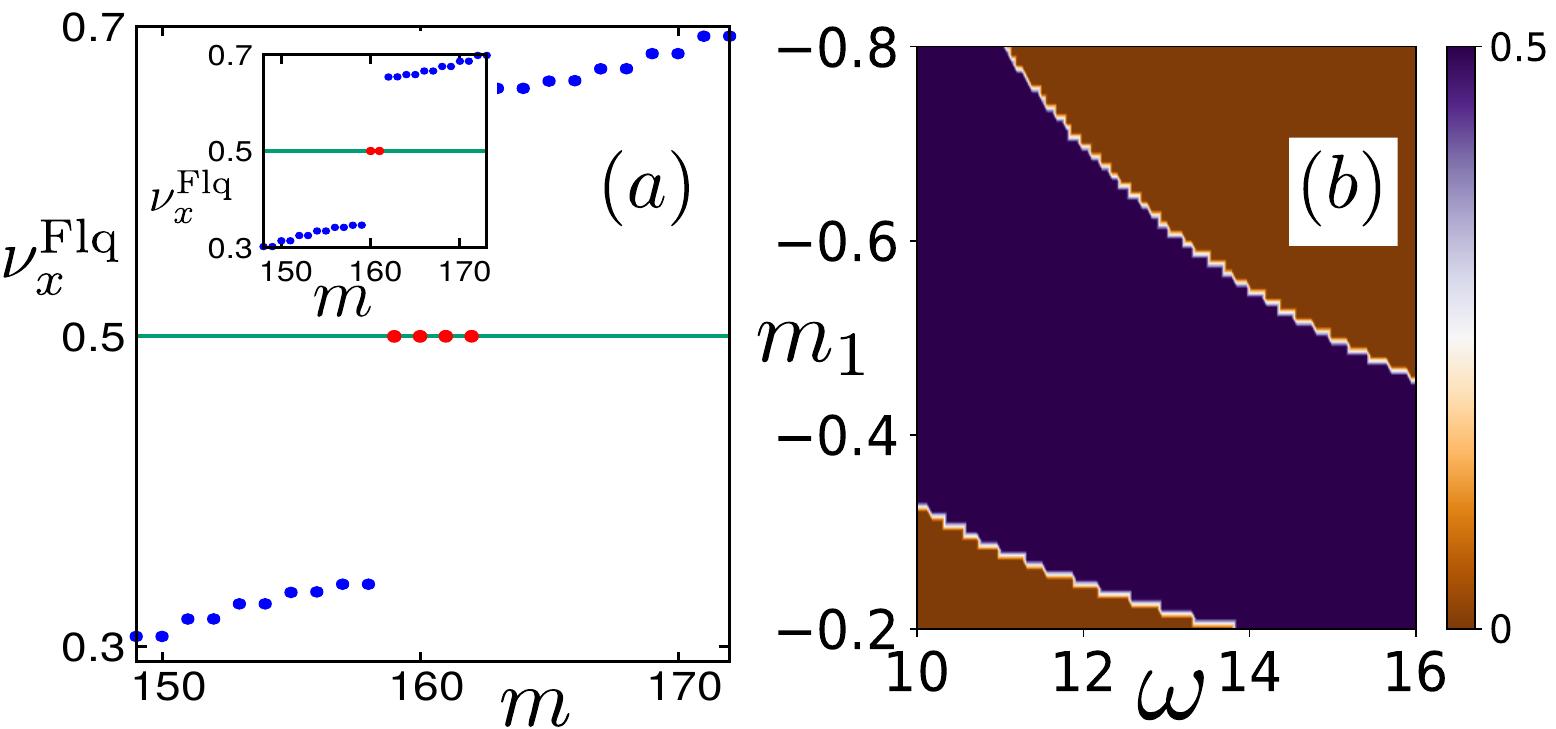}}
	\caption{(Color online) (a) Floquet Wannier Spectrum (FWS) is demonstrated for $B_x=0$ and the inset manifests the same for $B_x=0.3$. The values of all other parameters remain the same as in 
	Fig.\ \ref{LDOS}. Connecting with Fig.~\ref{LDOS}, we can comment that eight (four) MCMs correspond to four (two) FWS quantized at $0.5$. We refer the phase with eight and four MCMs as  
	\textit{weak} and \textit{strong} FSOTSC phases, respectively. (b) Floquet Quadrupole moment (FQM), which is only quantized at $0.5$ for the \textit{strong} FSOTSC phase, is demonstrated in 
	$m_1$-$\omega$ plane where $m_1$ and $\omega$ are the amplitude of the drive and the driving frequency respectively.
	}
	\label{Invariant}
\end{figure}

The quantized nature of Wannier Spectra at $0.5$ characterizes the SOTSC phase in our case. In the FSOTSC phase, one expects to obtain a quantized value pinned at $0.5$ similar to the static counterpart. In the first case for $B_x=0$, we obtain four eigenvalues at $0.5$ as shown in Fig.\ \ref{Invariant}(a), whereas for $B_x\neq 0$, one obtains such two eigenvalues only by diagonalizing 
${\mathcal H}^{\rm Flq}_{{\mathcal W}_x}$ (see the inset of Fig.\ \ref{Invariant}(a)). One can thus identify the FSOTSC phase, hosting two MCMs per corner in absence of the Zeeman field ($B_{x}=0$), 
as the \textit{weak} phase where there exist two pairs of FWS quantized at $0.5$. On the other hand, in presence of the Zeeman field ($B_{x}\neq 0$), the FSOTSC phase becomes a \textit{strong} one where a single MZM is localized per corner, leading to a single pair of FWS quantized at $0.5$.  Therefore, one can directly correlate the wave-function of MZMs at the corner with the topological signature of the invariant FWS. In this way, we can distinguish between the \textit{strong} and \textit{weak} FSOTSC phases that are respectively associated with two and four FWS eigenvalues stabilized at $0.5$. 
 
In order to calculate the FQM, being another invariant for the quantification of the topological phases, we numerically diagonalize the exact Floquet operator (Eq.(\ref{fo})). We then construct the Floquet many-body ground state $\Psi_{0,\rm F}$ by columnwise marshalling the quasi-states according to their quasi-energy $-\omega/2 \le \mu_m \le 0$: $\Psi_{0,\rm F}=\sum_{m \in \mu_m \le 0} |\phi_m \rangle \langle \phi_m |$~\cite{Nag19,Ghosh2020}. Now $Q^{\rm Flq}_{xy}(r)$ can be defined, considering the geometrical number operator ${\hat q}_{xy}={\hat n}(r) x y/L^2$ with ${\hat n}(r)$ being the number operator at $r=(x,y)$, as follows
\small
\begin{eqnarray}
Q^{\rm Flq}_{xy}&=& {\rm Re}\left[ -\frac{i}{2 \pi} {\rm Tr}\left( \ln \Big(\Psi_{0,\rm F}^\dagger \exp\Big[2\pi i \sum_r  {\hat q}_{xy} (r)\Big] \Psi_{0,\rm F} \Big)\right) \right]. \non \\
\label{qm_xy}
\end{eqnarray}
\normalsize
For $B_x\neq0$, we obtain $Q^{\rm Flq}_{xy}\equiv{\rm mod}(Q^{\rm Flq}_{xy},1)= 0.5$. On the other hand, when $B_x=0$, $Q^{\rm Flq}_{xy}$ turns out to be zero, depicting \textit{weak} topological nature of the phase. Therefore, similar to the FWS, we here also find finite (vanishing) FQM for \textit{strong} (\textit{weak}) FSOTSC phases. We then explore the dynamic FSOTSC phase for a range of the driving frequency $\omega$ and the driving amplitude $m_1$ where $Q^{\rm Flq}_{xy}=0.5$. This is shown in Fig.\ \ref{Invariant}(b). This clearly suggests that the emergent \textit{strong} phase is indeed an outcome of non-equilibrium dynamics as the underlying static model remains in a non-topological phase.

\section{Majorana Hinge modes in 3d}{\label{sec:III}}
In this section we generalize our earlier findings of FSOTSC phase in case of 3D.

\subsection{Model Hamiltonian and Driving Protocol}
\subsubsection{Model}
In 3D, we begin by writting down the Hamiltonian in the Bogoliubov-de Gennes (BdG) form as $H_{\rm BdG}=\sum_{{\vect{k}}}\Psi^\dagger_{\vect{k}} \mathcal{H}^{\rm 3D}_0({\vect{k}})\Psi_{\vect{k}}$, 
with $\Psi_{\vect{k}}=\left(c_{{\vect{k}},a\uparrow},- c_{-{\vect{k}},a\downarrow}^\dagger, c_{{\vect{k}},a\downarrow}, c_{-{\vect{k}},a\uparrow}^\dagger, c_{{\vect{k}},b\uparrow}, -c_{-{\vect{k}},b\downarrow}^\dagger, c_{{\vect{k}},b\downarrow}, c_{-{\vect{k}},b\uparrow}^\dagger\right)^{\tilde{T}}$ and $\mathcal{H}^{\rm 3D}_0({\vect{k}})$ is given by
\begin{eqnarray}
\mathcal{H}^{\rm 3D}_0&=&\epsilon(\vect{k}) \Gamma_1 +\lambda_x \sin k_x \Gamma_2+\lambda_y \sin k_y \Gamma_3 +\lambda_z \sin k_z \Gamma_4\non \\
&&+\Delta (\vect{k}) \Gamma_5 + B_x \Gamma_6+ B_y \Gamma_7 + B_z \Gamma_8 \equiv {\vect{N}}(\vect{k})\cdot {\bm \Gamma}\ , \non \\
\label{ham3D}
\end{eqnarray}
where, $\epsilon(\vect{k})=\left(m_0-t_x \cos k_x -t_y \cos k_y - t_z \cos k_z \right)$, $\Delta(\vect{k})=\Delta \left(\cos k_x-\cos k_y \right)$ and $\Gamma_1=\sigma_z \tau_z$, $\Gamma_2=\sigma_x s_x \tau_z$, $\Gamma_3=\sigma_x s_y \tau_z$, $\Gamma_4=\sigma_x s_z \tau_z$, $\Gamma_5= \tau_x$, $\Gamma_6= s_x$, $\Gamma_7= s_y$ and $\Gamma_8= s_z$. Similar to the Hamiltonian 
(Eq.(\ref{ham1})) in 2D case, $\mathcal{H}^{\rm 3D}_0$ respects both TRS $\mathcal{T}=is_y\mathcal{K}$ and PHS $\mathcal{C}=s_y\tau_y \mathcal{K}$ in absence of Zeeman field \ie $B_x=B_y=B_z=0$. In absence of the superconducting term $\Delta(\vect{k}) = 0$, the model supports zero-energy surface states for $|m_0|< t_x +t_y +t_z$. While for $|m_0|> t_x +t_y +t_z$, the models becomes non-topological (trivial band insulator). This model thus supports the first order topological phase in the absence of $\Delta(\vect{k})$. Interestingly, the model becomes a PHS protected SOTSC, hosting MKPs at the hinges along $z$ direction, in the presence of $\Delta(\vect{k})$ when $|m_0|< t_x +t_y +t_z$. In case of 2D system (Eq.~(\ref{ham1})), the SOTSC phase supports MCMs; here for 3D system (Eq.~(\ref{ham3D})), it hosts Majorana hinge modes (MHMs). Upon breaking TRS by introducing magnetic field $B_x \ne 0$, the degeneracy of MKPs get lifted and there exist only one MZM per hinge. We note that $B_y$ does the same job as done by $B_x$. In contrast, $B_z$ is not able to lift the degeneracy of MKPs. In general, the MHMs are observed along $c$ direction when the SC order has the form 
$\Delta(\vect{k})=\cos k_a -\cos k_b $ with $a,b,c=x,y,z$. The MKPs along the hinges in the $c$ direction remain unaffected by the magnetic field $B_c$. On the other hand, for $|m_0|> t_x +t_y +t_z$, the system continues to remain in the non-topological phases even with $\Delta(\vect{k}) \ne 0$ and $B_x \ne 0$. Therefore, it would be interesting to study the generation of FSOTSC phase in presence of 
$\Delta$ by kicking the mass term while the underlying static system remains in a non-topological phase. Our aim is to generate the Floquet MHMs (FMHMs) and their topological characterization in 3D geometry. 
\subsubsection{Driving Protocol and Floquet Operator}
We consider the same driving protocol in the form of periodic kick as followed in the 2D case where
\begin{eqnarray}
m(t)&=&m_1 \sum_{r=1}^{\infty} \delta(t-rT) \ .
\label{kick2}
\end{eqnarray}
Here, $T$ is the period of the drive and $m_1$ is the amplitude of the drive. With the periodic kick (see Eq.(\ref{kick2})), the Floquet operator reads  
\begin{eqnarray}
U(T)&=&{\rm \bf TO} \exp \left[-i\int_{0}^{T}dt\left(\mathcal{H}^{\rm 3D}_0(\vect{k})+m(t)\Gamma_1\right)\right] \nonumber \\
&=& \exp(-i \mathcal{H}^{\rm 3D}_0(\vect{k}) T)~\exp(-i m_1 \Gamma_1)\ .
\label{fo3d}
\end{eqnarray}
We can cast the Floquet Operator $U(T)$ in a more compact form as
\begin{eqnarray}
U(T)&=&C_T\left(p-i q \Gamma_1\right)-i S_T\sum_{j=1}^{8}\left(n_j\Gamma_j+m_j\Gamma_{j1}\right), \ \
\end{eqnarray}
where, $C_T=\cos(\left|{\vect{N}}(\vect{k})\right| T)$, $S_T=\sin(\left|{\vect{N}}(\vect{k})\right| T)$, $p=\cos m_1$, $q=\sin m_1$, $n_j=\frac{N_j(\vect{k})\cos m_1}{\left|{\vect{N}}(\vect{k})\right|}$, $m_j=\frac{N_j(\vect{k})\sin m_1}{\left|{\vect{N}}(\vect{k})\right|}$ and $\Gamma_{j1}=\frac{1}{2i}\left[\Gamma_j,\Gamma_1\right]$ with $j=2,3,4,5,6,7,8$. One can find the general form of the effective 
Hamiltonian as
\begin{widetext}
	\begin{eqnarray}
	H_{\rm Flq} =\frac{\xi_{\vect{k}}}{\sin\xi_{\vect{k}}T}\Bigg[\sin(\left|{\vect{N}}(\vect{k})\right| T) \cos m_1 \sum_{j=1}^{8}r_j \Gamma_j +\cos(\left|{\vect{N}}(\vect{k})\right| T) \sin m_1 \Gamma_1 +\sin(\left|{\vect{N}}(\vect{k})\right| T) \sin m_1 \sum_{j=2}^{8} r_j \Gamma_{j1} \Bigg]\ ,
	\end{eqnarray}
\end{widetext}
with $\xi_{\vect{k}}=\frac{1}{T}\cos^{-1} \left[\cos(\left|{\vect{N}}(\vect{k})\right| T) \cos m_1\right]$, $r_j=\frac{N_j(\vect{k})}{\left|{\vect{N}}(\vect{k})\right|}$. In the high-frequency limit \ie~$T \rightarrow 0$ and $m_1 \rightarrow 0$, neglecting the higher order terms in $T$ and $m_1$, we find the effective Hamiltonian as
\begin{equation}
H_{\rm Flq}\approx \mathcal{H}_0(\vect{k})+\frac{m_1}{T} \Gamma_1+ m_1 \sum_{j=2}^{8} r_j \Gamma_{j1}\ ,
\label{Highfreq3d}
\end{equation}
In Eq.(\ref{Highfreq3d}), terms associated with $\Gamma_{j1}$ are the new terms generated by the drive. Note that these new terms break TRS $\mathcal{T}$ present in the static Hamiltonian. Although $H_{\rm Flq}$ continues to preserve the anti-unitary PHS $\mathcal{C}$. Similar to the 2D case as described by Eq.(\ref{Highfreq2d}), here also the extra terms $r_j\Gamma_{j1}$ break the TRS of the model even when $B_x=B_y=B_z=0$. The Hamiltonian (Eq.(\ref{Highfreq3d})) shares similar characteristics as shown by the 2D Hamiltonian (Eq.(\ref{Highfreq2d})). Here also the mass term is renormalized to
$m_0 \to (m_0 + m_1/T)$ and the topological phase boundary becomes accordingly modified. The dynamical (explicit) breaking of TRS would lead to an interesting study as far as the \textit{weak} (\textit{strong}) FSOTSC phases are concerned.  As noted in the static case, the in-plane Zeeman field $B_x$ or $B_y$ can lead to an interesting effect for the Floquet case also.

\subsection{Floquet Majorana Hinge Mode}
\begin{figure}[H]
	\centering
	\subfigure{\includegraphics[width=0.48\textwidth]{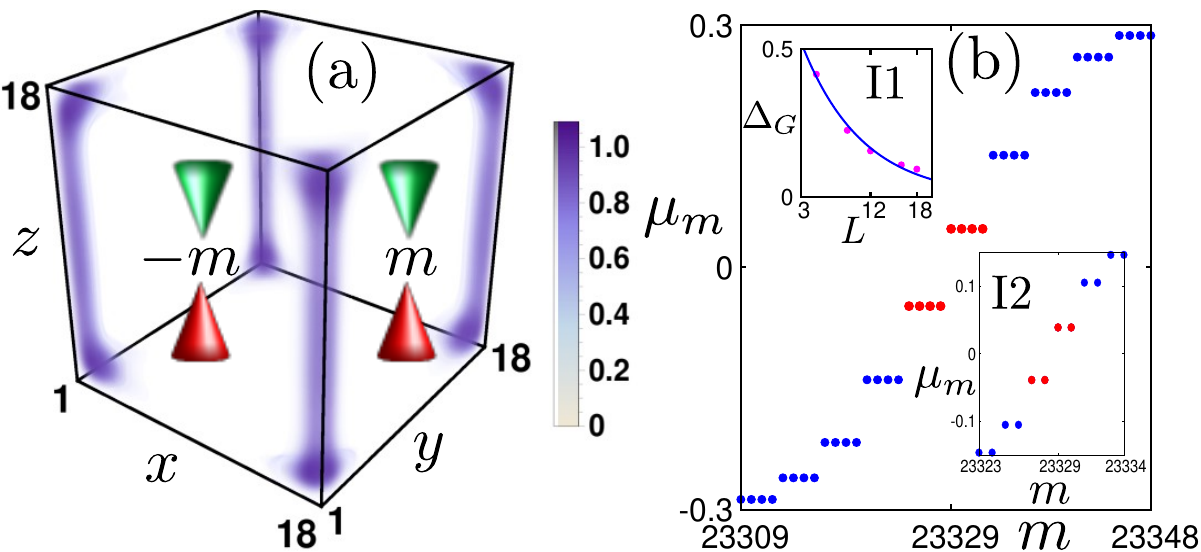}}
	\caption{(Color online) (a) The LDOS associated with the MZMs clearly exhibits the emergence of the 1D Floquet MHMs in finite geometry. The LDOS structure remains qualitatively invariant in the presence of an in-plane Zeeman field. We choose the other parameter values as $t_x=t_y=t_z=2.0$, $\lambda_x=\lambda_y=\lambda_z=1.0$, $m_0=5.5$, $m_1=-0.3$ and $T=0.393$. (b) The Floquet quasi-energy spectra is shown for the 3D FSOTSC in real space without the Zeeman field. There exist eight MHMs (due to the finite size effect, these modes do not appear exactly at zero energy) referring 
to the fact that this is a \textit{weak} phase where each hinge is occupied by two MZMs. The finite-size effect is investigated in the inset I1 showing the gap as a exponentially falling function of the system size in one real space direction as $\Delta_G \sim \beta \exp(-\alpha L)$ with $\alpha=0.127283$ and $\beta=0.769069$. With Zeeman field $B_x =0.4$, we depict (inset I2) the emergence of four MHMs i.e., one MZM per hinge. The Zeeman field can thus transform a \textit{weak} FSOTSC to a \textit{strong} FSOTSC phase.}
	\label{3DLDOS}
\end{figure}

Here, we numerically diagonalize the Floquet operator (Eq.(\ref{fo3d})) in real space geometry to manifest the signature of the 1D propagating MHMs along the $z$-direction in the LDOS spectrum. 
Like before, we consider two cases. When the Zeeman field $\vect{B}=0$, we obtain eight eigenvalues close to zero-energy like the 2D case as shown in Fig.\ \ref{3DLDOS}(b). 
However, when we incorporate $B_x \neq 0$ or $B_y \neq 0$, we obtain four modes near zero-energy (see Fig.\ \ref{3DLDOS}(b)I2). Interestingly, the transverse magnetic field $B_z$ does not give rise to these kinds of phenomena. However, the finite-size effect is very prominent here that we depict in the inset I1 of Fig.\ \ref{3DLDOS}(b). The finite-size gap $\Delta_G$ vanishes exponentially with $L$: 
$\Delta_G \sim \beta \exp(-\alpha L)$. This ensures the fact that MHMs are indeed zero-energy FSOTSC states in 3D. We show the LDOS explicitly in Fig.\ \ref{3DLDOS}(a) where one can observe that a finite spectral weight is uniformly distributed over the four hinges of the 3D cubic system. 

\vskip +1.0 cm

\vskip -2cm
\subsection{Low energy surface theory}
We here investigate the low-energy theory for 3D case to search for the existence of the hinge states, originated due to periodically kicked mass term, in the underlying static Hamiltonian (Eq.(\ref{ham3D})). For simplicity we choose $t_x=t_y=t_z=t$ and $\lambda_x=\lambda_y=\lambda_z=\lambda$. We expand the high-frequency effective Hamiltonian (Eq.(\ref{Highfreq3d})) around $\Gamma=(0,0,0)$ point
and obtain
\begin{eqnarray}
H_{\rm Flq,\Gamma} ^{\rm 3D}&=&\left(m'+\frac{t}{2}\partial_x^2 +\frac{t}{2}\partial_y^2 +\frac{t}{2}\partial_z^2   \right) \Gamma_1 +\lambda k_x \Gamma_2 \non \\
&&+\lambda k_y \Gamma_3 + \lambda k_z \Gamma_4 -\frac{\Delta}{2}\left(k_x^2-k_y^2\right)\Gamma_5 +\frac{m_1}{T} \Gamma_1\non \\
&&+m_1\lambda k_x \Gamma_{21}+m_1\lambda k_y \Gamma_{31}+m_1\lambda k_z \Gamma_{41}\non \\
&&-\frac{m_1 \Delta}{2}\left(k_x^2-k_y^2\right) \Gamma_{51}+ B_x \Gamma_6+ B_y \Gamma_7 + B_z \Gamma_8 ,\non \\
\label{3dlow1}
\end{eqnarray}
where, $m'=(m_0-3t)$ and $\Gamma_{21}=-\sigma_ys_x$, $\Gamma_{31}=-\sigma_ys_y$, $\Gamma_{41}=-\sigma_ys_z$ and $\Gamma_{51}=-\sigma_z \tau_y$. We choose a surface perpendicular to the
$xy$-plane, with a deviation from $yz$-plane by an angle $\theta$. In order to cast the above equations in a convenient form we translate to a rotated frame defined by $k_1=-\sin \theta~k_x +\cos \theta ~k_y$, $k_2=k_z$ and $k_3=\cos \theta~k_x +\sin \theta~k_y$. This rotation transforms $x,~y,~z \to ~y,~z,~x$ for $\theta=0$ while for $\theta=\pi/2$, $x,~y,~z \to -x,~z,~y$. We now consider OBC in the $x_3$ direction and  replace $k_3 \rightarrow -i \partial_3$.  One can hence divide Eq.(\ref{3dlow1}) into two parts as
\begin{eqnarray}
H_0&=&\left(m-\frac{t}{2}\partial_3^2 \right) \sigma_z \tau_z -i \lambda \partial_3 \ \sigma_x \left(s_x \cos \theta +s_y \sin \theta \right) \tau_z , \non \\
H_p&=&\lambda k_1 \sigma_x \left( -s_x \sin \theta + s_y \cos \theta \right) \tau_z + \lambda k_2 \sigma_x s_z \tau_z \non \\ &&-\frac{\Delta(\theta)}{2} \partial_3^2 \ \tau_x+im_1 \lambda \partial_3 \sigma_y \left(s_x \cos \theta +s_y \sin \theta \right) \non \\
&& -m_1 \lambda k_1 \sigma_y \left( -s_x \sin \theta + s_y \cos \theta \right) -m_1 \lambda k_2 \sigma_y s_z \non \\
&& +\frac{m_1\Delta(\theta)}{2} \partial_3^2 \ \sigma_z \tau_y+B_x s_x+B_y s_y+B_z s_z \ ,
\label{3dlow2}
\end{eqnarray}
where, $m=\left(m'+\frac{m_1}{T}\right)$ and $\Delta(\theta)=\Delta \left(\sin^2 \theta-\cos^2 \theta \right)$. We consider the following transformation for spins
\begin{equation}
\begin{pmatrix}
s_x \\
s_y \\
s_z 
\end{pmatrix}
=
\begin{pmatrix}
-\sin \theta & \cos \theta & 0 \\
\cos \theta & \sin \theta & 0 \\
0 & 0 & 1
\end{pmatrix}
\begin{pmatrix}
s_1 \\
s_3 \\
s_2 
\end{pmatrix}
\end{equation}
such that Eq.(\ref{3dlow2}) acquires a compact form as we notice for 2D case. Here,
$s_1$, $s_2$ and $s_3$ are Pauli matrices. Therefore, Eq.(\ref{3dlow2}) can be rewritten as 
\small
\begin{eqnarray}
H_0&=&\left(m-\frac{t}{2}\partial_3^2 \right) \sigma_z \tau_z -i \lambda \partial_3 \ \sigma_x s_3 \tau_z , \non \\
H_p&=&\lambda k_1 \sigma_x s_1 \tau_z + \lambda k_2 \sigma_x s_2 \tau_z-\frac{\Delta(\theta)}{2} \partial_3^2 \ \tau_x+im_1 \lambda \partial_3 \sigma_y s_3\non \\
&&-m_1 \lambda k_1 \sigma_y s_1 -m_1 \lambda k_2 \sigma_y s_2+\frac{m_1\Delta(\theta)}{2} \partial_3^2 \ \sigma_z \tau_y \non \\
&&+B_x \left(-s_1 \sin\theta+s_3 \cos \theta \right)+B_y \left(s_1 \cos\theta+s_3 \sin \theta \right) \non \\
&&+B_z s_2 \ \ ,
\label{3dlow3}
\end{eqnarray}
\normalsize
We assume $\Psi$ to be zero-energy solution of $H_0$ with the boundary condition $\Psi(x_3=0)=\Psi(x_3=\infty)=0$.  We finally obtain the following wave-function in the rotated frame as
\begin{equation}
\Psi_\alpha= A \ e^{-\mathcal{K}_1 x_3} \sin (\mathcal{K}_2 x_3) \ e^{ik_1 x_1+ik_2 x_2 } \ \Phi_\alpha,
\label{wf3D}
\end{equation}
where, $\mathcal{K}_1=\frac{\lambda}{t}$, $\mathcal{K}_2=\sqrt{\lvert \frac{2m}{t}\rvert-\mathcal{K}_1^2}$ and $A=\frac{4\mathcal{K}_1 \left(\mathcal{K}_1^2+\mathcal{K}_2^2\right)}{\mathcal{K}_2^2}$ 
and $\Phi_{\alpha}$ is $8$-component spinor satisfying $\sigma_y s_3 \Phi_{\alpha} = -\Phi_{\alpha}$. 

	\begin{eqnarray}
	\Phi_1&=&\ket{\sigma_y=+1}\otimes\ket{s_3=-1}\otimes\ket{\tau_z=+1},\non\\
	\Phi_2&=&\ket{\sigma_y=-1}\otimes\ket{s_3=+1}\otimes\ket{\tau_z=+1},\non \\
	\Phi_3&=&\ket{\sigma_y=+1}\otimes\ket{s_3=-1}\otimes\ket{\tau_z=-1},\non \\
	\Phi_4&=&\ket{\sigma_y=-1}\otimes\ket{s_3=+1}\otimes\ket{\tau_z=-1}\ .
	\end{eqnarray}

The matrix element of $H_p$ in the rotated frame within the above basis reads 
\begin{equation}
H_{\alpha\beta}^{\rm Surf}=\int_{0}^{\infty} dx_3 \ \Psi_{\alpha}^\dagger(x_3) H_p\Psi_{\beta}(x_3)\ .
\end{equation} 
Therefore, we obtain the Hamiltonian for the surface in the rotated frame to be
\begin{equation}
H^{\rm Surf}=\lambda k_1 s_3 \tau_y -\lambda k_2 s_3 \tau_x+ M(\theta)s_1-B(\theta)\tau_z \ .
\end{equation}
Here, $M(\theta)=\Delta(\theta) \lvert\frac{m}{t}\rvert$ and $B(\theta)=\left(B_x \cos \theta + B_y \sin \theta \right)$. The transverse magnetic field $B_z$ does not appear in the surface Hamiltonian referring to the fact that in-plane magnetic field plays the important role in determining the nature of the FSOTSC phase. Interestingly, $\Delta(\theta)=-\Delta(\theta+\pi/2)$ resulting in  $M(\theta)$ to change its sign between two adjacent surfaces under $C_4$ rotation around $z$ axis. Consequently,  one can get hinge mode in the junction between $xz$- and $yz$-plane. This sign change of the  mass term  is shown in Fig.~\ref{3DLDOS}~(a). We can explicitly write down the surface Hamiltonian for the $yz$ ($xz$) -surface by putting $\theta=0~(\frac{\pi}{2})$ as
\begin{eqnarray}
H^{yz}&=& \lambda k_y s_3 \tau_y -\lambda k_z s_3 \tau_x- \Delta  \left|\frac{m}{t}\right | s_1 -B_x\tau_z \ , \non \\
H^{xz}&=& -\lambda k_x s_3 \tau_y -\lambda k_z s_3 \tau_x+\Delta  \left|\frac{m}{t}\right | s_1 -B_y\tau_z \ . \quad
\end{eqnarray}
Let us now discuss the above surface Hamiltonian at length. As compared to the edge Hamiltonian for edge-$a$ with $k_a$ only, surface Hamiltonian for $ab$ surface consists of $k_a$ and $k_b$. In the absence of magnetic field, $\Delta  \left|\frac{m}{t}\right | s_1$ acts as a mass  in the Nambu space spanned by $s_3 \tau_{x,y}$. This mass changes its sign between two adjacent surfaces namely, $yz$ and $xz$. Both the blocks participate actively here as the mass term uniformly appears in both of them. Thus two MZMs per hinge are observed. On the other hand, in the presence of any of the in-plane Zeeman field, mass terms become different in the two blocks. This leads to a situation where one block can be made active keeping the other block inactive. As a result, one MZMs per hinge can be observed.  This behavior is again in resemblance with that of the 2D case. Therefore, the dynamical and explicit breaking of TRS imprints their signatures unanimously for 3D as well.  

\subsection{Hinge Mode Solution}
Having obtained low energy surface Hamiltonian, the hinge Hamiltonian can be estimated by considering the PBC (OBC) along (perpendicular to) hinge direction. We thus divide $H^{\rm Surf}$ into 
two parts as

\begin{eqnarray}
H_0^S&=&-i\lambda \partial_1 s_3 \tau_y + M(\theta) s_1 -B(\theta) \tau_z \ , \non \\
H^S_p&=&-\lambda k_2 s_3 \tau_x\ ,
\end{eqnarray}
Here the superconducting order parameter and Zeeman field are treated in the unperturbed Hamiltonian $H_0^S$. 
We solve $H_0^S$ exactly and expand $H^S_p$ in the basis of $H^S_0$. For $\vect{B}=0$, we obtain the Hamiltonian for the hinge mode (after replacing $k_2 \rightarrow k_z$) as 
\begin{equation}
H^{\rm Hinge}=-\lambda k_z \tau_y \ ,
\end{equation}
This manifests a propagating mode along $z$ direction. Therefore, it clear that dynamical breaking of TRS leads to two solutions of MHMs with $\tau_y= \pm 1$. 
Contrastingly for $B_x\neq 0$ or $B_y\neq 0$, we obtain the solution as 
\begin{equation}
H^{\rm Hinge}=-\lambda k_z \mathbb{I} \ , 
\end{equation}
which only hosts a single MHM as the hinge Hamiltonian is described by $\mathbb{I}$. Similar to the 2D case, here also the explicit breaking of TRS can lead to a situation different from the dynamical breaking of TRS by the periodic kick drive. Hence, one expects that without (with) magnetic field there exist two MHMs (one MHM) per hinge along $z$ direction as shown in the inset I2 of
Fig.\ \ref{3DLDOS}(b). 

\subsection{Floquet Wannier Spectrum}
\begin{figure}
	\centering
	\subfigure{\includegraphics[width=0.48\textwidth]{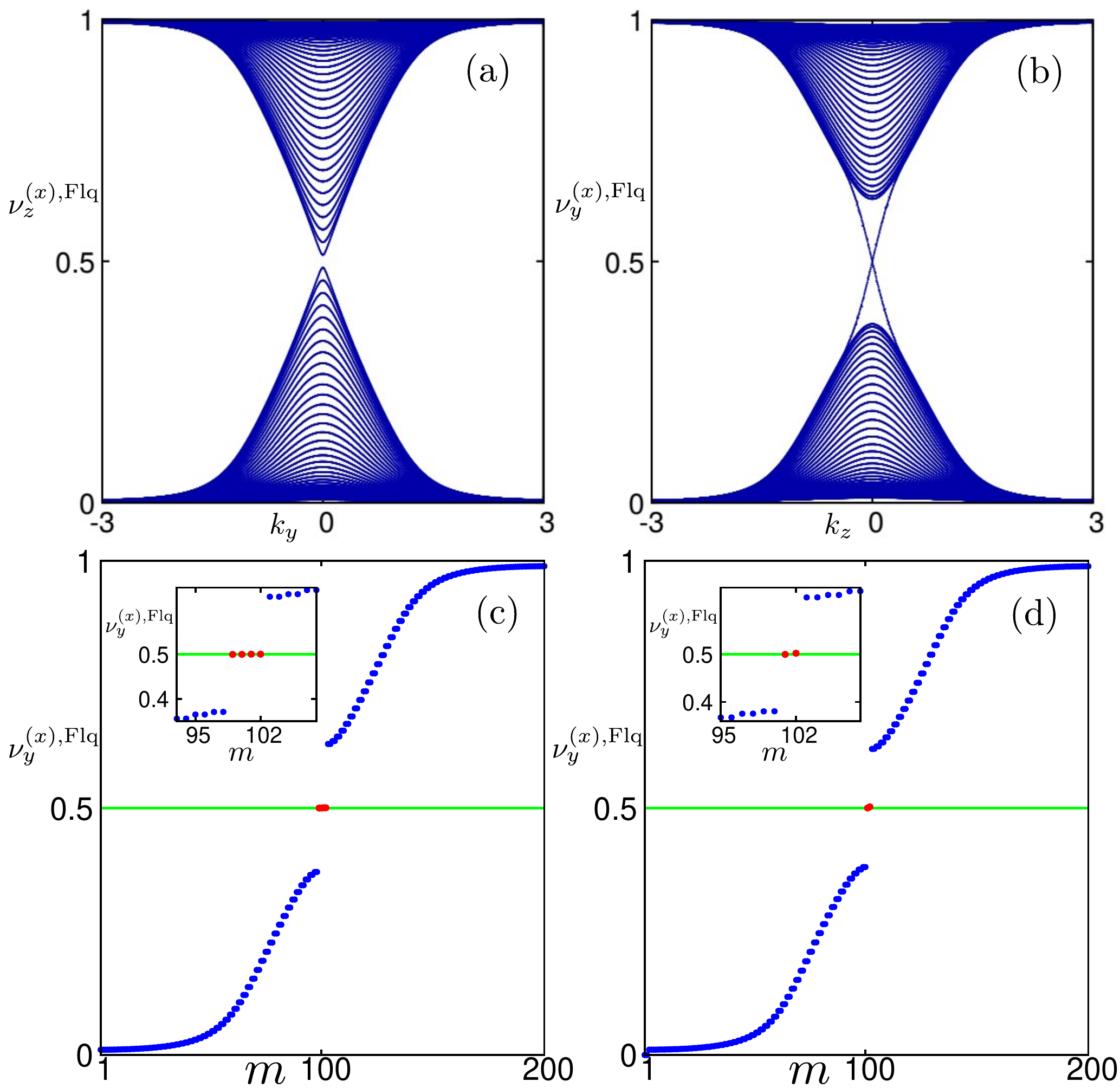}}
	\caption{(Color online) (a) We depict the FWS $\nu_z^{(x),\rm Flq}$ as a function of $k_y$ in the slab geometry with OBC along $x$ direction and PBC along $y$, $z$ directions. (b) We work in the same geometry as mentioned in panel (a) but show $\nu_y^{(x),\rm Flq}$ as a function of $k_z$. Since, the hinge mode propagates along $z$-direction, we obtain gapless FWS for $\nu_y^{(x), \rm Flq}$. 
	(c) We show the FWS $\nu_y^{(x),\rm Flq}$ as a function of state number at $k_z=0$ for $B_x=0$. In the inset one can clearly observe four eigenvalues at $0.5$ refer to the \textit{weak} FSOTSC phase. (d) We repeat (c) but with the in-plane Zeeman field $B_x\neq0$. We obtain two eigenvalue at $0.5$ as shown in the inset refer to \textit{strong} FSOTSC phase. The values of all the parameter remain same as in Fig.\ \ref{3DLDOS}.
	}
	\label{3DFWS}
\end{figure}

To calculate FWS using Eq.(\ref{WS}), we write down the Hamiltonian (Eq.(\ref{ham3D})) in slab geometry \ie we consider OBC in one direction while the other two directions continue to satisfy PBC. 
For a $x$-directed slab (OBC along $x$-direction; PBC along $y$ and $z$-direction), we can calculate $\nu_z^{(x),\rm Flq}$ ($\nu_y^{(x),\rm Flq}$) as a function of $k_y$ ($k_z$). Since we obtain propagating 1D hinge mode in $z$ direction only, the spectrum of $\nu_y^{(x),\rm Flq}$ and $\nu_x^{(y),\rm Flq}$ as a function of $k_z$ exhibits gapless nature, while all other FWS remain gapped. 
We illustrate the representative plots for FWS in Figs.\ \ref{3DFWS}(a)-(b). Focussing only at $k_z=0$ point, we show FWS as a function of the state-index at $k_z=0$ in Figs.\ \ref{3DFWS}(c)-(d). 
 In absence of the in-plane Zeeman field \ie for \textit{weak} FSOTSC phase, we obtain four eigenvalues at 0.5 correspondings to the two MHMs per hinge (see Fig. \ref{3DFWS}(c)). In contrast, when 
we turn on the in-plane Zeeman field, $B_x$ \ie for \textit{strong} FSOTSC, we obtain two eigenvalues at 0.5 (see Fig. \ref{3DFWS}(d)) corroborating one MHM per hinge. We also calculate the FQM to further distinguish between the \textit{weak} and \textit{strong} FSOTSC. To proceed, we write the Floquet operator $U(T)$ (Eq.(\ref{fo3d})) in rod geometry (considering PBC along $z$-direction, OBC along $x$ and $y$-direction). We then implement Eq.(\ref{qm_xy}) to calculate the FQM ($Q_{xy}^{\rm Flq}$) as a function of $k_z$. We find that $Q_{xy}^{\rm Flq}=0.5~(0)$ at $k_z=0$ for the \textit{strong} (\textit{weak}) phase~\cite{fu2020chiral}.

\section{Robustness of higher order Majorana modes against disorder}{\label{sec:IV}}

\begin{figure}[H]
	\centering
	\subfigure{\includegraphics[width=0.48\textwidth]{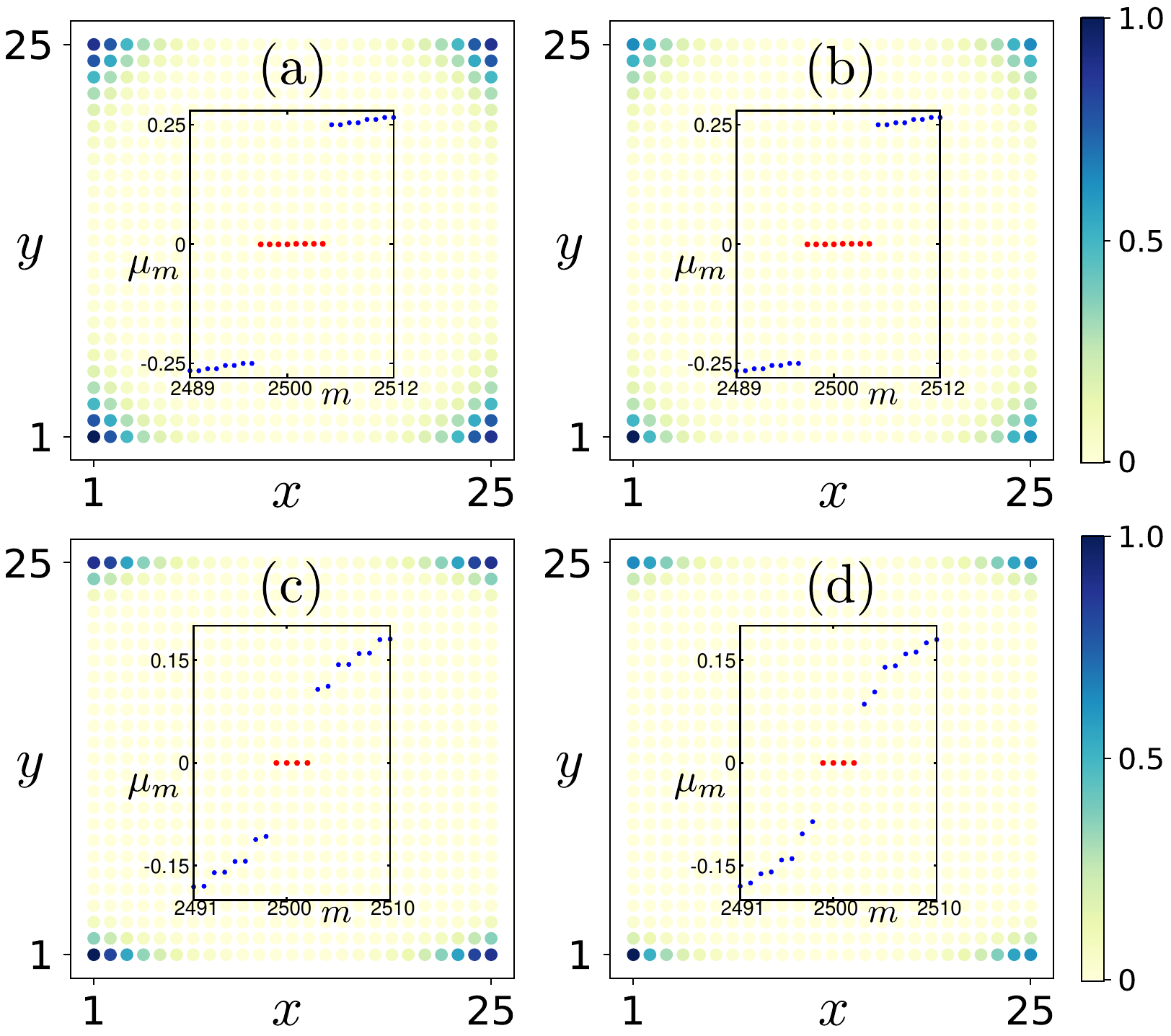}}
	\caption{(Color online)(a) LDOS, associated with zero quasi-energy states $\mu_m=0$, in finite geometry is demonstrated for $B_x=0$ and inset shows the Floquet quasi-energy spectrum for the same  
	for the disorder strength $W=0.1$.  (b) We repeat (a) with the disorder strength $W=0.4$. (c) and (d), we repeat (a) and (b) with $B_x=0.3$, respectively. One can clearly observe that the localization  
	properties become modified as the disorder strength increases substantially, otherwise for small disorder, they remain same as the clean case. Value of all other parameters remain same as in 
	Fig.\ \ref{LDOS}.
	}
	\label{LDOSDisorder}
\end{figure}

\begin{figure}[H]
	\centering
	\subfigure{\includegraphics[width=0.48\textwidth]{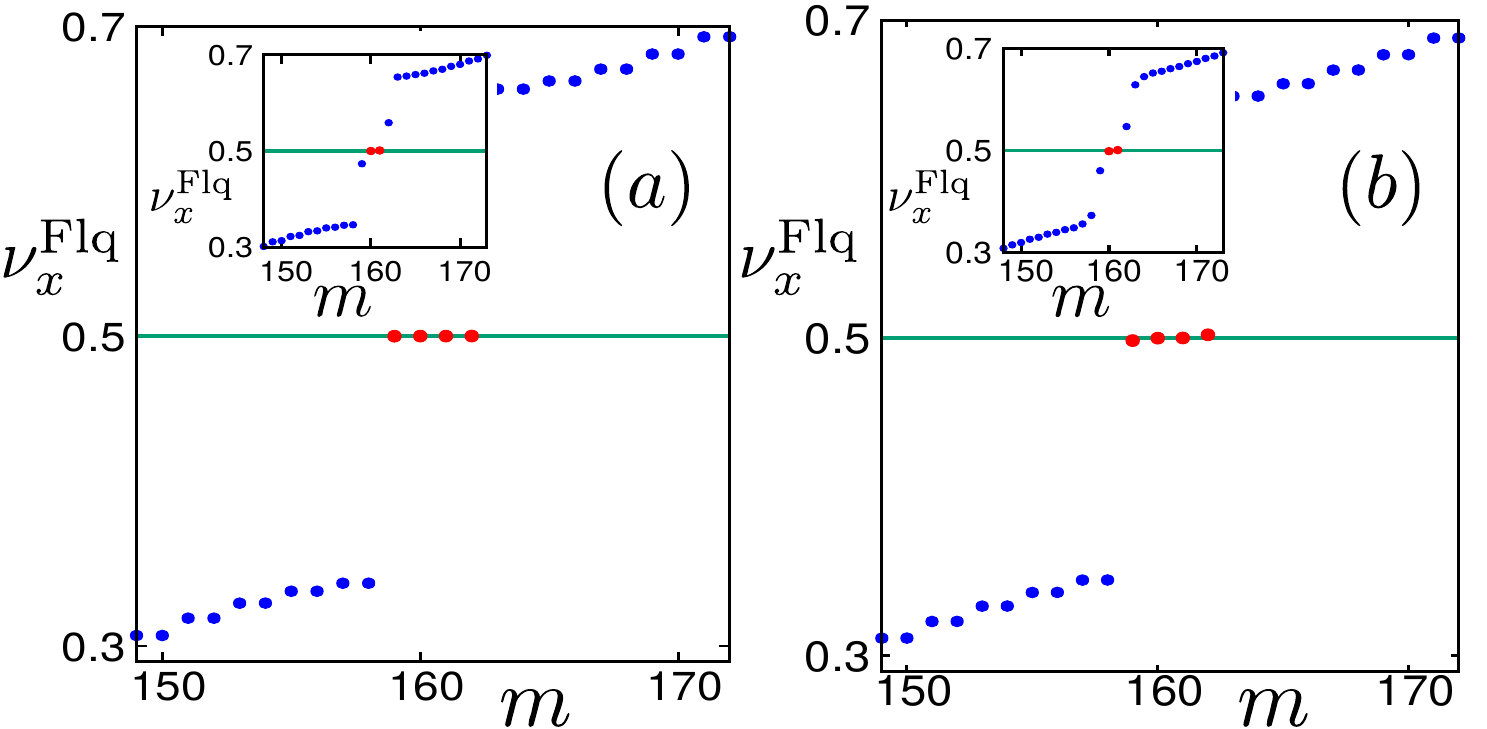}}
	\caption{(Color online) (a) FWS is depicted for the case $B_x=0$ when the disorder strength is chosen to be $W=0.1$, while the same for the case $B_x=0.3$ is shown in the inset.  
	(b) FWS is demonstrated with the disorder strength $W=0.4$ for $B_x=0$ and $B_x\neq 0$ (inset). One can obtain an indication that upon increasing the disorder strength substantially, the FWS  
	might deviate from the quantized value $0.5$. However, for our chosen disorder strength ($W<t_{x}, t_{y}, \lambda_{x}, \lambda_{y}$) the invariant behaves in robust manner referring to the fact 
	that these FSOTSC phases are stable against moderate disorder. Value of all the other parameters remain same as chosen in Fig.\ \ref{LDOS}.
	}
	\label{FWSDisorder}
\end{figure}

Having investigated the topological classification of the FSOTSC phases, we now focus on the robustness of these phases in presence of finite disorder. We first concentrate on the 2D case. Instead of choosing $m_1$ to be constant, we consider $m_1$ to be randomly distributed in between $\left[-\frac{W}{2},\frac{W}{2}\right]$ in an un-correlated manner. Here, $W$ being the strength of the disorder. 
 Note that, the disorder being random, we have taken average over $500$ disorder configurations in our numerical calculation. We analyze our results for two values of disorder strengths, $W=0.1$ and $0.4$ while hopping and spin-orbit coupling strength is fixed to unity. We first study the effect of disorder on the \textit{weak} phase hosting eight MCMs. For weak disorder strength, $W=0.1$, one does 
not observe any perceptible difference from the clean case (see Fig.\ \ref{LDOSDisorder}(a)). To be precise, the spectral weight of MCMs are uniformly distributed among all the four corners of the square lattice \ie the localization length remains almost unaltered in presence of weak disorder. While for $W=0.4$, we notice that the spectral weight of some corner modes in LDOS becomes higher compared to other as shown in Fig.\ \ref{LDOSDisorder}(b). This non-uniform distribution of spectral weight among the corners thus suggests that the strong disorder can substantially modify the localization properties of these MCMs. However, it is noteworthy that even strong disorder cannot lift the Majorana modes from the zero-energy as depicted in the inset of Figs.\ \ref{LDOSDisorder}(a) and (b). This results in the fact that FWS continues to exhibit similar behavior as compared to the clean case (see Figs.\ \ref{FWSDisorder}(a) and (b)). Therefore, \textit{weak} FSOTSC phase can preserve its signature in the presence of moderate disorder. 

 We also investigate the effect of moderate disorder in the \textit{strong} FSOTSC phase hosting four MZMs at the corners in the presence of explicit TRS breaking Zeeman field. Since the disorder does not break any further symmetry (except translational), we find that the effect of the disorder remains same as the earlier case with $B_x=0$. The uniform (non-uniform) distribution of spectral weight of MZMs 
is observed for $W=0.1$ ($0.4$) as depicted in Fig. \ \ref{LDOSDisorder}(c) and (d). These MCMs are localized at zero-energy always (see the inset of Fig.\ \ref{LDOSDisorder}(c) and (d)). As a result, 
their topological protection remains unaltered as noticed in the clean case. We find quantized FWS for weak as well as moderate disorder strength (see the inset of Figs.\ \ref{FWSDisorder}(a) and (b)).  
It is important to mention that the FQM is also quantised at a value $0.5$, for such strength of the disorder with $B_x \neq 0$.  This investigation with moderate disorder thus clearly suggests that both\textit{weak} and \textit{strong} FSOTSC phases are robust against moderate disorder. Having investigated the effect of disorder and protection of topological properties in 2D, we believe the same line 
of argument would also hold for 3D. 

\section{Summary and Conclusions}{\label{sec:V}}
To summarize, in this article, we provide a dynamical prescription to generate the FSOTSC phase starting from a 
2D/3D TI in close proximity to an unconventional $d$-wave superconductor. We periodically kick the mass term to generate MZMs at the corners and hinges in 2D square and 
3D cubic lattice geometry, respectively. Our aim here is to investigate the effect of TRS breaking magnetic field on these Floquet SOTSC phases as the initial (effective) Hamiltonian, describing the 
static (driven) system which preserves (breaks) the dynamical TRS. In 2D, we first consider our model in the presence of periodically kicked mass term when Zeeman field $B_x=0$. Here, we find 
FSOTSC phase harboring eight MZMs \ie two MZMs per corner. In order to characterize this FSOTSC phase, we compute two topological invariants namely, FWS and FQM. We find that this phase 
corresponds to four FWS quantized at $0.5$ while FQM vanishes; we identify this FSOTSC phase as the \textit{weak} one. Upon introduction of an explicit TRS breaking Zeeman field $B_x \neq 0$,  
we find four MZMs \ie one MZM per corner. We here obtain two quantized FWS at $0.5$ and quantized FQM referring to the \textit{strong} nature of the phase. We analytically support our numerical 
findings, as obtained by diagonalizing the Floquet operator in OBC, with the help of effective low-energy edge theory and MCMs solutions. The low-energy theory can successfully predict the nature 
of the SOTSC phase whether it is \textit{strong} or \textit{weak} in terms of the spinor states of MZMs. We further introduce moderate disorder in the driving amplitude to study the stability of the 
FSOTSC phase against disorder. We find that the FSOTSC is stable against the moderate strength of disorder. However, the localization property of the MZMs depends on the disorder strength. 
We also generalize our theory based on 3D model and identify the \textit{weak} (\textit{strong}) FSOTSC phase via FWS hosting MHMs. In this case, we derive the low energy surface theory and 
analytical solutions of the MHMs therein. The effect of the disorder remains similar as in 2D case.
 
As far as experimental feasibility of our setup is concerned, $d$-wave superconductivity in TI can be induced via the proximity effect (\eg $\rm Bi_{2}Sr_{2}CaCu_{2}O_{8+\delta}$)~\cite{ExperimentTI.dwave} with an induced gap amplitude $\Delta\sim 15~\rm meV$~\cite{ExperimentTI.dwave}. It has been theoretically~\cite{PhysRevB.81.041307} and experimentally~\cite{zhang2010crossover} demonstrated that the topological properties of (Bi,Sb)$_2$Te$_3$ thin films  can be tuned by the quantum confinement \ie varying the number of quintuple layers. It is indeed possible to tune the mass term in the underlying static model which might pave the way to realize the dynamical manipulation of mass term in the proximized topological superconductor.
In recent times, experimental advancements on the pump-probe techniques~\cite{Wang453,maczewsky2017observation,peng2016experimental} have enabled one to observe Floquet topological insulators~\cite{Wang453} and anomalous Hall effect in graphene~\cite{Experiment_LightInducedHall}.Therefore, we believe that the signature of MCMs and MHMs may be possible to achieve via pump-probe based time-resolved transport (\eg local scanning tunneling microscope (STM)) measurements~\cite{Experiment.MZM.STM, PhysRevB.99.214302, Tuovinen_2019} for an in-plane magnetic field $B_{x}\sim 7-8~\rm T$, amplitude of the drive $m_{1}\sim 100~\rm meV$ and period $T\sim 2~\rm fs$. 

At last, we would like to comment on robustness of these Floquet MZMs in HOTSC phases under periodic driving as far as heating and dissipation are concerned. Based on the recent theoretical and experimental investigations on quantum many-body systems~\cite{heating1,heating2}, the heating is suppressed in the prethermal window where our findings can be tested with dissipationless MZMs associated with the periodic steady state. We work in the high frequency regime away from the resonance points. This further enables us to minimize the heating 
effect~\cite{heating3}. We therefore believe that our theoretical findings do not suffer from heating issue and dissipation.

\acknowledgments{}
We acknowledge SAMKHYA: High-Performance Computing Facility provided by the Institute of Physics, Bhubaneswar, for our numerical computation. AKG thanks Atanu Jana for useful technical discussions.

\bibliography{bibfile}{}

\begin{thebibliography}{97}%
\makeatletter
\providecommand \@ifxundefined [1]{%
 \@ifx{#1\undefined}
}%
\providecommand \@ifnum [1]{%
 \ifnum #1\expandafter \@firstoftwo
 \else \expandafter \@secondoftwo
 \fi
}%
\providecommand \@ifx [1]{%
 \ifx #1\expandafter \@firstoftwo
 \else \expandafter \@secondoftwo
 \fi
}%
\providecommand \natexlab [1]{#1}%
\providecommand \enquote  [1]{``#1''}%
\providecommand \bibnamefont  [1]{#1}%
\providecommand \bibfnamefont [1]{#1}%
\providecommand \citenamefont [1]{#1}%
\providecommand \href@noop [0]{\@secondoftwo}%
\providecommand \href [0]{\begingroup \@sanitize@url \@href}%
\providecommand \@href[1]{\@@startlink{#1}\@@href}%
\providecommand \@@href[1]{\endgroup#1\@@endlink}%
\providecommand \@sanitize@url [0]{\catcode `\\12\catcode `\$12\catcode
  `\&12\catcode `\#12\catcode `\^12\catcode `\_12\catcode `\%12\relax}%
\providecommand \@@startlink[1]{}%
\providecommand \@@endlink[0]{}%
\providecommand \url  [0]{\begingroup\@sanitize@url \@url }%
\providecommand \@url [1]{\endgroup\@href {#1}{\urlprefix }}%
\providecommand \urlprefix  [0]{URL }%
\providecommand \Eprint [0]{\href }%
\providecommand \doibase [0]{http://dx.doi.org/}%
\providecommand \selectlanguage [0]{\@gobble}%
\providecommand \bibinfo  [0]{\@secondoftwo}%
\providecommand \bibfield  [0]{\@secondoftwo}%
\providecommand \translation [1]{[#1]}%
\providecommand \BibitemOpen [0]{}%
\providecommand \bibitemStop [0]{}%
\providecommand \bibitemNoStop [0]{.\EOS\space}%
\providecommand \EOS [0]{\spacefactor3000\relax}%
\providecommand \BibitemShut  [1]{\csname bibitem#1\endcsname}%
\let\auto@bib@innerbib\@empty
\bibitem [{\citenamefont {Kitaev}(2001)}]{Kitaev_2001}%
  \BibitemOpen
  \bibfield  {author} {\bibinfo {author} {\bibfnamefont {A~Yu}\ \bibnamefont
  {Kitaev}},\ }\bibfield  {title} {\enquote {\bibinfo {title} {Unpaired
  majorana fermions in quantum wires},}\ }\href {\doibase
  10.1070/1063-7869/44/10s/s29} {\bibfield  {journal} {\bibinfo  {journal}
  {Physics-Uspekhi}\ }\textbf {\bibinfo {volume} {44}},\ \bibinfo {pages}
  {131--136} (\bibinfo {year} {2001})}\BibitemShut {NoStop}%
\bibitem [{\citenamefont {Qi}\ and\ \citenamefont
  {Zhang}(2011)}]{qi2011topological}%
  \BibitemOpen
  \bibfield  {author} {\bibinfo {author} {\bibfnamefont {Xiao-Liang}\
  \bibnamefont {Qi}}\ and\ \bibinfo {author} {\bibfnamefont {Shou-Cheng}\
  \bibnamefont {Zhang}},\ }\bibfield  {title} {\enquote {\bibinfo {title}
  {Topological insulators and superconductors},}\ }\href {\doibase
  10.1103/RevModPhys.83.1057} {\bibfield  {journal} {\bibinfo  {journal} {Rev.
  Mod. Phys.}\ }\textbf {\bibinfo {volume} {83}},\ \bibinfo {pages} {1057}
  (\bibinfo {year} {2011})}\BibitemShut {NoStop}%
\bibitem [{\citenamefont {Hasan}\ and\ \citenamefont
  {Kane}(2010)}]{hasan2010colloquium}%
  \BibitemOpen
  \bibfield  {author} {\bibinfo {author} {\bibfnamefont {M~Zahid}\ \bibnamefont
  {Hasan}}\ and\ \bibinfo {author} {\bibfnamefont {Charles~L}\ \bibnamefont
  {Kane}},\ }\bibfield  {title} {\enquote {\bibinfo {title} {Colloquium:
  topological insulators},}\ }\href {\doibase 10.1103/RevModPhys.82.3045}
  {\bibfield  {journal} {\bibinfo  {journal} {Rev. Mod. Phys.}\ }\textbf
  {\bibinfo {volume} {82}},\ \bibinfo {pages} {3045} (\bibinfo {year}
  {2010})}\BibitemShut {NoStop}%
\bibitem [{\citenamefont {Das}\ \emph {et~al.}(2012)\citenamefont {Das},
  \citenamefont {Ronen}, \citenamefont {Most}, \citenamefont {Oreg},
  \citenamefont {Heiblum},\ and\ \citenamefont {Shtrikman}}]{das2012zero}%
  \BibitemOpen
  \bibfield  {author} {\bibinfo {author} {\bibfnamefont {Anindya}\ \bibnamefont
  {Das}}, \bibinfo {author} {\bibfnamefont {Yuval}\ \bibnamefont {Ronen}},
  \bibinfo {author} {\bibfnamefont {Yonatan}\ \bibnamefont {Most}}, \bibinfo
  {author} {\bibfnamefont {Yuval}\ \bibnamefont {Oreg}}, \bibinfo {author}
  {\bibfnamefont {Moty}\ \bibnamefont {Heiblum}}, \ and\ \bibinfo {author}
  {\bibfnamefont {Hadas}\ \bibnamefont {Shtrikman}},\ }\bibfield  {title}
  {\enquote {\bibinfo {title} {Zero-bias peaks and splitting in an al--inas
  nanowire topological superconductor as a signature of majorana fermions},}\
  }\href {\doibase https://doi.org/10.1038/nphys2479} {\bibfield  {journal}
  {\bibinfo  {journal} {Nature Physics}\ }\textbf {\bibinfo {volume} {8}},\
  \bibinfo {pages} {887--895} (\bibinfo {year} {2012})}\BibitemShut {NoStop}%
\bibitem [{\citenamefont {Deng}\ \emph {et~al.}(2016)\citenamefont {Deng},
  \citenamefont {Vaitiekenas}, \citenamefont {Hansen}, \citenamefont {Danon},
  \citenamefont {Leijnse}, \citenamefont {Flensberg}, \citenamefont {Nyg{\r
  a}rd}, \citenamefont {Krogstrup},\ and\ \citenamefont {Marcus}}]{Deng1557}%
  \BibitemOpen
  \bibfield  {author} {\bibinfo {author} {\bibfnamefont {M.~T.}\ \bibnamefont
  {Deng}}, \bibinfo {author} {\bibfnamefont {S.}~\bibnamefont {Vaitiekenas}},
  \bibinfo {author} {\bibfnamefont {E.~B.}\ \bibnamefont {Hansen}}, \bibinfo
  {author} {\bibfnamefont {J.}~\bibnamefont {Danon}}, \bibinfo {author}
  {\bibfnamefont {M.}~\bibnamefont {Leijnse}}, \bibinfo {author} {\bibfnamefont
  {K.}~\bibnamefont {Flensberg}}, \bibinfo {author} {\bibfnamefont
  {J.}~\bibnamefont {Nyg{\r a}rd}}, \bibinfo {author} {\bibfnamefont
  {P.}~\bibnamefont {Krogstrup}}, \ and\ \bibinfo {author} {\bibfnamefont
  {C.~M.}\ \bibnamefont {Marcus}},\ }\bibfield  {title} {\enquote {\bibinfo
  {title} {Majorana bound state in a coupled quantum-dot hybrid-nanowire
  system},}\ }\href {\doibase 10.1126/science.aaf3961} {\bibfield  {journal}
  {\bibinfo  {journal} {Science}\ }\textbf {\bibinfo {volume} {354}},\ \bibinfo
  {pages} {1557--1562} (\bibinfo {year} {2016})}\BibitemShut {NoStop}%
\bibitem [{\citenamefont {Ivanov}(2001)}]{Ivanov2001}%
  \BibitemOpen
  \bibfield  {author} {\bibinfo {author} {\bibfnamefont {D.~A.}\ \bibnamefont
  {Ivanov}},\ }\bibfield  {title} {\enquote {\bibinfo {title} {Non-abelian
  statistics of half-quantum vortices in $\mathit{p}$-wave superconductors},}\
  }\href {\doibase 10.1103/PhysRevLett.86.268} {\bibfield  {journal} {\bibinfo
  {journal} {Phys. Rev. Lett.}\ }\textbf {\bibinfo {volume} {86}},\ \bibinfo
  {pages} {268--271} (\bibinfo {year} {2001})}\BibitemShut {NoStop}%
\bibitem [{\citenamefont {Nayak}\ \emph {et~al.}(2008)\citenamefont {Nayak},
  \citenamefont {Simon}, \citenamefont {Stern}, \citenamefont {Freedman},\ and\
  \citenamefont {Das~Sarma}}]{nayak08}%
  \BibitemOpen
  \bibfield  {author} {\bibinfo {author} {\bibfnamefont {Chetan}\ \bibnamefont
  {Nayak}}, \bibinfo {author} {\bibfnamefont {Steven~H.}\ \bibnamefont
  {Simon}}, \bibinfo {author} {\bibfnamefont {Ady}\ \bibnamefont {Stern}},
  \bibinfo {author} {\bibfnamefont {Michael}\ \bibnamefont {Freedman}}, \ and\
  \bibinfo {author} {\bibfnamefont {Sankar}\ \bibnamefont {Das~Sarma}},\
  }\bibfield  {title} {\enquote {\bibinfo {title} {Non-abelian anyons and
  topological quantum computation},}\ }\href {\doibase
  10.1103/RevModPhys.80.1083} {\bibfield  {journal} {\bibinfo  {journal} {Rev.
  Mod. Phys.}\ }\textbf {\bibinfo {volume} {80}},\ \bibinfo {pages}
  {1083--1159} (\bibinfo {year} {2008})}\BibitemShut {NoStop}%
\bibitem [{\citenamefont {Fu}\ and\ \citenamefont {Kane}(2008)}]{Fu2008}%
  \BibitemOpen
  \bibfield  {author} {\bibinfo {author} {\bibfnamefont {Liang}\ \bibnamefont
  {Fu}}\ and\ \bibinfo {author} {\bibfnamefont {C.~L.}\ \bibnamefont {Kane}},\
  }\bibfield  {title} {\enquote {\bibinfo {title} {Superconducting proximity
  effect and majorana fermions at the surface of a topological insulator},}\
  }\href {\doibase 10.1103/PhysRevLett.100.096407} {\bibfield  {journal}
  {\bibinfo  {journal} {Phys. Rev. Lett.}\ }\textbf {\bibinfo {volume} {100}},\
  \bibinfo {pages} {096407} (\bibinfo {year} {2008})}\BibitemShut {NoStop}%
\bibitem [{\citenamefont {Sau}\ \emph {et~al.}(2010)\citenamefont {Sau},
  \citenamefont {Lutchyn}, \citenamefont {Tewari},\ and\ \citenamefont
  {Das~Sarma}}]{Sau2010}%
  \BibitemOpen
  \bibfield  {author} {\bibinfo {author} {\bibfnamefont {Jay~D.}\ \bibnamefont
  {Sau}}, \bibinfo {author} {\bibfnamefont {Roman~M.}\ \bibnamefont {Lutchyn}},
  \bibinfo {author} {\bibfnamefont {Sumanta}\ \bibnamefont {Tewari}}, \ and\
  \bibinfo {author} {\bibfnamefont {S.}~\bibnamefont {Das~Sarma}},\ }\bibfield
  {title} {\enquote {\bibinfo {title} {Generic new platform for topological
  quantum computation using semiconductor heterostructures},}\ }\href {\doibase
  10.1103/PhysRevLett.104.040502} {\bibfield  {journal} {\bibinfo  {journal}
  {Phys. Rev. Lett.}\ }\textbf {\bibinfo {volume} {104}},\ \bibinfo {pages}
  {040502} (\bibinfo {year} {2010})}\BibitemShut {NoStop}%
\bibitem [{\citenamefont {Lutchyn}\ \emph {et~al.}(2010)\citenamefont
  {Lutchyn}, \citenamefont {Sau},\ and\ \citenamefont {Das~Sarma}}]{Lutchyn10}%
  \BibitemOpen
  \bibfield  {author} {\bibinfo {author} {\bibfnamefont {Roman~M.}\
  \bibnamefont {Lutchyn}}, \bibinfo {author} {\bibfnamefont {Jay~D.}\
  \bibnamefont {Sau}}, \ and\ \bibinfo {author} {\bibfnamefont
  {S.}~\bibnamefont {Das~Sarma}},\ }\bibfield  {title} {\enquote {\bibinfo
  {title} {Majorana fermions and a topological phase transition in
  semiconductor-superconductor heterostructures},}\ }\href {\doibase
  10.1103/PhysRevLett.105.077001} {\bibfield  {journal} {\bibinfo  {journal}
  {Phys. Rev. Lett.}\ }\textbf {\bibinfo {volume} {105}},\ \bibinfo {pages}
  {077001} (\bibinfo {year} {2010})}\BibitemShut {NoStop}%
\bibitem [{\citenamefont {Qi}\ \emph {et~al.}(2010)\citenamefont {Qi},
  \citenamefont {Hughes},\ and\ \citenamefont {Zhang}}]{Hughes2010}%
  \BibitemOpen
  \bibfield  {author} {\bibinfo {author} {\bibfnamefont {Xiao-Liang}\
  \bibnamefont {Qi}}, \bibinfo {author} {\bibfnamefont {Taylor~L.}\
  \bibnamefont {Hughes}}, \ and\ \bibinfo {author} {\bibfnamefont {Shou-Cheng}\
  \bibnamefont {Zhang}},\ }\bibfield  {title} {\enquote {\bibinfo {title}
  {Chiral topological superconductor from the quantum hall state},}\ }\href
  {\doibase 10.1103/PhysRevB.82.184516} {\bibfield  {journal} {\bibinfo
  {journal} {Phys. Rev. B}\ }\textbf {\bibinfo {volume} {82}},\ \bibinfo
  {pages} {184516} (\bibinfo {year} {2010})}\BibitemShut {NoStop}%
\bibitem [{\citenamefont {Oreg}\ \emph {et~al.}(2010)\citenamefont {Oreg},
  \citenamefont {Refael},\ and\ \citenamefont {von Oppen}}]{Oreg2010}%
  \BibitemOpen
  \bibfield  {author} {\bibinfo {author} {\bibfnamefont {Yuval}\ \bibnamefont
  {Oreg}}, \bibinfo {author} {\bibfnamefont {Gil}\ \bibnamefont {Refael}}, \
  and\ \bibinfo {author} {\bibfnamefont {Felix}\ \bibnamefont {von Oppen}},\
  }\bibfield  {title} {\enquote {\bibinfo {title} {Helical liquids and majorana
  bound states in quantum wires},}\ }\href {\doibase
  10.1103/PhysRevLett.105.177002} {\bibfield  {journal} {\bibinfo  {journal}
  {Phys. Rev. Lett.}\ }\textbf {\bibinfo {volume} {105}},\ \bibinfo {pages}
  {177002} (\bibinfo {year} {2010})}\BibitemShut {NoStop}%
\bibitem [{\citenamefont {Benalcazar}\ \emph
  {et~al.}(2017{\natexlab{a}})\citenamefont {Benalcazar}, \citenamefont
  {Bernevig},\ and\ \citenamefont {Hughes}}]{benalcazar2017}%
  \BibitemOpen
  \bibfield  {author} {\bibinfo {author} {\bibfnamefont {Wladimir~A}\
  \bibnamefont {Benalcazar}}, \bibinfo {author} {\bibfnamefont {B~Andrei}\
  \bibnamefont {Bernevig}}, \ and\ \bibinfo {author} {\bibfnamefont {Taylor~L}\
  \bibnamefont {Hughes}},\ }\bibfield  {title} {\enquote {\bibinfo {title}
  {Quantized electric multipole insulators},}\ }\href {\doibase
  https://doi.org/10.1126/science.aah6442} {\bibfield  {journal} {\bibinfo
  {journal} {Science}\ }\textbf {\bibinfo {volume} {357}},\ \bibinfo {pages}
  {61--66} (\bibinfo {year} {2017}{\natexlab{a}})}\BibitemShut {NoStop}%
\bibitem [{\citenamefont {Benalcazar}\ \emph
  {et~al.}(2017{\natexlab{b}})\citenamefont {Benalcazar}, \citenamefont
  {Bernevig},\ and\ \citenamefont {Hughes}}]{benalcazarprb2017}%
  \BibitemOpen
  \bibfield  {author} {\bibinfo {author} {\bibfnamefont {Wladimir~A}\
  \bibnamefont {Benalcazar}}, \bibinfo {author} {\bibfnamefont {B~Andrei}\
  \bibnamefont {Bernevig}}, \ and\ \bibinfo {author} {\bibfnamefont {Taylor~L}\
  \bibnamefont {Hughes}},\ }\bibfield  {title} {\enquote {\bibinfo {title}
  {Electric multipole moments, topological multipole moment pumping, and chiral
  hinge states in crystalline insulators},}\ }\href {\doibase
  10.1103/PhysRevB.96.245115} {\bibfield  {journal} {\bibinfo  {journal} {Phys.
  Rev. B}\ }\textbf {\bibinfo {volume} {96}},\ \bibinfo {pages} {245115}
  (\bibinfo {year} {2017}{\natexlab{b}})}\BibitemShut {NoStop}%
\bibitem [{\citenamefont {Song}\ \emph {et~al.}(2017)\citenamefont {Song},
  \citenamefont {Fang},\ and\ \citenamefont {Fang}}]{Song2017}%
  \BibitemOpen
  \bibfield  {author} {\bibinfo {author} {\bibfnamefont {Zhida}\ \bibnamefont
  {Song}}, \bibinfo {author} {\bibfnamefont {Zhong}\ \bibnamefont {Fang}}, \
  and\ \bibinfo {author} {\bibfnamefont {Chen}\ \bibnamefont {Fang}},\
  }\bibfield  {title} {\enquote {\bibinfo {title}
  {$(d\ensuremath{-}2)$-dimensional edge states of rotation symmetry protected
  topological states},}\ }\href {\doibase 10.1103/PhysRevLett.119.246402}
  {\bibfield  {journal} {\bibinfo  {journal} {Phys. Rev. Lett.}\ }\textbf
  {\bibinfo {volume} {119}},\ \bibinfo {pages} {246402} (\bibinfo {year}
  {2017})}\BibitemShut {NoStop}%
\bibitem [{\citenamefont {Langbehn}\ \emph {et~al.}(2017)\citenamefont
  {Langbehn}, \citenamefont {Peng}, \citenamefont {Trifunovic}, \citenamefont
  {von Oppen},\ and\ \citenamefont {Brouwer}}]{Langbehn2017}%
  \BibitemOpen
  \bibfield  {author} {\bibinfo {author} {\bibfnamefont {Josias}\ \bibnamefont
  {Langbehn}}, \bibinfo {author} {\bibfnamefont {Yang}\ \bibnamefont {Peng}},
  \bibinfo {author} {\bibfnamefont {Luka}\ \bibnamefont {Trifunovic}}, \bibinfo
  {author} {\bibfnamefont {Felix}\ \bibnamefont {von Oppen}}, \ and\ \bibinfo
  {author} {\bibfnamefont {Piet~W.}\ \bibnamefont {Brouwer}},\ }\bibfield
  {title} {\enquote {\bibinfo {title} {Reflection-symmetric second-order
  topological insulators and superconductors},}\ }\href {\doibase
  10.1103/PhysRevLett.119.246401} {\bibfield  {journal} {\bibinfo  {journal}
  {Phys. Rev. Lett.}\ }\textbf {\bibinfo {volume} {119}},\ \bibinfo {pages}
  {246401} (\bibinfo {year} {2017})}\BibitemShut {NoStop}%
\bibitem [{\citenamefont {Schindler}\ \emph {et~al.}(2018)\citenamefont
  {Schindler}, \citenamefont {Cook}, \citenamefont {Vergniory}, \citenamefont
  {Wang}, \citenamefont {Parkin}, \citenamefont {Bernevig},\ and\ \citenamefont
  {Neupert}}]{schindler2018}%
  \BibitemOpen
  \bibfield  {author} {\bibinfo {author} {\bibfnamefont {Frank}\ \bibnamefont
  {Schindler}}, \bibinfo {author} {\bibfnamefont {Ashley~M}\ \bibnamefont
  {Cook}}, \bibinfo {author} {\bibfnamefont {Maia~G}\ \bibnamefont
  {Vergniory}}, \bibinfo {author} {\bibfnamefont {Zhijun}\ \bibnamefont
  {Wang}}, \bibinfo {author} {\bibfnamefont {Stuart~SP}\ \bibnamefont
  {Parkin}}, \bibinfo {author} {\bibfnamefont {B~Andrei}\ \bibnamefont
  {Bernevig}}, \ and\ \bibinfo {author} {\bibfnamefont {Titus}\ \bibnamefont
  {Neupert}},\ }\bibfield  {title} {\enquote {\bibinfo {title} {Higher-order
  topological insulators},}\ }\href {\doibase
  https://doi.org/10.1126/sciadv.aat0346} {\bibfield  {journal} {\bibinfo
  {journal} {Science adv.}\ }\textbf {\bibinfo {volume} {4}},\ \bibinfo {pages}
  {eaat0346} (\bibinfo {year} {2018})}\BibitemShut {NoStop}%
\bibitem [{\citenamefont {Khalaf}(2018)}]{Khalaf2018}%
  \BibitemOpen
  \bibfield  {author} {\bibinfo {author} {\bibfnamefont {Eslam}\ \bibnamefont
  {Khalaf}},\ }\bibfield  {title} {\enquote {\bibinfo {title} {Higher-order
  topological insulators and superconductors protected by inversion
  symmetry},}\ }\href {\doibase 10.1103/PhysRevB.97.205136} {\bibfield
  {journal} {\bibinfo  {journal} {Phys. Rev. B}\ }\textbf {\bibinfo {volume}
  {97}},\ \bibinfo {pages} {205136} (\bibinfo {year} {2018})}\BibitemShut
  {NoStop}%
\bibitem [{\citenamefont {Geier}\ \emph {et~al.}(2018)\citenamefont {Geier},
  \citenamefont {Trifunovic}, \citenamefont {Hoskam},\ and\ \citenamefont
  {Brouwer}}]{Geier2018}%
  \BibitemOpen
  \bibfield  {author} {\bibinfo {author} {\bibfnamefont {Max}\ \bibnamefont
  {Geier}}, \bibinfo {author} {\bibfnamefont {Luka}\ \bibnamefont
  {Trifunovic}}, \bibinfo {author} {\bibfnamefont {Max}\ \bibnamefont
  {Hoskam}}, \ and\ \bibinfo {author} {\bibfnamefont {Piet~W.}\ \bibnamefont
  {Brouwer}},\ }\bibfield  {title} {\enquote {\bibinfo {title} {Second-order
  topological insulators and superconductors with an order-two crystalline
  symmetry},}\ }\href {\doibase 10.1103/PhysRevB.97.205135} {\bibfield
  {journal} {\bibinfo  {journal} {Phys. Rev. B}\ }\textbf {\bibinfo {volume}
  {97}},\ \bibinfo {pages} {205135} (\bibinfo {year} {2018})}\BibitemShut
  {NoStop}%
\bibitem [{\citenamefont {Franca}\ \emph {et~al.}(2018)\citenamefont {Franca},
  \citenamefont {van~den Brink},\ and\ \citenamefont {Fulga}}]{Franca2018}%
  \BibitemOpen
  \bibfield  {author} {\bibinfo {author} {\bibfnamefont {S.}~\bibnamefont
  {Franca}}, \bibinfo {author} {\bibfnamefont {J.}~\bibnamefont {van~den
  Brink}}, \ and\ \bibinfo {author} {\bibfnamefont {I.~C.}\ \bibnamefont
  {Fulga}},\ }\bibfield  {title} {\enquote {\bibinfo {title} {An anomalous
  higher-order topological insulator},}\ }\href {\doibase
  10.1103/PhysRevB.98.201114} {\bibfield  {journal} {\bibinfo  {journal} {Phys.
  Rev. B}\ }\textbf {\bibinfo {volume} {98}},\ \bibinfo {pages} {201114}
  (\bibinfo {year} {2018})}\BibitemShut {NoStop}%
\bibitem [{\citenamefont {Zhu}(2018)}]{Zhu2018}%
  \BibitemOpen
  \bibfield  {author} {\bibinfo {author} {\bibfnamefont {Xiaoyu}\ \bibnamefont
  {Zhu}},\ }\bibfield  {title} {\enquote {\bibinfo {title} {Tunable majorana
  corner states in a two-dimensional second-order topological superconductor
  induced by magnetic fields},}\ }\href {\doibase 10.1103/PhysRevB.97.205134}
  {\bibfield  {journal} {\bibinfo  {journal} {Phys. Rev. B}\ }\textbf {\bibinfo
  {volume} {97}},\ \bibinfo {pages} {205134} (\bibinfo {year}
  {2018})}\BibitemShut {NoStop}%
\bibitem [{\citenamefont {Liu}\ \emph {et~al.}(2018)\citenamefont {Liu},
  \citenamefont {He},\ and\ \citenamefont {Nori}}]{Liu2018}%
  \BibitemOpen
  \bibfield  {author} {\bibinfo {author} {\bibfnamefont {Tao}\ \bibnamefont
  {Liu}}, \bibinfo {author} {\bibfnamefont {James~Jun}\ \bibnamefont {He}}, \
  and\ \bibinfo {author} {\bibfnamefont {Franco}\ \bibnamefont {Nori}},\
  }\bibfield  {title} {\enquote {\bibinfo {title} {Majorana corner states in a
  two-dimensional magnetic topological insulator on a high-temperature
  superconductor},}\ }\href {\doibase 10.1103/PhysRevB.98.245413} {\bibfield
  {journal} {\bibinfo  {journal} {Phys. Rev. B}\ }\textbf {\bibinfo {volume}
  {98}},\ \bibinfo {pages} {245413} (\bibinfo {year} {2018})}\BibitemShut
  {NoStop}%
\bibitem [{\citenamefont {Yan}\ \emph {et~al.}(2018)\citenamefont {Yan},
  \citenamefont {Song},\ and\ \citenamefont {Wang}}]{Yan2018}%
  \BibitemOpen
  \bibfield  {author} {\bibinfo {author} {\bibfnamefont {Zhongbo}\ \bibnamefont
  {Yan}}, \bibinfo {author} {\bibfnamefont {Fei}\ \bibnamefont {Song}}, \ and\
  \bibinfo {author} {\bibfnamefont {Zhong}\ \bibnamefont {Wang}},\ }\bibfield
  {title} {\enquote {\bibinfo {title} {Majorana corner modes in a
  high-temperature platform},}\ }\href {\doibase
  10.1103/PhysRevLett.121.096803} {\bibfield  {journal} {\bibinfo  {journal}
  {Phys. Rev. Lett.}\ }\textbf {\bibinfo {volume} {121}},\ \bibinfo {pages}
  {096803} (\bibinfo {year} {2018})}\BibitemShut {NoStop}%
\bibitem [{\citenamefont {Wang}\ \emph {et~al.}(2019)\citenamefont {Wang},
  \citenamefont {Wieder}, \citenamefont {Li}, \citenamefont {Yan},\ and\
  \citenamefont {Bernevig}}]{wang2018higher}%
  \BibitemOpen
  \bibfield  {author} {\bibinfo {author} {\bibfnamefont {Zhijun}\ \bibnamefont
  {Wang}}, \bibinfo {author} {\bibfnamefont {Benjamin~J.}\ \bibnamefont
  {Wieder}}, \bibinfo {author} {\bibfnamefont {Jian}\ \bibnamefont {Li}},
  \bibinfo {author} {\bibfnamefont {Binghai}\ \bibnamefont {Yan}}, \ and\
  \bibinfo {author} {\bibfnamefont {B.~Andrei}\ \bibnamefont {Bernevig}},\
  }\bibfield  {title} {\enquote {\bibinfo {title} {Higher-order topology,
  monopole nodal lines, and the origin of large fermi arcs in transition metal
  dichalcogenides $x{\mathrm{te}}_{2}$ ($x=\mathrm{Mo},\mathrm{W}$)},}\ }\href
  {\doibase 10.1103/PhysRevLett.123.186401} {\bibfield  {journal} {\bibinfo
  {journal} {Phys. Rev. Lett.}\ }\textbf {\bibinfo {volume} {123}},\ \bibinfo
  {pages} {186401} (\bibinfo {year} {2019})}\BibitemShut {NoStop}%
\bibitem [{\citenamefont {Wang}\ \emph {et~al.}(2018)\citenamefont {Wang},
  \citenamefont {Lin},\ and\ \citenamefont {Hughes}}]{WangWeak2018}%
  \BibitemOpen
  \bibfield  {author} {\bibinfo {author} {\bibfnamefont {Yuxuan}\ \bibnamefont
  {Wang}}, \bibinfo {author} {\bibfnamefont {Mao}\ \bibnamefont {Lin}}, \ and\
  \bibinfo {author} {\bibfnamefont {Taylor~L.}\ \bibnamefont {Hughes}},\
  }\bibfield  {title} {\enquote {\bibinfo {title} {Weak-pairing higher order
  topological superconductors},}\ }\href {\doibase 10.1103/PhysRevB.98.165144}
  {\bibfield  {journal} {\bibinfo  {journal} {Phys. Rev. B}\ }\textbf {\bibinfo
  {volume} {98}},\ \bibinfo {pages} {165144} (\bibinfo {year}
  {2018})}\BibitemShut {NoStop}%
\bibitem [{\citenamefont {Ezawa}(2018)}]{Ezawakagome}%
  \BibitemOpen
  \bibfield  {author} {\bibinfo {author} {\bibfnamefont {Motohiko}\
  \bibnamefont {Ezawa}},\ }\bibfield  {title} {\enquote {\bibinfo {title}
  {Higher-order topological insulators and semimetals on the breathing kagome
  and pyrochlore lattices},}\ }\href {\doibase 10.1103/PhysRevLett.120.026801}
  {\bibfield  {journal} {\bibinfo  {journal} {Phys. Rev. Lett.}\ }\textbf
  {\bibinfo {volume} {120}},\ \bibinfo {pages} {026801} (\bibinfo {year}
  {2018})}\BibitemShut {NoStop}%
\bibitem [{\citenamefont {C\ifmmode \u{a}\else \u{a}\fi{}lug\ifmmode~\u{a}\else
  \u{a}\fi{}ru}\ \emph {et~al.}(2019)\citenamefont {C\ifmmode \u{a}\else
  \u{a}\fi{}lug\ifmmode~\u{a}\else \u{a}\fi{}ru}, \citenamefont {Juri\ifmmode
  \check{c}\else \v{c}\fi{}i\ifmmode~\acute{c}\else \'{c}\fi{}},\ and\
  \citenamefont {Roy}}]{Roy2019}%
  \BibitemOpen
  \bibfield  {author} {\bibinfo {author} {\bibfnamefont {Dumitru}\ \bibnamefont
  {C\ifmmode \u{a}\else \u{a}\fi{}lug\ifmmode~\u{a}\else \u{a}\fi{}ru}},
  \bibinfo {author} {\bibfnamefont {Vladimir}\ \bibnamefont {Juri\ifmmode
  \check{c}\else \v{c}\fi{}i\ifmmode~\acute{c}\else \'{c}\fi{}}}, \ and\
  \bibinfo {author} {\bibfnamefont {Bitan}\ \bibnamefont {Roy}},\ }\bibfield
  {title} {\enquote {\bibinfo {title} {Higher-order topological phases: A
  general principle of construction},}\ }\href {\doibase
  10.1103/PhysRevB.99.041301} {\bibfield  {journal} {\bibinfo  {journal} {Phys.
  Rev. B}\ }\textbf {\bibinfo {volume} {99}},\ \bibinfo {pages} {041301}
  (\bibinfo {year} {2019})}\BibitemShut {NoStop}%
\bibitem [{\citenamefont {Trifunovic}\ and\ \citenamefont
  {Brouwer}(2019)}]{Trifunovic2019}%
  \BibitemOpen
  \bibfield  {author} {\bibinfo {author} {\bibfnamefont {Luka}\ \bibnamefont
  {Trifunovic}}\ and\ \bibinfo {author} {\bibfnamefont {Piet~W.}\ \bibnamefont
  {Brouwer}},\ }\bibfield  {title} {\enquote {\bibinfo {title} {Higher-order
  bulk-boundary correspondence for topological crystalline phases},}\ }\href
  {\doibase 10.1103/PhysRevX.9.011012} {\bibfield  {journal} {\bibinfo
  {journal} {Phys. Rev. X}\ }\textbf {\bibinfo {volume} {9}},\ \bibinfo {pages}
  {011012} (\bibinfo {year} {2019})}\BibitemShut {NoStop}%
\bibitem [{\citenamefont {Zeng}\ \emph {et~al.}(2019)\citenamefont {Zeng},
  \citenamefont {Stanescu}, \citenamefont {Zhang}, \citenamefont {Scarola},\
  and\ \citenamefont {Tewari}}]{ZengPRL2019}%
  \BibitemOpen
  \bibfield  {author} {\bibinfo {author} {\bibfnamefont {Chuanchang}\
  \bibnamefont {Zeng}}, \bibinfo {author} {\bibfnamefont {T.~D.}\ \bibnamefont
  {Stanescu}}, \bibinfo {author} {\bibfnamefont {Chuanwei}\ \bibnamefont
  {Zhang}}, \bibinfo {author} {\bibfnamefont {V.~W.}\ \bibnamefont {Scarola}},
  \ and\ \bibinfo {author} {\bibfnamefont {Sumanta}\ \bibnamefont {Tewari}},\
  }\bibfield  {title} {\enquote {\bibinfo {title} {Majorana corner modes with
  solitons in an attractive hubbard-hofstadter model of cold atom optical
  lattices},}\ }\href {\doibase 10.1103/PhysRevLett.123.060402} {\bibfield
  {journal} {\bibinfo  {journal} {Phys. Rev. Lett.}\ }\textbf {\bibinfo
  {volume} {123}},\ \bibinfo {pages} {060402} (\bibinfo {year}
  {2019})}\BibitemShut {NoStop}%
\bibitem [{\citenamefont {Zhang}\ \emph
  {et~al.}(2019{\natexlab{a}})\citenamefont {Zhang}, \citenamefont {Cole},\
  and\ \citenamefont {Das~Sarma}}]{Zhang2019}%
  \BibitemOpen
  \bibfield  {author} {\bibinfo {author} {\bibfnamefont {Rui-Xing}\
  \bibnamefont {Zhang}}, \bibinfo {author} {\bibfnamefont {William~S.}\
  \bibnamefont {Cole}}, \ and\ \bibinfo {author} {\bibfnamefont
  {S.}~\bibnamefont {Das~Sarma}},\ }\bibfield  {title} {\enquote {\bibinfo
  {title} {Helical hinge majorana modes in iron-based superconductors},}\
  }\href {\doibase 10.1103/PhysRevLett.122.187001} {\bibfield  {journal}
  {\bibinfo  {journal} {Phys. Rev. Lett.}\ }\textbf {\bibinfo {volume} {122}},\
  \bibinfo {pages} {187001} (\bibinfo {year} {2019}{\natexlab{a}})}\BibitemShut
  {NoStop}%
\bibitem [{\citenamefont {Volpez}\ \emph {et~al.}(2019)\citenamefont {Volpez},
  \citenamefont {Loss},\ and\ \citenamefont {Klinovaja}}]{Volpez2019}%
  \BibitemOpen
  \bibfield  {author} {\bibinfo {author} {\bibfnamefont {Yanick}\ \bibnamefont
  {Volpez}}, \bibinfo {author} {\bibfnamefont {Daniel}\ \bibnamefont {Loss}}, \
  and\ \bibinfo {author} {\bibfnamefont {Jelena}\ \bibnamefont {Klinovaja}},\
  }\bibfield  {title} {\enquote {\bibinfo {title} {Second-order topological
  superconductivity in $\ensuremath{\pi}$-junction rashba layers},}\ }\href
  {\doibase 10.1103/PhysRevLett.122.126402} {\bibfield  {journal} {\bibinfo
  {journal} {Phys. Rev. Lett.}\ }\textbf {\bibinfo {volume} {122}},\ \bibinfo
  {pages} {126402} (\bibinfo {year} {2019})}\BibitemShut {NoStop}%
\bibitem [{\citenamefont {Yan}(2019)}]{YanPRB2019}%
  \BibitemOpen
  \bibfield  {author} {\bibinfo {author} {\bibfnamefont {Zhongbo}\ \bibnamefont
  {Yan}},\ }\bibfield  {title} {\enquote {\bibinfo {title} {Majorana corner and
  hinge modes in second-order topological insulator/superconductor
  heterostructures},}\ }\href {\doibase 10.1103/PhysRevB.100.205406} {\bibfield
   {journal} {\bibinfo  {journal} {Phys. Rev. B}\ }\textbf {\bibinfo {volume}
  {100}},\ \bibinfo {pages} {205406} (\bibinfo {year} {2019})}\BibitemShut
  {NoStop}%
\bibitem [{\citenamefont {Ghorashi}\ \emph {et~al.}(2019)\citenamefont
  {Ghorashi}, \citenamefont {Hu}, \citenamefont {Hughes},\ and\ \citenamefont
  {Rossi}}]{Ghorashi2019}%
  \BibitemOpen
  \bibfield  {author} {\bibinfo {author} {\bibfnamefont {Sayed Ali~Akbar}\
  \bibnamefont {Ghorashi}}, \bibinfo {author} {\bibfnamefont {Xiang}\
  \bibnamefont {Hu}}, \bibinfo {author} {\bibfnamefont {Taylor~L.}\
  \bibnamefont {Hughes}}, \ and\ \bibinfo {author} {\bibfnamefont {Enrico}\
  \bibnamefont {Rossi}},\ }\bibfield  {title} {\enquote {\bibinfo {title}
  {Second-order dirac superconductors and magnetic field induced majorana hinge
  modes},}\ }\href {\doibase 10.1103/PhysRevB.100.020509} {\bibfield  {journal}
  {\bibinfo  {journal} {Phys. Rev. B}\ }\textbf {\bibinfo {volume} {100}},\
  \bibinfo {pages} {020509} (\bibinfo {year} {2019})}\BibitemShut {NoStop}%
\bibitem [{\citenamefont {Ghorashi}\ \emph {et~al.}(2020)\citenamefont
  {Ghorashi}, \citenamefont {Hughes},\ and\ \citenamefont
  {Rossi}}]{GhorashiPRL2020}%
  \BibitemOpen
  \bibfield  {author} {\bibinfo {author} {\bibfnamefont {Sayed Ali~Akbar}\
  \bibnamefont {Ghorashi}}, \bibinfo {author} {\bibfnamefont {Taylor~L.}\
  \bibnamefont {Hughes}}, \ and\ \bibinfo {author} {\bibfnamefont {Enrico}\
  \bibnamefont {Rossi}},\ }\bibfield  {title} {\enquote {\bibinfo {title}
  {Vortex and surface phase transitions in superconducting higher-order
  topological insulators},}\ }\href {\doibase 10.1103/PhysRevLett.125.037001}
  {\bibfield  {journal} {\bibinfo  {journal} {Phys. Rev. Lett.}\ }\textbf
  {\bibinfo {volume} {125}},\ \bibinfo {pages} {037001} (\bibinfo {year}
  {2020})}\BibitemShut {NoStop}%
\bibitem [{\citenamefont {De}\ \emph {et~al.}(2020)\citenamefont {De},
  \citenamefont {Khanna},\ and\ \citenamefont {Rao}}]{Sumathi2020}%
  \BibitemOpen
  \bibfield  {author} {\bibinfo {author} {\bibfnamefont {Suman~Jyoti}\
  \bibnamefont {De}}, \bibinfo {author} {\bibfnamefont {Udit}\ \bibnamefont
  {Khanna}}, \ and\ \bibinfo {author} {\bibfnamefont {Sumathi}\ \bibnamefont
  {Rao}},\ }\bibfield  {title} {\enquote {\bibinfo {title} {Magnetic flux
  periodicity in second order topological superconductors},}\ }\href {\doibase
  10.1103/PhysRevB.101.125429} {\bibfield  {journal} {\bibinfo  {journal}
  {Phys. Rev. B}\ }\textbf {\bibinfo {volume} {101}},\ \bibinfo {pages}
  {125429} (\bibinfo {year} {2020})}\BibitemShut {NoStop}%
\bibitem [{\citenamefont {Wu}\ \emph {et~al.}(2020)\citenamefont {Wu},
  \citenamefont {Hou}, \citenamefont {Li}, \citenamefont {Luo}, \citenamefont
  {Shi},\ and\ \citenamefont {Zhang}}]{Wu2020}%
  \BibitemOpen
  \bibfield  {author} {\bibinfo {author} {\bibfnamefont {Ya-Jie}\ \bibnamefont
  {Wu}}, \bibinfo {author} {\bibfnamefont {Junpeng}\ \bibnamefont {Hou}},
  \bibinfo {author} {\bibfnamefont {Yun-Mei}\ \bibnamefont {Li}}, \bibinfo
  {author} {\bibfnamefont {Xi-Wang}\ \bibnamefont {Luo}}, \bibinfo {author}
  {\bibfnamefont {Xiaoyan}\ \bibnamefont {Shi}}, \ and\ \bibinfo {author}
  {\bibfnamefont {Chuanwei}\ \bibnamefont {Zhang}},\ }\bibfield  {title}
  {\enquote {\bibinfo {title} {In-plane zeeman-field-induced majorana corner
  and hinge modes in an $s$-wave superconductor heterostructure},}\ }\href
  {\doibase 10.1103/PhysRevLett.124.227001} {\bibfield  {journal} {\bibinfo
  {journal} {Phys. Rev. Lett.}\ }\textbf {\bibinfo {volume} {124}},\ \bibinfo
  {pages} {227001} (\bibinfo {year} {2020})}\BibitemShut {NoStop}%
\bibitem [{\citenamefont {Laubscher}\ \emph {et~al.}()\citenamefont
  {Laubscher}, \citenamefont {Chughtai}, \citenamefont {Loss},\ and\
  \citenamefont {Klinovaja}}]{jelena2020HOTSC}%
  \BibitemOpen
  \bibfield  {author} {\bibinfo {author} {\bibfnamefont {K.}~\bibnamefont
  {Laubscher}}, \bibinfo {author} {\bibfnamefont {D.}~\bibnamefont {Chughtai}},
  \bibinfo {author} {\bibfnamefont {D.}~\bibnamefont {Loss}}, \ and\ \bibinfo
  {author} {\bibfnamefont {J}~\bibnamefont {Klinovaja}},\ }\bibfield  {title}
  {\enquote {\bibinfo {title} {Kramers pairs of majorana corner states in a
  topological insulator bilayer},}\ }\href@noop {} {\ }\Eprint
  {http://arxiv.org/abs/2007.13579} {arXiv:2007.13579 [cond-mat.mes-hall]}
  \BibitemShut {NoStop}%
\bibitem [{\citenamefont {Roy}(2020)}]{BitanTSC2020}%
  \BibitemOpen
  \bibfield  {author} {\bibinfo {author} {\bibfnamefont {Bitan}\ \bibnamefont
  {Roy}},\ }\bibfield  {title} {\enquote {\bibinfo {title} {Higher-order
  topological superconductors in $\mathcal{P}$-, $\mathcal{T}$-odd quadrupolar
  dirac materials},}\ }\href {\doibase 10.1103/PhysRevB.101.220506} {\bibfield
  {journal} {\bibinfo  {journal} {Phys. Rev. B}\ }\textbf {\bibinfo {volume}
  {101}},\ \bibinfo {pages} {220506} (\bibinfo {year} {2020})}\BibitemShut
  {NoStop}%
\bibitem [{\citenamefont {Zhang}\ and\ \citenamefont
  {Trauzettel}(2020)}]{SongboPRR12020}%
  \BibitemOpen
  \bibfield  {author} {\bibinfo {author} {\bibfnamefont {Song-Bo}\ \bibnamefont
  {Zhang}}\ and\ \bibinfo {author} {\bibfnamefont {Bj\"orn}\ \bibnamefont
  {Trauzettel}},\ }\bibfield  {title} {\enquote {\bibinfo {title} {Detection of
  second-order topological superconductors by josephson junctions},}\ }\href
  {\doibase 10.1103/PhysRevResearch.2.012018} {\bibfield  {journal} {\bibinfo
  {journal} {Phys. Rev. Research}\ }\textbf {\bibinfo {volume} {2}},\ \bibinfo
  {pages} {012018} (\bibinfo {year} {2020})}\BibitemShut {NoStop}%
\bibitem [{\citenamefont {Zhang}\ \emph
  {et~al.}(2020{\natexlab{a}})\citenamefont {Zhang}, \citenamefont {Rui},
  \citenamefont {Calzona}, \citenamefont {Choi}, \citenamefont {Schnyder},\
  and\ \citenamefont {Trauzettel}}]{SongboPRR22020}%
  \BibitemOpen
  \bibfield  {author} {\bibinfo {author} {\bibfnamefont {Song-Bo}\ \bibnamefont
  {Zhang}}, \bibinfo {author} {\bibfnamefont {W.~B.}\ \bibnamefont {Rui}},
  \bibinfo {author} {\bibfnamefont {Alessio}\ \bibnamefont {Calzona}}, \bibinfo
  {author} {\bibfnamefont {Sang-Jun}\ \bibnamefont {Choi}}, \bibinfo {author}
  {\bibfnamefont {Andreas~P.}\ \bibnamefont {Schnyder}}, \ and\ \bibinfo
  {author} {\bibfnamefont {Bj\"orn}\ \bibnamefont {Trauzettel}},\ }\bibfield
  {title} {\enquote {\bibinfo {title} {Topological and holonomic quantum
  computation based on second-order topological superconductors},}\ }\href
  {\doibase 10.1103/PhysRevResearch.2.043025} {\bibfield  {journal} {\bibinfo
  {journal} {Phys. Rev. Research}\ }\textbf {\bibinfo {volume} {2}},\ \bibinfo
  {pages} {043025} (\bibinfo {year} {2020}{\natexlab{a}})}\BibitemShut
  {NoStop}%
\bibitem [{\citenamefont {Zhang}\ \emph
  {et~al.}(2020{\natexlab{b}})\citenamefont {Zhang}, \citenamefont {Calzona},\
  and\ \citenamefont {Trauzettel}}]{SongboPRB2020}%
  \BibitemOpen
  \bibfield  {author} {\bibinfo {author} {\bibfnamefont {Song-Bo}\ \bibnamefont
  {Zhang}}, \bibinfo {author} {\bibfnamefont {Alessio}\ \bibnamefont
  {Calzona}}, \ and\ \bibinfo {author} {\bibfnamefont {Bj\"orn}\ \bibnamefont
  {Trauzettel}},\ }\bibfield  {title} {\enquote {\bibinfo {title}
  {All-electrically tunable networks of majorana bound states},}\ }\href
  {\doibase 10.1103/PhysRevB.102.100503} {\bibfield  {journal} {\bibinfo
  {journal} {Phys. Rev. B}\ }\textbf {\bibinfo {volume} {102}},\ \bibinfo
  {pages} {100503} (\bibinfo {year} {2020}{\natexlab{b}})}\BibitemShut
  {NoStop}%
\bibitem [{\citenamefont {Kheirkhah}\ \emph {et~al.}()\citenamefont
  {Kheirkhah}, \citenamefont {Yan},\ and\ \citenamefont
  {Marsiglio}}]{kheirkhah2020vortex}%
  \BibitemOpen
  \bibfield  {author} {\bibinfo {author} {\bibfnamefont {Majid}\ \bibnamefont
  {Kheirkhah}}, \bibinfo {author} {\bibfnamefont {Zhongbo}\ \bibnamefont
  {Yan}}, \ and\ \bibinfo {author} {\bibfnamefont {Frank}\ \bibnamefont
  {Marsiglio}},\ }\bibfield  {title} {\enquote {\bibinfo {title} {Vortex line
  topology in iron-based superconductors with and without second-order
  topology},}\ }\href@noop {} {\ }\Eprint {http://arxiv.org/abs/2007.10326}
  {arXiv:2007.10326 [cond-mat.supr-con]} \BibitemShut {NoStop}%
\bibitem [{\citenamefont {Plekhanov}\ \emph {et~al.}()\citenamefont
  {Plekhanov}, \citenamefont {M{\"u}ller}, \citenamefont {Volpez},
  \citenamefont {Kennes}, \citenamefont {Schoeller}, \citenamefont {Loss},\
  and\ \citenamefont {Klinovaja}}]{PlekhanovArxiv2020}%
  \BibitemOpen
  \bibfield  {author} {\bibinfo {author} {\bibfnamefont {K.}~\bibnamefont
  {Plekhanov}}, \bibinfo {author} {\bibfnamefont {N.}~\bibnamefont
  {M{\"u}ller}}, \bibinfo {author} {\bibfnamefont {Y.}~\bibnamefont {Volpez}},
  \bibinfo {author} {\bibfnamefont {D.~M.}\ \bibnamefont {Kennes}}, \bibinfo
  {author} {\bibfnamefont {H.}~\bibnamefont {Schoeller}}, \bibinfo {author}
  {\bibfnamefont {D.}~\bibnamefont {Loss}}, \ and\ \bibinfo {author}
  {\bibfnamefont {J.}~\bibnamefont {Klinovaja}},\ }\bibfield  {title} {\enquote
  {\bibinfo {title} {Quadrupole spin polarization as signature of second-order
  topological superconductors},}\ }\href@noop {} {\ }\Eprint
  {http://arxiv.org/abs/2008.03611} {arXiv:2008.03611 [cond-mat.mes-hall]}
  \BibitemShut {NoStop}%
\bibitem [{\citenamefont {Xue}\ \emph {et~al.}(2019)\citenamefont {Xue},
  \citenamefont {Yang}, \citenamefont {Gao}, \citenamefont {Chong},\ and\
  \citenamefont {Zhang}}]{XueAcousticKagome}%
  \BibitemOpen
  \bibfield  {author} {\bibinfo {author} {\bibfnamefont {Haoran}\ \bibnamefont
  {Xue}}, \bibinfo {author} {\bibfnamefont {Yahui}\ \bibnamefont {Yang}},
  \bibinfo {author} {\bibfnamefont {Fei}\ \bibnamefont {Gao}}, \bibinfo
  {author} {\bibfnamefont {Yidong}\ \bibnamefont {Chong}}, \ and\ \bibinfo
  {author} {\bibfnamefont {Baile}\ \bibnamefont {Zhang}},\ }\bibfield  {title}
  {\enquote {\bibinfo {title} {Acoustic higher-order topological insulator on a
  kagome lattice},}\ }\href {\doibase 10.1038/s41563-018-0251-x} {\bibfield
  {journal} {\bibinfo  {journal} {Nature Materials}\ }\textbf {\bibinfo
  {volume} {18}},\ \bibinfo {pages} {108--112} (\bibinfo {year}
  {2019})}\BibitemShut {NoStop}%
\bibitem [{\citenamefont {Chen}\ \emph {et~al.}(2019)\citenamefont {Chen},
  \citenamefont {Deng}, \citenamefont {Shi}, \citenamefont {Zhao},
  \citenamefont {Chen},\ and\ \citenamefont {Dong}}]{PhotonicChen}%
  \BibitemOpen
  \bibfield  {author} {\bibinfo {author} {\bibfnamefont {Xiao-Dong}\
  \bibnamefont {Chen}}, \bibinfo {author} {\bibfnamefont {Wei-Min}\
  \bibnamefont {Deng}}, \bibinfo {author} {\bibfnamefont {Fu-Long}\
  \bibnamefont {Shi}}, \bibinfo {author} {\bibfnamefont {Fu-Li}\ \bibnamefont
  {Zhao}}, \bibinfo {author} {\bibfnamefont {Min}\ \bibnamefont {Chen}}, \ and\
  \bibinfo {author} {\bibfnamefont {Jian-Wen}\ \bibnamefont {Dong}},\
  }\bibfield  {title} {\enquote {\bibinfo {title} {Direct observation of corner
  states in second-order topological photonic crystal slabs},}\ }\href
  {\doibase 10.1103/PhysRevLett.122.233902} {\bibfield  {journal} {\bibinfo
  {journal} {Phys. Rev. Lett.}\ }\textbf {\bibinfo {volume} {122}},\ \bibinfo
  {pages} {233902} (\bibinfo {year} {2019})}\BibitemShut {NoStop}%
\bibitem [{\citenamefont {Xie}\ \emph {et~al.}(2019)\citenamefont {Xie},
  \citenamefont {Su}, \citenamefont {Wang}, \citenamefont {Su}, \citenamefont
  {Shen}, \citenamefont {Zhan}, \citenamefont {Lu}, \citenamefont {Wang},\ and\
  \citenamefont {Chen}}]{PhotonicXie}%
  \BibitemOpen
  \bibfield  {author} {\bibinfo {author} {\bibfnamefont {Bi-Ye}\ \bibnamefont
  {Xie}}, \bibinfo {author} {\bibfnamefont {Guang-Xu}\ \bibnamefont {Su}},
  \bibinfo {author} {\bibfnamefont {Hong-Fei}\ \bibnamefont {Wang}}, \bibinfo
  {author} {\bibfnamefont {Hai}\ \bibnamefont {Su}}, \bibinfo {author}
  {\bibfnamefont {Xiao-Peng}\ \bibnamefont {Shen}}, \bibinfo {author}
  {\bibfnamefont {Peng}\ \bibnamefont {Zhan}}, \bibinfo {author} {\bibfnamefont
  {Ming-Hui}\ \bibnamefont {Lu}}, \bibinfo {author} {\bibfnamefont {Zhen-Lin}\
  \bibnamefont {Wang}}, \ and\ \bibinfo {author} {\bibfnamefont {Yan-Feng}\
  \bibnamefont {Chen}},\ }\bibfield  {title} {\enquote {\bibinfo {title}
  {Visualization of higher-order topological insulating phases in
  two-dimensional dielectric photonic crystals},}\ }\href {\doibase
  10.1103/PhysRevLett.122.233903} {\bibfield  {journal} {\bibinfo  {journal}
  {Phys. Rev. Lett.}\ }\textbf {\bibinfo {volume} {122}},\ \bibinfo {pages}
  {233903} (\bibinfo {year} {2019})}\BibitemShut {NoStop}%
\bibitem [{\citenamefont {Imhof}\ \emph {et~al.}(2018)\citenamefont {Imhof},
  \citenamefont {Berger}, \citenamefont {Bayer}, \citenamefont {Brehm},
  \citenamefont {Molenkamp}, \citenamefont {Kiessling}, \citenamefont
  {Schindler}, \citenamefont {Lee}, \citenamefont {Greiter}, \citenamefont
  {Neupert},\ and\ \citenamefont {Thomale}}]{Imhof2018}%
  \BibitemOpen
  \bibfield  {author} {\bibinfo {author} {\bibfnamefont {Stefan}\ \bibnamefont
  {Imhof}}, \bibinfo {author} {\bibfnamefont {Christian}\ \bibnamefont
  {Berger}}, \bibinfo {author} {\bibfnamefont {Florian}\ \bibnamefont {Bayer}},
  \bibinfo {author} {\bibfnamefont {Johannes}\ \bibnamefont {Brehm}}, \bibinfo
  {author} {\bibfnamefont {Laurens~W.}\ \bibnamefont {Molenkamp}}, \bibinfo
  {author} {\bibfnamefont {Tobias}\ \bibnamefont {Kiessling}}, \bibinfo
  {author} {\bibfnamefont {Frank}\ \bibnamefont {Schindler}}, \bibinfo {author}
  {\bibfnamefont {Ching~Hua}\ \bibnamefont {Lee}}, \bibinfo {author}
  {\bibfnamefont {Martin}\ \bibnamefont {Greiter}}, \bibinfo {author}
  {\bibfnamefont {Titus}\ \bibnamefont {Neupert}}, \ and\ \bibinfo {author}
  {\bibfnamefont {Ronny}\ \bibnamefont {Thomale}},\ }\bibfield  {title}
  {\enquote {\bibinfo {title} {Topolectrical-circuit realization of topological
  corner modes},}\ }\href {\doibase 10.1038/s41567-018-0246-1} {\bibfield
  {journal} {\bibinfo  {journal} {Nature Phys.}\ }\textbf {\bibinfo {volume}
  {14}},\ \bibinfo {pages} {925--929} (\bibinfo {year} {2018})}\BibitemShut
  {NoStop}%
\bibitem [{\citenamefont {Lindner}\ \emph {et~al.}(2011)\citenamefont
  {Lindner}, \citenamefont {Refael},\ and\ \citenamefont
  {Galitski}}]{lindner2011floquet}%
  \BibitemOpen
  \bibfield  {author} {\bibinfo {author} {\bibfnamefont {Netanel~H}\
  \bibnamefont {Lindner}}, \bibinfo {author} {\bibfnamefont {Gil}\ \bibnamefont
  {Refael}}, \ and\ \bibinfo {author} {\bibfnamefont {Victor}\ \bibnamefont
  {Galitski}},\ }\bibfield  {title} {\enquote {\bibinfo {title} {Floquet
  topological insulator in semiconductor quantum wells},}\ }\href {\doibase
  https://doi.org/10.1038/nphys1926} {\bibfield  {journal} {\bibinfo  {journal}
  {Nature Physics}\ }\textbf {\bibinfo {volume} {7}},\ \bibinfo {pages}
  {490--495} (\bibinfo {year} {2011})}\BibitemShut {NoStop}%
\bibitem [{\citenamefont {D\'ora}\ \emph {et~al.}(2012)\citenamefont {D\'ora},
  \citenamefont {Cayssol}, \citenamefont {Simon},\ and\ \citenamefont
  {Moessner}}]{Dora12}%
  \BibitemOpen
  \bibfield  {author} {\bibinfo {author} {\bibfnamefont {Bal\'azs}\
  \bibnamefont {D\'ora}}, \bibinfo {author} {\bibfnamefont {J\'er\^ome}\
  \bibnamefont {Cayssol}}, \bibinfo {author} {\bibfnamefont {Ferenc}\
  \bibnamefont {Simon}}, \ and\ \bibinfo {author} {\bibfnamefont {Roderich}\
  \bibnamefont {Moessner}},\ }\bibfield  {title} {\enquote {\bibinfo {title}
  {Optically engineering the topological properties of a spin hall
  insulator},}\ }\href {\doibase 10.1103/PhysRevLett.108.056602} {\bibfield
  {journal} {\bibinfo  {journal} {Phys. Rev. Lett.}\ }\textbf {\bibinfo
  {volume} {108}},\ \bibinfo {pages} {056602} (\bibinfo {year}
  {2012})}\BibitemShut {NoStop}%
\bibitem [{\citenamefont {Rudner}\ \emph {et~al.}(2013)\citenamefont {Rudner},
  \citenamefont {Lindner}, \citenamefont {Berg},\ and\ \citenamefont
  {Levin}}]{Rudner2013}%
  \BibitemOpen
  \bibfield  {author} {\bibinfo {author} {\bibfnamefont {Mark~S.}\ \bibnamefont
  {Rudner}}, \bibinfo {author} {\bibfnamefont {Netanel~H.}\ \bibnamefont
  {Lindner}}, \bibinfo {author} {\bibfnamefont {Erez}\ \bibnamefont {Berg}}, \
  and\ \bibinfo {author} {\bibfnamefont {Michael}\ \bibnamefont {Levin}},\
  }\bibfield  {title} {\enquote {\bibinfo {title} {Anomalous edge states and
  the bulk-edge correspondence for periodically driven two-dimensional
  systems},}\ }\href {\doibase 10.1103/PhysRevX.3.031005} {\bibfield  {journal}
  {\bibinfo  {journal} {Phys. Rev. X}\ }\textbf {\bibinfo {volume} {3}},\
  \bibinfo {pages} {031005} (\bibinfo {year} {2013})}\BibitemShut {NoStop}%
\bibitem [{\citenamefont {Thakurathi}\ \emph {et~al.}(2013)\citenamefont
  {Thakurathi}, \citenamefont {Patel}, \citenamefont {Sen},\ and\ \citenamefont
  {Dutta}}]{Thakurathi2013}%
  \BibitemOpen
  \bibfield  {author} {\bibinfo {author} {\bibfnamefont {Manisha}\ \bibnamefont
  {Thakurathi}}, \bibinfo {author} {\bibfnamefont {Aavishkar~A.}\ \bibnamefont
  {Patel}}, \bibinfo {author} {\bibfnamefont {Diptiman}\ \bibnamefont {Sen}}, \
  and\ \bibinfo {author} {\bibfnamefont {Amit}\ \bibnamefont {Dutta}},\
  }\bibfield  {title} {\enquote {\bibinfo {title} {Floquet generation of
  majorana end modes and topological invariants},}\ }\href {\doibase
  10.1103/PhysRevB.88.155133} {\bibfield  {journal} {\bibinfo  {journal} {Phys.
  Rev. B}\ }\textbf {\bibinfo {volume} {88}},\ \bibinfo {pages} {155133}
  (\bibinfo {year} {2013})}\BibitemShut {NoStop}%
\bibitem [{\citenamefont {Rechtsman}\ \emph {et~al.}(2013)\citenamefont
  {Rechtsman}, \citenamefont {Zeuner}, \citenamefont {Plotnik}, \citenamefont
  {Lumer}, \citenamefont {Podolsky}, \citenamefont {Dreisow}, \citenamefont
  {Nolte}, \citenamefont {Segev},\ and\ \citenamefont
  {Szameit}}]{rechtsman2013photonic}%
  \BibitemOpen
  \bibfield  {author} {\bibinfo {author} {\bibfnamefont {Mikael~C}\
  \bibnamefont {Rechtsman}}, \bibinfo {author} {\bibfnamefont {Julia~M}\
  \bibnamefont {Zeuner}}, \bibinfo {author} {\bibfnamefont {Yonatan}\
  \bibnamefont {Plotnik}}, \bibinfo {author} {\bibfnamefont {Yaakov}\
  \bibnamefont {Lumer}}, \bibinfo {author} {\bibfnamefont {Daniel}\
  \bibnamefont {Podolsky}}, \bibinfo {author} {\bibfnamefont {Felix}\
  \bibnamefont {Dreisow}}, \bibinfo {author} {\bibfnamefont {Stefan}\
  \bibnamefont {Nolte}}, \bibinfo {author} {\bibfnamefont {Mordechai}\
  \bibnamefont {Segev}}, \ and\ \bibinfo {author} {\bibfnamefont {Alexander}\
  \bibnamefont {Szameit}},\ }\bibfield  {title} {\enquote {\bibinfo {title}
  {Photonic floquet topological insulators},}\ }\href {\doibase
  https://doi.org/10.1038/nature12066} {\bibfield  {journal} {\bibinfo
  {journal} {Nature}\ }\textbf {\bibinfo {volume} {496}},\ \bibinfo {pages}
  {196--200} (\bibinfo {year} {2013})}\BibitemShut {NoStop}%
\bibitem [{\citenamefont {Maczewsky}\ \emph {et~al.}(2017)\citenamefont
  {Maczewsky}, \citenamefont {Zeuner}, \citenamefont {Nolte},\ and\
  \citenamefont {Szameit}}]{maczewsky2017observation}%
  \BibitemOpen
  \bibfield  {author} {\bibinfo {author} {\bibfnamefont {Lukas~J}\ \bibnamefont
  {Maczewsky}}, \bibinfo {author} {\bibfnamefont {Julia~M}\ \bibnamefont
  {Zeuner}}, \bibinfo {author} {\bibfnamefont {Stefan}\ \bibnamefont {Nolte}},
  \ and\ \bibinfo {author} {\bibfnamefont {Alexander}\ \bibnamefont
  {Szameit}},\ }\bibfield  {title} {\enquote {\bibinfo {title} {Observation of
  photonic anomalous floquet topological insulators},}\ }\href {\doibase
  https://doi.org/10.1038/ncomms13756} {\bibfield  {journal} {\bibinfo
  {journal} {Nature communications}\ }\textbf {\bibinfo {volume} {8}},\
  \bibinfo {pages} {13756} (\bibinfo {year} {2017})}\BibitemShut {NoStop}%
\bibitem [{\citenamefont {Eckardt}(2017)}]{Eckardt2017}%
  \BibitemOpen
  \bibfield  {author} {\bibinfo {author} {\bibfnamefont {Andr\'e}\ \bibnamefont
  {Eckardt}},\ }\bibfield  {title} {\enquote {\bibinfo {title} {Colloquium:
  Atomic quantum gases in periodically driven optical lattices},}\ }\href
  {\doibase 10.1103/RevModPhys.89.011004} {\bibfield  {journal} {\bibinfo
  {journal} {Rev. Mod. Phys.}\ }\textbf {\bibinfo {volume} {89}},\ \bibinfo
  {pages} {011004} (\bibinfo {year} {2017})}\BibitemShut {NoStop}%
\bibitem [{\citenamefont {Bomantara}\ \emph {et~al.}(2019)\citenamefont
  {Bomantara}, \citenamefont {Zhou}, \citenamefont {Pan},\ and\ \citenamefont
  {Gong}}]{Bomantara2019}%
  \BibitemOpen
  \bibfield  {author} {\bibinfo {author} {\bibfnamefont {Raditya~Weda}\
  \bibnamefont {Bomantara}}, \bibinfo {author} {\bibfnamefont {Longwen}\
  \bibnamefont {Zhou}}, \bibinfo {author} {\bibfnamefont {Jiaxin}\ \bibnamefont
  {Pan}}, \ and\ \bibinfo {author} {\bibfnamefont {Jiangbin}\ \bibnamefont
  {Gong}},\ }\bibfield  {title} {\enquote {\bibinfo {title} {Coupled-wire
  construction of static and floquet second-order topological insulators},}\
  }\href {\doibase 10.1103/PhysRevB.99.045441} {\bibfield  {journal} {\bibinfo
  {journal} {Phys. Rev. B}\ }\textbf {\bibinfo {volume} {99}},\ \bibinfo
  {pages} {045441} (\bibinfo {year} {2019})}\BibitemShut {NoStop}%
\bibitem [{\citenamefont {Nag}\ \emph {et~al.}(2019)\citenamefont {Nag},
  \citenamefont {Juri\ifmmode \check{c}\else \v{c}\fi{}i\ifmmode~\acute{c}\else
  \'{c}\fi{}},\ and\ \citenamefont {Roy}}]{Nag19}%
  \BibitemOpen
  \bibfield  {author} {\bibinfo {author} {\bibfnamefont {Tanay}\ \bibnamefont
  {Nag}}, \bibinfo {author} {\bibfnamefont {Vladimir}\ \bibnamefont
  {Juri\ifmmode \check{c}\else \v{c}\fi{}i\ifmmode~\acute{c}\else \'{c}\fi{}}},
  \ and\ \bibinfo {author} {\bibfnamefont {Bitan}\ \bibnamefont {Roy}},\
  }\bibfield  {title} {\enquote {\bibinfo {title} {Out of equilibrium
  higher-order topological insulator: Floquet engineering and quench
  dynamics},}\ }\href {\doibase 10.1103/PhysRevResearch.1.032045} {\bibfield
  {journal} {\bibinfo  {journal} {Phys. Rev. Research}\ }\textbf {\bibinfo
  {volume} {1}},\ \bibinfo {pages} {032045} (\bibinfo {year}
  {2019})}\BibitemShut {NoStop}%
\bibitem [{\citenamefont {Peng}\ and\ \citenamefont
  {Refael}(2019)}]{YangPRL2019}%
  \BibitemOpen
  \bibfield  {author} {\bibinfo {author} {\bibfnamefont {Yang}\ \bibnamefont
  {Peng}}\ and\ \bibinfo {author} {\bibfnamefont {Gil}\ \bibnamefont
  {Refael}},\ }\bibfield  {title} {\enquote {\bibinfo {title} {Floquet
  second-order topological insulators from nonsymmorphic space-time
  symmetries},}\ }\href {\doibase 10.1103/PhysRevLett.123.016806} {\bibfield
  {journal} {\bibinfo  {journal} {Phys. Rev. Lett.}\ }\textbf {\bibinfo
  {volume} {123}},\ \bibinfo {pages} {016806} (\bibinfo {year}
  {2019})}\BibitemShut {NoStop}%
\bibitem [{\citenamefont {Seshadri}\ \emph {et~al.}(2019)\citenamefont
  {Seshadri}, \citenamefont {Dutta},\ and\ \citenamefont {Sen}}]{Seshadri2019}%
  \BibitemOpen
  \bibfield  {author} {\bibinfo {author} {\bibfnamefont {Ranjani}\ \bibnamefont
  {Seshadri}}, \bibinfo {author} {\bibfnamefont {Anirban}\ \bibnamefont
  {Dutta}}, \ and\ \bibinfo {author} {\bibfnamefont {Diptiman}\ \bibnamefont
  {Sen}},\ }\bibfield  {title} {\enquote {\bibinfo {title} {Generating a
  second-order topological insulator with multiple corner states by periodic
  driving},}\ }\href {\doibase 10.1103/PhysRevB.100.115403} {\bibfield
  {journal} {\bibinfo  {journal} {Phys. Rev. B}\ }\textbf {\bibinfo {volume}
  {100}},\ \bibinfo {pages} {115403} (\bibinfo {year} {2019})}\BibitemShut
  {NoStop}%
\bibitem [{\citenamefont {Chaudhary}\ \emph {et~al.}()\citenamefont
  {Chaudhary}, \citenamefont {Haim}, \citenamefont {Peng},\ and\ \citenamefont
  {Refael}}]{chaudhary2019phononinduced}%
  \BibitemOpen
  \bibfield  {author} {\bibinfo {author} {\bibfnamefont {Swati}\ \bibnamefont
  {Chaudhary}}, \bibinfo {author} {\bibfnamefont {Arbel}\ \bibnamefont {Haim}},
  \bibinfo {author} {\bibfnamefont {Yang}\ \bibnamefont {Peng}}, \ and\
  \bibinfo {author} {\bibfnamefont {Gil}\ \bibnamefont {Refael}},\ }\bibfield
  {title} {\enquote {\bibinfo {title} {Phonon-induced floquet second-order
  topological phases protected by space-time symmetries},}\ }\href@noop {} {\
  }\Eprint {http://arxiv.org/abs/1911.07892} {arXiv:1911.07892
  [cond-mat.mes-hall]} \BibitemShut {NoStop}%
\bibitem [{\citenamefont {Rodriguez-Vega}\ \emph {et~al.}(2019)\citenamefont
  {Rodriguez-Vega}, \citenamefont {Kumar},\ and\ \citenamefont
  {Seradjeh}}]{Martin2019}%
  \BibitemOpen
  \bibfield  {author} {\bibinfo {author} {\bibfnamefont {Martin}\ \bibnamefont
  {Rodriguez-Vega}}, \bibinfo {author} {\bibfnamefont {Abhishek}\ \bibnamefont
  {Kumar}}, \ and\ \bibinfo {author} {\bibfnamefont {Babak}\ \bibnamefont
  {Seradjeh}},\ }\bibfield  {title} {\enquote {\bibinfo {title} {Higher-order
  floquet topological phases with corner and bulk bound states},}\ }\href
  {\doibase 10.1103/PhysRevB.100.085138} {\bibfield  {journal} {\bibinfo
  {journal} {Phys. Rev. B}\ }\textbf {\bibinfo {volume} {100}},\ \bibinfo
  {pages} {085138} (\bibinfo {year} {2019})}\BibitemShut {NoStop}%
\bibitem [{\citenamefont {Plekhanov}\ \emph {et~al.}(2019)\citenamefont
  {Plekhanov}, \citenamefont {Thakurathi}, \citenamefont {Loss},\ and\
  \citenamefont {Klinovaja}}]{Jelena2019}%
  \BibitemOpen
  \bibfield  {author} {\bibinfo {author} {\bibfnamefont {Kirill}\ \bibnamefont
  {Plekhanov}}, \bibinfo {author} {\bibfnamefont {Manisha}\ \bibnamefont
  {Thakurathi}}, \bibinfo {author} {\bibfnamefont {Daniel}\ \bibnamefont
  {Loss}}, \ and\ \bibinfo {author} {\bibfnamefont {Jelena}\ \bibnamefont
  {Klinovaja}},\ }\bibfield  {title} {\enquote {\bibinfo {title} {Floquet
  second-order topological superconductor driven via ferromagnetic
  resonance},}\ }\href {\doibase 10.1103/PhysRevResearch.1.032013} {\bibfield
  {journal} {\bibinfo  {journal} {Phys. Rev. Research}\ }\textbf {\bibinfo
  {volume} {1}},\ \bibinfo {pages} {032013} (\bibinfo {year}
  {2019})}\BibitemShut {NoStop}%
\bibitem [{\citenamefont {Ghosh}\ \emph {et~al.}(2020)\citenamefont {Ghosh},
  \citenamefont {Paul},\ and\ \citenamefont {Saha}}]{Ghosh2020}%
  \BibitemOpen
  \bibfield  {author} {\bibinfo {author} {\bibfnamefont {Arnob~Kumar}\
  \bibnamefont {Ghosh}}, \bibinfo {author} {\bibfnamefont {Ganesh~C.}\
  \bibnamefont {Paul}}, \ and\ \bibinfo {author} {\bibfnamefont {Arijit}\
  \bibnamefont {Saha}},\ }\bibfield  {title} {\enquote {\bibinfo {title}
  {Higher order topological insulator via periodic driving},}\ }\href {\doibase
  10.1103/PhysRevB.101.235403} {\bibfield  {journal} {\bibinfo  {journal}
  {Phys. Rev. B}\ }\textbf {\bibinfo {volume} {101}},\ \bibinfo {pages}
  {235403} (\bibinfo {year} {2020})}\BibitemShut {NoStop}%
\bibitem [{\citenamefont {Huang}\ and\ \citenamefont {Liu}(2020)}]{Huang2020}%
  \BibitemOpen
  \bibfield  {author} {\bibinfo {author} {\bibfnamefont {Biao}\ \bibnamefont
  {Huang}}\ and\ \bibinfo {author} {\bibfnamefont {W.~Vincent}\ \bibnamefont
  {Liu}},\ }\bibfield  {title} {\enquote {\bibinfo {title} {Floquet
  higher-order topological insulators with anomalous dynamical polarization},}\
  }\href {\doibase 10.1103/PhysRevLett.124.216601} {\bibfield  {journal}
  {\bibinfo  {journal} {Phys. Rev. Lett.}\ }\textbf {\bibinfo {volume} {124}},\
  \bibinfo {pages} {216601} (\bibinfo {year} {2020})}\BibitemShut {NoStop}%
\bibitem [{\citenamefont {Hu}\ \emph {et~al.}(2020)\citenamefont {Hu},
  \citenamefont {Huang}, \citenamefont {Zhao},\ and\ \citenamefont
  {Liu}}]{HuPRL2020}%
  \BibitemOpen
  \bibfield  {author} {\bibinfo {author} {\bibfnamefont {Haiping}\ \bibnamefont
  {Hu}}, \bibinfo {author} {\bibfnamefont {Biao}\ \bibnamefont {Huang}},
  \bibinfo {author} {\bibfnamefont {Erhai}\ \bibnamefont {Zhao}}, \ and\
  \bibinfo {author} {\bibfnamefont {W.~Vincent}\ \bibnamefont {Liu}},\
  }\bibfield  {title} {\enquote {\bibinfo {title} {Dynamical singularities of
  floquet higher-order topological insulators},}\ }\href {\doibase
  10.1103/PhysRevLett.124.057001} {\bibfield  {journal} {\bibinfo  {journal}
  {Phys. Rev. Lett.}\ }\textbf {\bibinfo {volume} {124}},\ \bibinfo {pages}
  {057001} (\bibinfo {year} {2020})}\BibitemShut {NoStop}%
\bibitem [{\citenamefont {Bomantara}\ and\ \citenamefont
  {Gong}(2020)}]{BomantaraPRB2020}%
  \BibitemOpen
  \bibfield  {author} {\bibinfo {author} {\bibfnamefont {Raditya~Weda}\
  \bibnamefont {Bomantara}}\ and\ \bibinfo {author} {\bibfnamefont {Jiangbin}\
  \bibnamefont {Gong}},\ }\bibfield  {title} {\enquote {\bibinfo {title}
  {Measurement-only quantum computation with floquet majorana corner modes},}\
  }\href {\doibase 10.1103/PhysRevB.101.085401} {\bibfield  {journal} {\bibinfo
   {journal} {Phys. Rev. B}\ }\textbf {\bibinfo {volume} {101}},\ \bibinfo
  {pages} {085401} (\bibinfo {year} {2020})}\BibitemShut {NoStop}%
\bibitem [{\citenamefont {Peng}(2020)}]{YangPRR2020}%
  \BibitemOpen
  \bibfield  {author} {\bibinfo {author} {\bibfnamefont {Yang}\ \bibnamefont
  {Peng}},\ }\bibfield  {title} {\enquote {\bibinfo {title} {Floquet
  higher-order topological insulators and superconductors with space-time
  symmetries},}\ }\href {\doibase 10.1103/PhysRevResearch.2.013124} {\bibfield
  {journal} {\bibinfo  {journal} {Phys. Rev. Research}\ }\textbf {\bibinfo
  {volume} {2}},\ \bibinfo {pages} {013124} (\bibinfo {year}
  {2020})}\BibitemShut {NoStop}%
\bibitem [{\citenamefont {Nag}\ \emph {et~al.}()\citenamefont {Nag},
  \citenamefont {Juri\ifmmode \check{c}\else \v{c}\fi{}i\ifmmode~\acute{c}\else
  \'{c}\fi{}},\ and\ \citenamefont {Roy}}]{Nag2020}%
  \BibitemOpen
  \bibfield  {author} {\bibinfo {author} {\bibfnamefont {T.}~\bibnamefont
  {Nag}}, \bibinfo {author} {\bibfnamefont {V.}~\bibnamefont {Juri\ifmmode
  \check{c}\else \v{c}\fi{}i\ifmmode~\acute{c}\else \'{c}\fi{}}}, \ and\
  \bibinfo {author} {\bibfnamefont {B.}~\bibnamefont {Roy}},\ }\bibfield
  {title} {\enquote {\bibinfo {title} {Hierarchy of higher-order floquet
  topological phases in three dimensions},}\ }\href@noop {} {\ }\Eprint
  {http://arxiv.org/abs/2009.10719} {arXiv:2009.10719 [cond-mat.mes-hall]}
  \BibitemShut {NoStop}%
\bibitem [{\citenamefont {Tiwari}\ \emph {et~al.}()\citenamefont {Tiwari},
  \citenamefont {Jahin},\ and\ \citenamefont {Wang}}]{ApoorvTiwari2020}%
  \BibitemOpen
  \bibfield  {author} {\bibinfo {author} {\bibfnamefont {A.}~\bibnamefont
  {Tiwari}}, \bibinfo {author} {\bibfnamefont {A.}~\bibnamefont {Jahin}}, \
  and\ \bibinfo {author} {\bibfnamefont {Y.}~\bibnamefont {Wang}},\ }\bibfield
  {title} {\enquote {\bibinfo {title} {Chiral dirac superconductors:
  Second-order and boundary-obstructed topology},}\ }\href@noop {} {\ }\Eprint
  {http://arxiv.org/abs/2005.12291} {arXiv:2005.12291 [cond-mat.mes-hall]}
  \BibitemShut {NoStop}%
\bibitem [{\citenamefont {Zhang}\ and\ \citenamefont {Yang}()}]{ZhangYang2020}%
  \BibitemOpen
  \bibfield  {author} {\bibinfo {author} {\bibfnamefont {R.~X.}\ \bibnamefont
  {Zhang}}\ and\ \bibinfo {author} {\bibfnamefont {Z.~C.}\ \bibnamefont
  {Yang}},\ }\bibfield  {title} {\enquote {\bibinfo {title} {Tunable fragile
  topology in floquet systems},}\ }\href@noop {} {\ }\Eprint
  {http://arxiv.org/abs/2005.08970} {arXiv:2005.08970 [cond-mat.mes-hall]}
  \BibitemShut {NoStop}%
\bibitem [{\citenamefont {Ghosh}\ \emph {et~al.}()\citenamefont {Ghosh},
  \citenamefont {Nag},\ and\ \citenamefont {Saha}}]{ghosh2020floquet}%
  \BibitemOpen
  \bibfield  {author} {\bibinfo {author} {\bibfnamefont {Arnob~Kumar}\
  \bibnamefont {Ghosh}}, \bibinfo {author} {\bibfnamefont {Tanay}\ \bibnamefont
  {Nag}}, \ and\ \bibinfo {author} {\bibfnamefont {Arijit}\ \bibnamefont
  {Saha}},\ }\bibfield  {title} {\enquote {\bibinfo {title} {Floquet generation
  of second order topological superconductor},}\ }\href@noop {} {\ }\Eprint
  {http://arxiv.org/abs/2009.11220} {arXiv:2009.11220 [cond-mat.mes-hall]}
  \BibitemShut {NoStop}%
\bibitem [{\citenamefont {Bhat}\ and\ \citenamefont
  {Bera}()}]{bhat2020equilibrium}%
  \BibitemOpen
  \bibfield  {author} {\bibinfo {author} {\bibfnamefont {Ruchira~V}\
  \bibnamefont {Bhat}}\ and\ \bibinfo {author} {\bibfnamefont {Soumya}\
  \bibnamefont {Bera}},\ }\bibfield  {title} {\enquote {\bibinfo {title} {Out
  of equilibrium chiral higher order topological insulator on a $\pi$-flux
  square lattice},}\ }\href@noop {} {\ }\Eprint
  {http://arxiv.org/abs/2011.01742} {arXiv:2011.01742 [cond-mat.mes-hall]}
  \BibitemShut {NoStop}%
\bibitem [{\citenamefont {Zhu}\ \emph {et~al.}()\citenamefont {Zhu},
  \citenamefont {Chong},\ and\ \citenamefont {Gong}}]{GongArxiv2020}%
  \BibitemOpen
  \bibfield  {author} {\bibinfo {author} {\bibfnamefont {Weiwei}\ \bibnamefont
  {Zhu}}, \bibinfo {author} {\bibfnamefont {Y.~D.}\ \bibnamefont {Chong}}, \
  and\ \bibinfo {author} {\bibfnamefont {Jiangbin}\ \bibnamefont {Gong}},\
  }\bibfield  {title} {\enquote {\bibinfo {title} {Floquet higher order
  topological insulator in a periodically driven bipartite lattice},}\
  }\href@noop {} {\ }\Eprint {http://arxiv.org/abs/2010.03879}
  {arXiv:2010.03879 [cond-mat.mes-hall]} \BibitemShut {NoStop}%
\bibitem [{\citenamefont {Zhang}\ \emph
  {et~al.}(2019{\natexlab{b}})\citenamefont {Zhang}, \citenamefont {Cole},
  \citenamefont {Wu},\ and\ \citenamefont
  {Das~Sarma}}]{PhysRevLett.123.167001}%
  \BibitemOpen
  \bibfield  {author} {\bibinfo {author} {\bibfnamefont {Rui-Xing}\
  \bibnamefont {Zhang}}, \bibinfo {author} {\bibfnamefont {William~S.}\
  \bibnamefont {Cole}}, \bibinfo {author} {\bibfnamefont {Xianxin}\
  \bibnamefont {Wu}}, \ and\ \bibinfo {author} {\bibfnamefont {S.}~\bibnamefont
  {Das~Sarma}},\ }\bibfield  {title} {\enquote {\bibinfo {title} {Higher-order
  topology and nodal topological superconductivity in fe(se,te)
  heterostructures},}\ }\href {\doibase 10.1103/PhysRevLett.123.167001}
  {\bibfield  {journal} {\bibinfo  {journal} {Phys. Rev. Lett.}\ }\textbf
  {\bibinfo {volume} {123}},\ \bibinfo {pages} {167001} (\bibinfo {year}
  {2019}{\natexlab{b}})}\BibitemShut {NoStop}%
\bibitem [{\citenamefont {Bomantara}(2020)}]{RWBomantaraPRR2020}%
  \BibitemOpen
  \bibfield  {author} {\bibinfo {author} {\bibfnamefont {Raditya~Weda}\
  \bibnamefont {Bomantara}},\ }\bibfield  {title} {\enquote {\bibinfo {title}
  {Time-induced second-order topological superconductors},}\ }\href {\doibase
  10.1103/PhysRevResearch.2.033495} {\bibfield  {journal} {\bibinfo  {journal}
  {Phys. Rev. Research}\ }\textbf {\bibinfo {volume} {2}},\ \bibinfo {pages}
  {033495} (\bibinfo {year} {2020})}\BibitemShut {NoStop}%
\bibitem [{\citenamefont {Wang}\ \emph
  {et~al.}(2013{\natexlab{a}})\citenamefont {Wang}, \citenamefont {Steinberg},
  \citenamefont {Jarillo-Herrero},\ and\ \citenamefont {Gedik}}]{Wang453}%
  \BibitemOpen
  \bibfield  {author} {\bibinfo {author} {\bibfnamefont {Y.~H.}\ \bibnamefont
  {Wang}}, \bibinfo {author} {\bibfnamefont {H.}~\bibnamefont {Steinberg}},
  \bibinfo {author} {\bibfnamefont {P.}~\bibnamefont {Jarillo-Herrero}}, \ and\
  \bibinfo {author} {\bibfnamefont {N.}~\bibnamefont {Gedik}},\ }\bibfield
  {title} {\enquote {\bibinfo {title} {Observation of floquet-bloch states on
  the surface of a topological insulator},}\ }\href {\doibase
  10.1126/science.1239834} {\bibfield  {journal} {\bibinfo  {journal}
  {Science}\ }\textbf {\bibinfo {volume} {342}},\ \bibinfo {pages} {453--457}
  (\bibinfo {year} {2013}{\natexlab{a}})}\BibitemShut {NoStop}%
\bibitem [{\citenamefont {Peng}\ \emph {et~al.}(2016)\citenamefont {Peng},
  \citenamefont {Qin}, \citenamefont {Zhao}, \citenamefont {Shen},
  \citenamefont {Xu}, \citenamefont {Bao}, \citenamefont {Jia},\ and\
  \citenamefont {Zhu}}]{peng2016experimental}%
  \BibitemOpen
  \bibfield  {author} {\bibinfo {author} {\bibfnamefont {Yu-Gui}\ \bibnamefont
  {Peng}}, \bibinfo {author} {\bibfnamefont {Cheng-Zhi}\ \bibnamefont {Qin}},
  \bibinfo {author} {\bibfnamefont {De-Gang}\ \bibnamefont {Zhao}}, \bibinfo
  {author} {\bibfnamefont {Ya-Xi}\ \bibnamefont {Shen}}, \bibinfo {author}
  {\bibfnamefont {Xiang-Yuan}\ \bibnamefont {Xu}}, \bibinfo {author}
  {\bibfnamefont {Ming}\ \bibnamefont {Bao}}, \bibinfo {author} {\bibfnamefont
  {Han}\ \bibnamefont {Jia}}, \ and\ \bibinfo {author} {\bibfnamefont
  {Xue-Feng}\ \bibnamefont {Zhu}},\ }\bibfield  {title} {\enquote {\bibinfo
  {title} {Experimental demonstration of anomalous floquet topological
  insulator for sound},}\ }\href {\doibase https://doi.org/10.1038/ncomms13368}
  {\bibfield  {journal} {\bibinfo  {journal} {Nature communications}\ }\textbf
  {\bibinfo {volume} {7}},\ \bibinfo {pages} {13368} (\bibinfo {year}
  {2016})}\BibitemShut {NoStop}%
\bibitem [{\citenamefont {Lejman}\ \emph {et~al.}(2014)\citenamefont {Lejman},
  \citenamefont {Vaudel}, \citenamefont {Infante}, \citenamefont {Gemeiner},
  \citenamefont {Gusev}, \citenamefont {Dkhil},\ and\ \citenamefont
  {Ruello}}]{Experiment2014}%
  \BibitemOpen
  \bibfield  {author} {\bibinfo {author} {\bibfnamefont {M.}~\bibnamefont
  {Lejman}}, \bibinfo {author} {\bibfnamefont {G.}~\bibnamefont {Vaudel}},
  \bibinfo {author} {\bibfnamefont {I.~C.}\ \bibnamefont {Infante}}, \bibinfo
  {author} {\bibfnamefont {P.}~\bibnamefont {Gemeiner}}, \bibinfo {author}
  {\bibfnamefont {V.~E.}\ \bibnamefont {Gusev}}, \bibinfo {author}
  {\bibfnamefont {B.}~\bibnamefont {Dkhil}}, \ and\ \bibinfo {author}
  {\bibfnamefont {P.}~\bibnamefont {Ruello}},\ }\bibfield  {title} {\enquote
  {\bibinfo {title} {Giant ultrafast photo-induced shear strain in
  ferroelectric bifeo3},}\ }\href {https://doi.org/10.1038/ncomms5301}
  {\bibfield  {journal} {\bibinfo  {journal} {Nature Communications}\ }\textbf
  {\bibinfo {volume} {5}},\ \bibinfo {pages} {4301} (\bibinfo {year}
  {2014})}\BibitemShut {NoStop}%
\bibitem [{\citenamefont {Fleury}\ \emph {et~al.}(2016)\citenamefont {Fleury},
  \citenamefont {Khanikaev},\ and\ \citenamefont {Alu}}]{fleury2016floquet}%
  \BibitemOpen
  \bibfield  {author} {\bibinfo {author} {\bibfnamefont {Romain}\ \bibnamefont
  {Fleury}}, \bibinfo {author} {\bibfnamefont {Alexander~B}\ \bibnamefont
  {Khanikaev}}, \ and\ \bibinfo {author} {\bibfnamefont {Andrea}\ \bibnamefont
  {Alu}},\ }\bibfield  {title} {\enquote {\bibinfo {title} {Floquet topological
  insulators for sound},}\ }\href@noop {} {\bibfield  {journal} {\bibinfo
  {journal} {Nature communications}\ }\textbf {\bibinfo {volume} {7}},\
  \bibinfo {pages} {11744} (\bibinfo {year} {2016})}\BibitemShut {NoStop}%
\bibitem [{\citenamefont {\ifmmode \check{C}\else
  \v{C}\fi{}ade\ifmmode~\check{z}\else \v{z}\fi{}}\ \emph
  {et~al.}(2019)\citenamefont {\ifmmode \check{C}\else
  \v{C}\fi{}ade\ifmmode~\check{z}\else \v{z}\fi{}}, \citenamefont {Mondaini},\
  and\ \citenamefont {Sacramento}}]{mass_drive_FTS}%
  \BibitemOpen
  \bibfield  {author} {\bibinfo {author} {\bibfnamefont {Tilen}\ \bibnamefont
  {\ifmmode \check{C}\else \v{C}\fi{}ade\ifmmode~\check{z}\else \v{z}\fi{}}},
  \bibinfo {author} {\bibfnamefont {Rubem}\ \bibnamefont {Mondaini}}, \ and\
  \bibinfo {author} {\bibfnamefont {Pedro~D.}\ \bibnamefont {Sacramento}},\
  }\bibfield  {title} {\enquote {\bibinfo {title} {Edge and bulk localization
  of floquet topological superconductors},}\ }\href {\doibase
  10.1103/PhysRevB.99.014301} {\bibfield  {journal} {\bibinfo  {journal} {Phys.
  Rev. B}\ }\textbf {\bibinfo {volume} {99}},\ \bibinfo {pages} {014301}
  (\bibinfo {year} {2019})}\BibitemShut {NoStop}%
\bibitem [{\citenamefont {Nag}\ and\ \citenamefont {Roy}()}]{nag20}%
  \BibitemOpen
  \bibfield  {author} {\bibinfo {author} {\bibfnamefont {Tanay}\ \bibnamefont
  {Nag}}\ and\ \bibinfo {author} {\bibfnamefont {Bitan}\ \bibnamefont {Roy}},\
  }\bibfield  {title} {\enquote {\bibinfo {title} {Anomalous and normal
  dislocation modes in floquet topological insulators},}\ }\href@noop {} {\
  }\Eprint {http://arxiv.org/abs/2010.11952} {arXiv:2010.11952
  [cond-mat.meso-hall]} \BibitemShut {NoStop}%
\bibitem [{\citenamefont {Nag}\ \emph {et~al.}(2014)\citenamefont {Nag},
  \citenamefont {Roy}, \citenamefont {Dutta},\ and\ \citenamefont
  {Sen}}]{mass_kick_DL1}%
  \BibitemOpen
  \bibfield  {author} {\bibinfo {author} {\bibfnamefont {Tanay}\ \bibnamefont
  {Nag}}, \bibinfo {author} {\bibfnamefont {Sthitadhi}\ \bibnamefont {Roy}},
  \bibinfo {author} {\bibfnamefont {Amit}\ \bibnamefont {Dutta}}, \ and\
  \bibinfo {author} {\bibfnamefont {Diptiman}\ \bibnamefont {Sen}},\ }\bibfield
   {title} {\enquote {\bibinfo {title} {Dynamical localization in a chain of
  hard core bosons under periodic driving},}\ }\href {\doibase
  10.1103/PhysRevB.89.165425} {\bibfield  {journal} {\bibinfo  {journal} {Phys.
  Rev. B}\ }\textbf {\bibinfo {volume} {89}},\ \bibinfo {pages} {165425}
  (\bibinfo {year} {2014})}\BibitemShut {NoStop}%
\bibitem [{\citenamefont {Agarwala}\ \emph {et~al.}(2016)\citenamefont
  {Agarwala}, \citenamefont {Bhattacharya}, \citenamefont {Dutta},\ and\
  \citenamefont {Sen}}]{mass_kick_DL2}%
  \BibitemOpen
  \bibfield  {author} {\bibinfo {author} {\bibfnamefont {Adhip}\ \bibnamefont
  {Agarwala}}, \bibinfo {author} {\bibfnamefont {Utso}\ \bibnamefont
  {Bhattacharya}}, \bibinfo {author} {\bibfnamefont {Amit}\ \bibnamefont
  {Dutta}}, \ and\ \bibinfo {author} {\bibfnamefont {Diptiman}\ \bibnamefont
  {Sen}},\ }\bibfield  {title} {\enquote {\bibinfo {title} {Effects of periodic
  kicking on dispersion and wave packet dynamics in graphene},}\ }\href
  {\doibase 10.1103/PhysRevB.93.174301} {\bibfield  {journal} {\bibinfo
  {journal} {Phys. Rev. B}\ }\textbf {\bibinfo {volume} {93}},\ \bibinfo
  {pages} {174301} (\bibinfo {year} {2016})}\BibitemShut {NoStop}%
\bibitem [{\citenamefont {{Patel, Aavishkar A.}}\ \emph
  {et~al.}(2013)\citenamefont {{Patel, Aavishkar A.}}, \citenamefont {{Sharma,
  Shraddha}},\ and\ \citenamefont {{Dutta, Amit}}}]{refId0}%
  \BibitemOpen
  \bibfield  {author} {\bibinfo {author} {\bibnamefont {{Patel, Aavishkar
  A.}}}, \bibinfo {author} {\bibnamefont {{Sharma, Shraddha}}}, \ and\ \bibinfo
  {author} {\bibnamefont {{Dutta, Amit}}},\ }\bibfield  {title} {\enquote
  {\bibinfo {title} {Quench dynamics of edge states in 2-d topological
  insulator ribbons},}\ }\href {\doibase 10.1140/epjb/e2013-40657-2} {\bibfield
   {journal} {\bibinfo  {journal} {Eur. Phys. J. B}\ }\textbf {\bibinfo
  {volume} {86}},\ \bibinfo {pages} {367} (\bibinfo {year} {2013})}\BibitemShut
  {NoStop}%
\bibitem [{\citenamefont {Rajak}\ and\ \citenamefont {Dutta}(2014)}]{ps2}%
  \BibitemOpen
  \bibfield  {author} {\bibinfo {author} {\bibfnamefont {Atanu}\ \bibnamefont
  {Rajak}}\ and\ \bibinfo {author} {\bibfnamefont {Amit}\ \bibnamefont
  {Dutta}},\ }\bibfield  {title} {\enquote {\bibinfo {title} {Survival
  probability of an edge majorana in a one-dimensional $p$-wave superconducting
  chain under sudden quenching of parameters},}\ }\href {\doibase
  10.1103/PhysRevE.89.042125} {\bibfield  {journal} {\bibinfo  {journal} {Phys.
  Rev. E}\ }\textbf {\bibinfo {volume} {89}},\ \bibinfo {pages} {042125}
  (\bibinfo {year} {2014})}\BibitemShut {NoStop}%
\bibitem [{\citenamefont {Wu}\ \emph {et~al.}(2018)\citenamefont {Wu},
  \citenamefont {Fatemi}, \citenamefont {Gibson}, \citenamefont {Watanabe},
  \citenamefont {Taniguchi}, \citenamefont {Cava},\ and\ \citenamefont
  {Jarillo-Herrero}}]{Wu76}%
  \BibitemOpen
  \bibfield  {author} {\bibinfo {author} {\bibfnamefont {Sanfeng}\ \bibnamefont
  {Wu}}, \bibinfo {author} {\bibfnamefont {Valla}\ \bibnamefont {Fatemi}},
  \bibinfo {author} {\bibfnamefont {Quinn~D.}\ \bibnamefont {Gibson}}, \bibinfo
  {author} {\bibfnamefont {Kenji}\ \bibnamefont {Watanabe}}, \bibinfo {author}
  {\bibfnamefont {Takashi}\ \bibnamefont {Taniguchi}}, \bibinfo {author}
  {\bibfnamefont {Robert~J.}\ \bibnamefont {Cava}}, \ and\ \bibinfo {author}
  {\bibfnamefont {Pablo}\ \bibnamefont {Jarillo-Herrero}},\ }\bibfield  {title}
  {\enquote {\bibinfo {title} {Observation of the quantum spin hall effect up
  to 100 kelvin in a monolayer crystal},}\ }\href {\doibase
  10.1126/science.aan6003} {\bibfield  {journal} {\bibinfo  {journal}
  {Science}\ }\textbf {\bibinfo {volume} {359}},\ \bibinfo {pages} {76--79}
  (\bibinfo {year} {2018})}\BibitemShut {NoStop}%
\bibitem [{\citenamefont {Qian}\ \emph {et~al.}(2014)\citenamefont {Qian},
  \citenamefont {Liu}, \citenamefont {Fu},\ and\ \citenamefont
  {Li}}]{Qian1344}%
  \BibitemOpen
  \bibfield  {author} {\bibinfo {author} {\bibfnamefont {Xiaofeng}\
  \bibnamefont {Qian}}, \bibinfo {author} {\bibfnamefont {Junwei}\ \bibnamefont
  {Liu}}, \bibinfo {author} {\bibfnamefont {Liang}\ \bibnamefont {Fu}}, \ and\
  \bibinfo {author} {\bibfnamefont {Ju}~\bibnamefont {Li}},\ }\bibfield
  {title} {\enquote {\bibinfo {title} {Quantum spin hall effect in
  two-dimensional transition metal dichalcogenides},}\ }\href {\doibase
  10.1126/science.1256815} {\bibfield  {journal} {\bibinfo  {journal}
  {Science}\ }\textbf {\bibinfo {volume} {346}},\ \bibinfo {pages} {1344--1347}
  (\bibinfo {year} {2014})}\BibitemShut {NoStop}%
\bibitem [{\citenamefont {Fu}\ \emph {et~al.}()\citenamefont {Fu},
  \citenamefont {Hu}, \citenamefont {Li}, \citenamefont {Li},\ and\
  \citenamefont {Shen}}]{fu2020chiral}%
  \BibitemOpen
  \bibfield  {author} {\bibinfo {author} {\bibfnamefont {Bo}~\bibnamefont
  {Fu}}, \bibinfo {author} {\bibfnamefont {Zi-Ang}\ \bibnamefont {Hu}},
  \bibinfo {author} {\bibfnamefont {Chang-An}\ \bibnamefont {Li}}, \bibinfo
  {author} {\bibfnamefont {Jian}\ \bibnamefont {Li}}, \ and\ \bibinfo {author}
  {\bibfnamefont {Shun-Qing}\ \bibnamefont {Shen}},\ }\bibfield  {title}
  {\enquote {\bibinfo {title} {Chiral majorana hinge modes in superconducting
  dirac materials},}\ }\href@noop {} {\ }\Eprint
  {http://arxiv.org/abs/2010.15633} {arXiv:2010.15633 [cond-mat.supr-con]}
  \BibitemShut {NoStop}%
\bibitem [{\citenamefont {Wang}\ \emph
  {et~al.}(2013{\natexlab{b}})\citenamefont {Wang} \emph
  {et~al.}}]{ExperimentTI.dwave}%
  \BibitemOpen
  \bibfield  {author} {\bibinfo {author} {\bibfnamefont {Ding H. Fedorov~A.}\
  \bibnamefont {Wang}, \bibfnamefont {E.}} \emph {et~al.},\ }\bibfield  {title}
  {\enquote {\bibinfo {title} {Fully gapped topological surface states in
  bi2se3 films induced by a d-wave high-temperature superconductor},}\ }\href
  {https://doi.org/10.1038/nphys2744} {\bibfield  {journal} {\bibinfo
  {journal} {Nature Phys.}\ }\textbf {\bibinfo {volume} {9}},\ \bibinfo {pages}
  {621--625} (\bibinfo {year} {2013}{\natexlab{b}})}\BibitemShut {NoStop}%
\bibitem [{\citenamefont {Liu}\ \emph {et~al.}(2010)\citenamefont {Liu},
  \citenamefont {Zhang}, \citenamefont {Yan}, \citenamefont {Qi}, \citenamefont
  {Frauenheim}, \citenamefont {Dai}, \citenamefont {Fang},\ and\ \citenamefont
  {Zhang}}]{PhysRevB.81.041307}%
  \BibitemOpen
  \bibfield  {author} {\bibinfo {author} {\bibfnamefont {Chao-Xing}\
  \bibnamefont {Liu}}, \bibinfo {author} {\bibfnamefont {HaiJun}\ \bibnamefont
  {Zhang}}, \bibinfo {author} {\bibfnamefont {Binghai}\ \bibnamefont {Yan}},
  \bibinfo {author} {\bibfnamefont {Xiao-Liang}\ \bibnamefont {Qi}}, \bibinfo
  {author} {\bibfnamefont {Thomas}\ \bibnamefont {Frauenheim}}, \bibinfo
  {author} {\bibfnamefont {Xi}~\bibnamefont {Dai}}, \bibinfo {author}
  {\bibfnamefont {Zhong}\ \bibnamefont {Fang}}, \ and\ \bibinfo {author}
  {\bibfnamefont {Shou-Cheng}\ \bibnamefont {Zhang}},\ }\bibfield  {title}
  {\enquote {\bibinfo {title} {Oscillatory crossover from two-dimensional to
  three-dimensional topological insulators},}\ }\href {\doibase
  10.1103/PhysRevB.81.041307} {\bibfield  {journal} {\bibinfo  {journal} {Phys.
  Rev. B}\ }\textbf {\bibinfo {volume} {81}},\ \bibinfo {pages} {041307}
  (\bibinfo {year} {2010})}\BibitemShut {NoStop}%
\bibitem [{\citenamefont {Zhang}\ \emph {et~al.}(2010)\citenamefont {Zhang},
  \citenamefont {He}, \citenamefont {Chang}, \citenamefont {Song},
  \citenamefont {Wang}, \citenamefont {Chen}, \citenamefont {Jia},
  \citenamefont {Fang}, \citenamefont {Dai}, \citenamefont {Shan} \emph
  {et~al.}}]{zhang2010crossover}%
  \BibitemOpen
  \bibfield  {author} {\bibinfo {author} {\bibfnamefont {Yi}~\bibnamefont
  {Zhang}}, \bibinfo {author} {\bibfnamefont {Ke}~\bibnamefont {He}}, \bibinfo
  {author} {\bibfnamefont {Cui-Zu}\ \bibnamefont {Chang}}, \bibinfo {author}
  {\bibfnamefont {Can-Li}\ \bibnamefont {Song}}, \bibinfo {author}
  {\bibfnamefont {Li-Li}\ \bibnamefont {Wang}}, \bibinfo {author}
  {\bibfnamefont {Xi}~\bibnamefont {Chen}}, \bibinfo {author} {\bibfnamefont
  {Jin-Feng}\ \bibnamefont {Jia}}, \bibinfo {author} {\bibfnamefont {Zhong}\
  \bibnamefont {Fang}}, \bibinfo {author} {\bibfnamefont {Xi}~\bibnamefont
  {Dai}}, \bibinfo {author} {\bibfnamefont {Wen-Yu}\ \bibnamefont {Shan}},
  \emph {et~al.},\ }\bibfield  {title} {\enquote {\bibinfo {title} {Crossover
  of the three-dimensional topological insulator bi 2 se 3 to the
  two-dimensional limit},}\ }\href {\doibase 10.1038/nphys1689} {\bibfield
  {journal} {\bibinfo  {journal} {Nature Physics}\ }\textbf {\bibinfo {volume}
  {6}},\ \bibinfo {pages} {584--588} (\bibinfo {year} {2010})}\BibitemShut
  {NoStop}%
\bibitem [{\citenamefont {McIver}\ \emph {et~al.}(2020)\citenamefont {McIver},
  \citenamefont {Schulte}, \citenamefont {Stein} \emph
  {et~al.}}]{Experiment_LightInducedHall}%
  \BibitemOpen
  \bibfield  {author} {\bibinfo {author} {\bibfnamefont {J.~W.}\ \bibnamefont
  {McIver}}, \bibinfo {author} {\bibfnamefont {B.}~\bibnamefont {Schulte}},
  \bibinfo {author} {\bibfnamefont {FU.}\ \bibnamefont {Stein}},  \emph
  {et~al.},\ }\bibfield  {title} {\enquote {\bibinfo {title} {Light-induced
  anomalous hall effect in graphene},}\ }\href
  {https://doi.org/10.1038/s41567-019-0698-y} {\bibfield  {journal} {\bibinfo
  {journal} {Nature Phys.}\ }\textbf {\bibinfo {volume} {16}},\ \bibinfo
  {pages} {38--41} (\bibinfo {year} {2020})}\BibitemShut {NoStop}%
\bibitem [{\citenamefont {Nadj-Perge}\ \emph {et~al.}(2014)\citenamefont
  {Nadj-Perge}, \citenamefont {Drozdov}, \citenamefont {Li}, \citenamefont
  {Chen}, \citenamefont {Jeon}, \citenamefont {Seo}, \citenamefont {MacDonald},
  \citenamefont {Andrei~Bernevig},\ and\ \citenamefont
  {Yazdani}}]{Experiment.MZM.STM}%
  \BibitemOpen
  \bibfield  {author} {\bibinfo {author} {\bibfnamefont {S.}~\bibnamefont
  {Nadj-Perge}}, \bibinfo {author} {\bibfnamefont {I.~K.}\ \bibnamefont
  {Drozdov}}, \bibinfo {author} {\bibfnamefont {J.}~\bibnamefont {Li}},
  \bibinfo {author} {\bibfnamefont {H.}~\bibnamefont {Chen}}, \bibinfo {author}
  {\bibfnamefont {S.}~\bibnamefont {Jeon}}, \bibinfo {author} {\bibfnamefont
  {J.}~\bibnamefont {Seo}}, \bibinfo {author} {\bibfnamefont {A.~H.}\
  \bibnamefont {MacDonald}}, \bibinfo {author} {\bibfnamefont {B.}~\bibnamefont
  {Andrei~Bernevig}}, \ and\ \bibinfo {author} {\bibfnamefont {A.}~\bibnamefont
  {Yazdani}},\ }\bibfield  {title} {\enquote {\bibinfo {title} {Observation of
  majorana fermions in ferromagnetic atomic chains on a superconductor},}\
  }\href {https://doi.org/10.1126/science.1259327} {\bibfield  {journal}
  {\bibinfo  {journal} {Science}\ }\textbf {\bibinfo {volume} {346}},\ \bibinfo
  {pages} {602} (\bibinfo {year} {2014})}\BibitemShut {NoStop}%
\bibitem [{\citenamefont {Sato}\ \emph {et~al.}(2019)\citenamefont {Sato},
  \citenamefont {McIver}, \citenamefont {Nuske}, \citenamefont {Tang},
  \citenamefont {Jotzu}, \citenamefont {Schulte}, \citenamefont {H\"ubener},
  \citenamefont {De~Giovannini}, \citenamefont {Mathey}, \citenamefont
  {Sentef}, \citenamefont {Cavalleri},\ and\ \citenamefont
  {Rubio}}]{PhysRevB.99.214302}%
  \BibitemOpen
  \bibfield  {author} {\bibinfo {author} {\bibfnamefont {S.~A.}\ \bibnamefont
  {Sato}}, \bibinfo {author} {\bibfnamefont {J.~W.}\ \bibnamefont {McIver}},
  \bibinfo {author} {\bibfnamefont {M.}~\bibnamefont {Nuske}}, \bibinfo
  {author} {\bibfnamefont {P.}~\bibnamefont {Tang}}, \bibinfo {author}
  {\bibfnamefont {G.}~\bibnamefont {Jotzu}}, \bibinfo {author} {\bibfnamefont
  {B.}~\bibnamefont {Schulte}}, \bibinfo {author} {\bibfnamefont
  {H.}~\bibnamefont {H\"ubener}}, \bibinfo {author} {\bibfnamefont
  {U.}~\bibnamefont {De~Giovannini}}, \bibinfo {author} {\bibfnamefont
  {L.}~\bibnamefont {Mathey}}, \bibinfo {author} {\bibfnamefont {M.~A.}\
  \bibnamefont {Sentef}}, \bibinfo {author} {\bibfnamefont {A.}~\bibnamefont
  {Cavalleri}}, \ and\ \bibinfo {author} {\bibfnamefont {A.}~\bibnamefont
  {Rubio}},\ }\bibfield  {title} {\enquote {\bibinfo {title} {Microscopic
  theory for the light-induced anomalous hall effect in graphene},}\ }\href
  {\doibase 10.1103/PhysRevB.99.214302} {\bibfield  {journal} {\bibinfo
  {journal} {Phys. Rev. B}\ }\textbf {\bibinfo {volume} {99}},\ \bibinfo
  {pages} {214302} (\bibinfo {year} {2019})}\BibitemShut {NoStop}%
\bibitem [{\citenamefont {Tuovinen}\ \emph {et~al.}(2019)\citenamefont
  {Tuovinen}, \citenamefont {Perfetto}, \citenamefont {van Leeuwen},
  \citenamefont {Stefanucci},\ and\ \citenamefont {Sentef}}]{Tuovinen_2019}%
  \BibitemOpen
  \bibfield  {author} {\bibinfo {author} {\bibfnamefont {Riku}\ \bibnamefont
  {Tuovinen}}, \bibinfo {author} {\bibfnamefont {Enrico}\ \bibnamefont
  {Perfetto}}, \bibinfo {author} {\bibfnamefont {Robert}\ \bibnamefont {van
  Leeuwen}}, \bibinfo {author} {\bibfnamefont {Gianluca}\ \bibnamefont
  {Stefanucci}}, \ and\ \bibinfo {author} {\bibfnamefont {Michael~A}\
  \bibnamefont {Sentef}},\ }\bibfield  {title} {\enquote {\bibinfo {title}
  {Distinguishing majorana zero modes from impurity states through
  time-resolved transport},}\ }\href {\doibase 10.1088/1367-2630/ab4ab7}
  {\bibfield  {journal} {\bibinfo  {journal} {New Journal of Physics}\ }\textbf
  {\bibinfo {volume} {21}},\ \bibinfo {pages} {103038} (\bibinfo {year}
  {2019})}\BibitemShut {NoStop}%
\bibitem [{\citenamefont {Abanin}\ \emph {et~al.}(2015)\citenamefont {Abanin},
  \citenamefont {De~Roeck},\ and\ \citenamefont {Huveneers}}]{heating1}%
  \BibitemOpen
  \bibfield  {author} {\bibinfo {author} {\bibfnamefont {Dmitry~A.}\
  \bibnamefont {Abanin}}, \bibinfo {author} {\bibfnamefont {Wojciech}\
  \bibnamefont {De~Roeck}}, \ and\ \bibinfo {author} {\bibfnamefont
  {Fran\ifmmode \mbox{\c{c}}\else~\c{c}\fi{}ois}\ \bibnamefont {Huveneers}},\
  }\bibfield  {title} {\enquote {\bibinfo {title} {Exponentially slow heating
  in periodically driven many-body systems},}\ }\href {\doibase
  10.1103/PhysRevLett.115.256803} {\bibfield  {journal} {\bibinfo  {journal}
  {Phys. Rev. Lett.}\ }\textbf {\bibinfo {volume} {115}},\ \bibinfo {pages}
  {256803} (\bibinfo {year} {2015})}\BibitemShut {NoStop}%
\bibitem [{\citenamefont {Messer}\ \emph {et~al.}(2018)\citenamefont {Messer},
  \citenamefont {Sandholzer}, \citenamefont {G\"org}, \citenamefont {Minguzzi},
  \citenamefont {Desbuquois},\ and\ \citenamefont {Esslinger}}]{heating2}%
  \BibitemOpen
  \bibfield  {author} {\bibinfo {author} {\bibfnamefont {Michael}\ \bibnamefont
  {Messer}}, \bibinfo {author} {\bibfnamefont {Kilian}\ \bibnamefont
  {Sandholzer}}, \bibinfo {author} {\bibfnamefont {Frederik}\ \bibnamefont
  {G\"org}}, \bibinfo {author} {\bibfnamefont {Joaqu\'{\i}n}\ \bibnamefont
  {Minguzzi}}, \bibinfo {author} {\bibfnamefont {R\'emi}\ \bibnamefont
  {Desbuquois}}, \ and\ \bibinfo {author} {\bibfnamefont {Tilman}\ \bibnamefont
  {Esslinger}},\ }\bibfield  {title} {\enquote {\bibinfo {title} {Floquet
  dynamics in driven fermi-hubbard systems},}\ }\href {\doibase
  10.1103/PhysRevLett.121.233603} {\bibfield  {journal} {\bibinfo  {journal}
  {Phys. Rev. Lett.}\ }\textbf {\bibinfo {volume} {121}},\ \bibinfo {pages}
  {233603} (\bibinfo {year} {2018})}\BibitemShut {NoStop}%
\bibitem [{\citenamefont {Bukov}\ \emph {et~al.}(2015)\citenamefont {Bukov},
  \citenamefont {D'Alessio},\ and\ \citenamefont {Polkovnikov}}]{heating3}%
  \BibitemOpen
  \bibfield  {author} {\bibinfo {author} {\bibfnamefont {Marin}\ \bibnamefont
  {Bukov}}, \bibinfo {author} {\bibfnamefont {Luca}\ \bibnamefont {D'Alessio}},
  \ and\ \bibinfo {author} {\bibfnamefont {Anatoli}\ \bibnamefont
  {Polkovnikov}},\ }\bibfield  {title} {\enquote {\bibinfo {title} {Universal
  high-frequency behavior of periodically driven systems: from dynamical
  stabilization to floquet engineering},}\ }\href {\doibase
  10.1080/00018732.2015.1055918} {\bibfield  {journal} {\bibinfo  {journal}
  {Advances in Physics}\ }\textbf {\bibinfo {volume} {64}},\ \bibinfo {pages}
  {139--226} (\bibinfo {year} {2015})}\BibitemShut {NoStop}%
\end{thebibliography}%

\end{document}